\documentclass[manuscript]{aastex}

\shorttitle{S5 0716+714}
 \shortauthors{Poon et al.}

\begin{document}


\title{The optical microvariability and spectral changes of the\\
     BL Lacertae object S5 0716+714}


\author{H. Poon}
\affil{Astronomy Department, Beijing Normal University,
    Beijing 100875, China}\email{china\_108@yahoo.com}

\author{J. H. Fan}
\affil{Center for Astrophysics, Guangzhou University, Guangzhou
510006, China}

\and

\author{J. N. Fu}
\affil{Astronomy Department, Beijing Normal University,
    Beijing 100875, China}



\begin{abstract}
We monitored the BL Lac object S5 0716+714 in the optical band
during October 2008, December 2008 and February 2009 with a best
temporal resolution of about 5 minutes in the \emph{BVRI} bands.
Four fast flares were observed with amplitudes ranging from 0.3 to
0.75 mag. The source remained active during the whole monitoring
campaign, showing microvariability in all days except for one. The
overall variability amplitudes are $\Delta$\emph{B} $\sim$
0$^{m}$.89, $\Delta$\emph{V} $\sim$ 0$^{m}$.80, $\Delta$\emph{R}
$\sim$ 0$^{m}$.73 and $\Delta$\emph{I} $\sim$ 0$^{m}$.51. Typical
timescales of microvariability range from 2 to 8 hours. The overall
\emph{V} - \emph{R} color index ranges from 0.37 to 0.59. Strong
bluer-when-brighter chromatism was found on internight timescales.
However, different spectral behavior was found on intranight
timescales. A possible time lag of $\sim$ 11 mins between \emph{B}
and \emph{I} bands was found on one night. The shock-in-jet model
and geometric effects can be applied to explain the source's
intranight behavior.
\end{abstract}





\section{Introduction}
Blazars represent an extreme subclass of active galactic nuclei.
They are characterized by rapid and strong variability throughout
the entire electromagnetic wavebands, high and variable
polarization($>$3\%) from radio to optical wavelengths. In the
unified model of AGNs, blazars make an angle of less than
$10\,^{\circ}$
 from the line of sight (Urry \& Padovani 1995). For low-energy peaked (or radio-selected) blazars
 the continuum from radio through the UV or soft X-rays is mainly contributed by synchrotron radiation,
 while a second hump in the spectrum at higher energies usually peaks in the GeV $\gamma$-ray
 band.¡±\
 Blazars exhibit variability on different timescales, from years
down to hours or less (See Fan et al. 2005). Understanding blazar
variability is one of the major issues of active galactic nuclei
studies. There may be different behavior on different timescales.
Periodicity may be found on the long term, as is the case of OJ 287
(Sillanpaa et al. 1988) and other blazars (Fan et al. 2002, 2007).
The periodicity of OJ 287 can be explained by a binary black hole
model, with the secondary black hole passing through the accretion
disk of the primary black hole twice per revolution (Sillanpaa et.
al 1988; Lehto \& Valtonen 1996). On shorter timescales, 3C 66A was
claimed to have an optical period of $\sim$ 65 days during its
bright state (Lainela et al. 1999), while Mkn 501 may have displayed
a period of 23 days in high energy data (Osone 2006). Recently,
analyses of X-ray data have yielded excellent evidence for a
quasi-period of about an hour for 3C 273 (Espaillat et al. 2008) and
very good evidence for near periods of $\sim$ 16 days for AO
0235+164 and $\sim$ 420 days for 1ES 2321+419 (Rani et al. 2009).
This may be due to a shock wave moving along a helical path in the
relativistic jet (e.g., Marscher 1996) or the unstability in the
accretion disk (Fan et al. 2008) for Mkn 501. Different viewing
angles may also result in different level of brightness, with the
source being dimmer at larger viewing angles. BL Lacertae shows
variability on both intranight and internight timescales with
different spectral behaviors. It shows an intranight strongly
chromatic trend and an internight midly chromatic trend, indicating
two different components operating in the engine (Villata et al.
2004; Hu et al. 2006). Understanding the shortest variability
timescale is of special importance. Brightness changes of up to a
few tenths of a magnitude over the course of a night or less is
known as intra-night optical variability (INOV) (Wagner \& Witzel
1995) or the so-called microvariability. It can bring new insight
into the understanding of the nature of blazars, probing into the
innermost structure down to microparsecs, putting constraints on the
size of the source and its physical environments.

Microvariability was first discovered in the sixties by Matthews and
Sandage (1963), who found 3C 48 changed $0^{m}$.04 in the \emph{V}
band in 15 minutes.
 But their results were not taken seriously and were considered due to instrumental errors. However, with the development of CCD, microvariability
  was confirmed to be the intrinsic nature of active galactic nuclei, especially for blazars (e.g., Miller et al. 1989). From a complete sample of BL Lac objects
  (Stickel et al. 1991), Heidt and Wagner (1996) found 80\% exhibited microvariability, proving it to be the nature of BL Lac objects.
Romero et al. (1999) detected microvariability in 60\% of their
selected sample of 23 southern AGNs. Microvariability became
extensively observed since the eighties. Up to now, the reasons for
it are still unclear. Many different models have been proposed to
explain this phenomenon. It conceivably may be due to eclipses if
there is a binary black hole system (Xie et al. 2002). The
interaction of relativistic shock waves and inhomogeneous jet can
also cause microvariability in the jet (Maraschi et al. 1989; Qian
et al. 1991; Marscher 1992; Marscher 1996). Other models like
instabilities and perturbations in the accretion disk can explain
some of the microvariability phenomena (e.g., Mangalam \& Wiita
1993; Wiita 1996; Fan et al. 2008). In order to understand the
radiation mechanism and the structure of the radiating region,
long-term and multiwavelength observations are needed. The BL Lac
object S5 0716+714 is one the brightest BL Lac objects noted for its
microvariability. Its high duty cycle means that it is always in an
active state (Wagner \& Witzel 1995). It has been the target of many
monitoring programs (e.g., Wagner et al. 1996; Qian et al. 2001;
Raiteri et al. 2003). Montagni et al. (2006) reported the fastest
variability rate of 0.1-0.12mag/hr. Five major outbursts have been
observed so far, occurring at the beginning of 1995, in late 1997,
in the fall of 2001, in March 2004 and at the beginning of 2007
(Raiteri et al. 2003; Foschini et al. 2006; Gupta et al. 2008).
These five outbursts indicate a long-term variability timescale of
$\sim$ 3.0$\pm$0.3 years (e.g., Raiteri et al. 2003). Correlated
radio/optical variability has been observed for the source. In a
4-week monitoring program, Quirrenbach et al. (1991) discovered a
period change of  1 day to 7 days in both radio and optical bands.
Heidt \& Wagner (1996) reported a period of 4 days in the optical
band while Qian, Tao \& Fan (2002) derived a 10-day period from
their 5.3 yr optical monitoring. Recently Gupta et al. (2009) have
found good evidence for quasi-periods in five of the 20 best quality
nightly data sets of Montagni et al. (2006); these ranged from
$\sim$ 23 to $\sim$ 75 minutes.

The spectral change of S5 0716+714 has been observed extensively
(e.g., Ghisellini et al. 1997; Raiteri et al. 2003; Villata et al.
2004; Gu et al. 2006; Wu et al. 2005). Different behaviors have been
reported. Raiteri et al. (2003) reported different chromatism in
different timescales in their 8-year monitoring program. Sometimes
the source was bluer when brighter, sometimes the opposite,
sometimes no spectral change was seen despite changes in brightness.
Ghisellini et al. (1997) and Wu et al. (2005) found the source
exhibited a bluer-when-brighter chromatism when it was in fast
flares. However, Stalin et al. (2006) found even if the source was
in fast flare, there was no color change with brightness. In this
paper we concentrate on the microvariability and spectral changes of
S5 0716+714. We monitored the source from October 25 to 30 2008,
December 23 to 29 2008 and February 3 to 10 2009. The temporal
resolution was around 5 to 8 minutes in the four optical bands
(\emph{BVRI}). Because of the high temporal resolution, we can
provide high quality data with accurate results.

The paper is organized as follows: Section 2 describes observations
and data reduction procedures. Section 3 presents the results.
Discussions are reported in Section 4. A summary is given is Section
5.

\section{Observations and Data Reductions}


Our photometric observations were carried out at the 85-cm telescope
which is located at the Xinglong Station of the National
Astronomical Observatories of China (NAOC). This telescope is
equipped with a standard Johnson-Cousin-Bessel multicolor CCD
photometric system built on the primary focus (Zhou et al. 2008).
The PI1024 BFT camera has 1024 $\times$ 1024 square pixels, a field
of view of  16$'$.5 $\times$ 16$'$.5 at a focal ratio of 3.27 (f =
2780mm) with a scale of 0.96 arcsec per pixel.

After flat-fielding, bias and dark corrections, aperture photometry
was performed using the apphot task of IRAF. The instrumental
magnitudes of the source and comparison stars were then collected
and then processed in order to obtain the standard magnitudes of S5
0716+71 and the relevant errors (Note that data in the \emph{I} band
were not calibrated due to large photometric errors). Two standard
stars in the blazar field, 5 and 6 in the finding chart of Villata
et al. (1988), were compared to check that any reported variations
were intrinsic to the blazar and that each standard star was
non-variable. The magnitude difference between the two stars in the
\emph{B}, \emph{V,} \emph{R}, \emph{I} bands are 0$^{m}$.09,
0$^{m}$.08, 0$^{m}$.08 and 0$^{m}$.12 respectively. The
observational log is presented in Tables 1 $-$ 4 with columns being
Julian date, magnitude and uncertainty. We first determined the
differential instrumental magnitude of the blazar, star 5 and star
6. We determined observational scatter from the blazar $-$ star 5
($\sigma$ (BL $-$ Star 5)) and star 5 $-$ star 6 ($\sigma$ (Star 5
$-$ Star 6)). The rms errors are calculated from the two stars using
the formula:

\begin{equation}
 \sigma  = \sqrt{\frac{\Sigma(m_i - \bar{m})^2}{N-1}}
\end{equation}
where m$_{i}$ = (m$_{5}$ $-$ m$_{6}$)$_{i}$ is the differential
magnitude of stars 5 and 6 while $\bar{m}$ = m$_{5}$ $-$ m$_{6}$ is
the differential magnitude averaged over the entire dataset, and N
is the number of observations on a given night. The variability of
the source is then investigated by means of the variability
parameter, C, introduced by Romero et al. (1999). This parameter is
defined as: C =  ${\frac{\sigma_{(BL - 5)}}{\sigma_{(5 - 6)}}}$. If
C $>$ 2.576, the source is variable at 99\% confidence level. The
object is decided to be variable only when C $\geq$ 2.576 at least
in two bands if it is monitored in three or more bands (Jang \&
Miller 1997). The intranight variability is calculated (Heidt \&
Wagner 1996):
\begin{equation}
 Amp = \sqrt{(A_{max} - A_{min})^2 - 2\sigma ^2},
\end{equation}
where A$_{max}$ and A$_{min}$ are the maximum and minimum values of
each light curve and $\sigma$ is the same as what is described
above. The results are given in Table 5 for filters \emph{B},
\emph{V}, \emph{R} and \emph{I}. Column (1) is the universal date of
observation, column (2) the band, column (3) the intranight
variability amplitude, column (4) the number of data points, column
(5) the value of the C parameter (V/N indicates whether the source
is variable or not), column (6) the rms errors, column (7) the
Pearson correlation coefficients of the \emph{V} - \emph{R} color
index vs. R magnitude.

Our monitoring program of S5 0716+714 was divided into three
periods. The first period was from October 25 to 30 2008, the second
from December 23 to 29 2008 and the third from February 3 to 10
2009. As a result of weather conditions, 14 nights of data were
obtained. In the first period of our observations, only filters
\emph{V} and \emph{R} were used. In the second and third periods
longer observation times allowed us to usually use four filters:
\emph{B}, \emph{V}, \emph{R} and \emph{I}. The typical exposure
times for \emph{BVRI }measurements were 100s, 70s, 50s and 30s
respectively. The readout time is about 10s, resulting in a temporal
resolution of around 5 minutes for the best weather conditions.
Because of the dense sampling and high quality (photometric error
$\sim$ 0.004 mag), the data set can provide highly reliable results.




\section{Results}
\subsection{Light Curves}
The overall light curves in bands \emph{V }and \emph{R }are
displayed in Fig. 1. The light curves of the two bands for
individual periods are displayed in Fig. 2 - 4. The source remained
active during the whole period of monitoring. The variations in
\emph{BVRI} are $\Delta$\emph{B} $\sim$ 0$^{m}$.89, $\Delta$\emph{V}
$\sim$ 0$^{m}$.80, $\Delta$\emph{R} $\sim$ 0$^{m}$.73 mag,
$\Delta$\emph{I} $\sim$ 0$^{m}$.51. One can see four flares during
our period of observation for four months (from JD 2454765 to JD
2454873) from Fig. 1. The first flare peaked on JD 2454766, with
\emph{V} $\sim$ 13$^{m}$.41 $\pm$ 0$^{m}$.01, \emph{R} $\sim$
13$^{m}$.01 $\pm$ 0$^{m}$.01. It is quite likely that the actual
peak was even higher since the blazar's brightness was declining
monotonically throughout the period observed that night.  After
that, the magnitude dropped about 0.4 mag in 4 days. In the second
flare, the source brightness rose since the first day (JD 2454824)
of our second period of observation, reaching a maximum on JD
2454825 (\emph{V} $\sim$ 13$^{m}$.59 $\pm$ 0$^{m}$.01, \emph{R}
$\sim$ 13$^{m}$.16 $\pm$ 0$^{m}$.01) and then quickly dropped about
0.5 mag in the following 3 days, reaching a minimum on JD 2454828
(\emph{V} $\sim$ 14$^{m}$.11 $\pm$ 0$^{m}$.04, \emph{R} $\sim$
13$^{m}$.65 $\pm$ 0$^{m}$.03). The source brightness then rose again
until the end of our second period of observation, resulting in
another flare with a magnitude change of about 0.3 mag in 2 days. In
the third period of observations the fourth flare showed the source
reaching the maximum brightness seen in our entire data set, with
\emph{R} $\sim$ 12$^{m}$.95 $\pm$ 0$^{m}$.01 and \emph{V} $\sim$
13$^{m}$.35 $\pm$ 0$^{m}$.01 on JD 2454867. The source then rapidly
dropped about 0.8 mag in 6 days until the end of our observation on
JD 2454873.

Clear intranight variations can be seen during the whole monitoring
campaign. We detected significant brightness changes in a large
fraction of nights: only in 1 night (marked with an asterisk in
Table 5) did the source remain stable during the observation time
window. The intranight variability amplitude ranges from $\sim$ 0.04
- 0.28 mag with a typical duration of a few hours. The lightcurves
of microvariability are generally irregular except for the ones of
JD 2454825 and JD 2454826 of period 2 which are symmetric, and may
show one possible cycle of $\sim$ 8 hr, respectively. In the first
period, the source usually displayed microvariability of large
amplitude ($>$0.1 mag). Some examples of the lightcurves are shown
in Fig. 5. On JD 2454765, the source brightened gradually by 0.183
$\pm$ 0.01 mag in 192 mins in the \emph{R} band. On the next day,
after reaching the peak of the first flare, the brightness of the
source began to fall. The magnitude change was 0.157 $\pm$ 0.01 mag
in 248 mins in the \emph{R} band. On the last day of period one,
which was JD 2454770, the brightness rose 0.116 $\pm$ 0.01 mag in
185 mins in the \emph{R} band.

In the second period of our observations, noticeable
microvariability was observed on JD 2454824 and JD 2454825. On the
former day, the brightness first dropped 0.110 $\pm$ 0.01 mag in 149
mins in the \emph{R} band and then became somewhat steady (Fig. 6).
On JD 2454825, the lightcurves are symmetric. The source first
brightened by 0.105 $\pm$ 0.008 mag in 259 minutes, reaching the
peak of the second flare, and then gradually dropped 0.137 mag $\pm$
0.01 in 301 minutes (\emph{R} band, see Fig. 7). The lightcurves on
JD 2454826 are sawtooth-like and have sharp turnoffs, showing a
complete period. The brightness of the source first dropped 0.054
$\pm$ 0.01 mag in 172 mins and then rose 0.07 $\pm$ 0.01 mag in 268
mins (\emph{R} band). It then dropped again with an amplitude of
0.0461 $\pm$ 0.01 mag, a timescale of 117 mins, forming a complete
cycle (Fig. 8). The lightcurves of JD 2454825 and JD 2454826 are of
particular interest and are discussed further in section 4.

In the third period of our observations, the source showed
substantial brightness changes in all the four days of observations.
On JD 2454865 - 2454866, it first dropped 0.112 $\pm$ 0.02 mag in
258 mins and then rose 0.073 $\pm$ 0.01 in 206 mins (\emph{R} band,
see Fig. 9). On 2454866 - 2454867, in the \emph{R} band, the source
brightened steadily by 0.152 $\pm$ 0.02 mag in 145 mins, reaching
the peak of the fourth flare, and then it dropped 0.07 $\pm$ 0.01
mag in 463 mins (Fig. 10). The source displayed the greatest
magnitude change on JD 2454871 - 2454872. The brightness dropped
0.245 $\pm$ 0.03 mag over the course of the night in 480 mins
(\emph{R} band, see Fig. 11). On the last night of our whole
observation campaign, JD 2454872 - 2454873, the source displayed a
rather regular lightcurve. At the beginning, the brightness dropped
0.082 $\pm$ 0.01 mag in 274 mins and then rose 0.113 $\pm$ 0.01 mag
in 240 mins (\emph{R} band, see Fig. 12). From Fig. 5 - 12, one can
see the lightcurves of microvariability of different bands are
consistent with each other.

\subsection{Spectral Variability}
Based on our observations of high temporal resolution, the spectral
variability with brightness was investigated in this section.
Magnitudes have been dereddened by using a galactic extinction
coefficient A$_{B}$ = 0.132, A$_{V}$=0.102, A$_{R}$=0.082 (Schlegel
et al. 1998). We concentrate on the \emph{V}-\emph{R} index, the
best sampled one.  The overall \emph{V} - \emph{R} color index
ranges from 0.380 to 0.485. For individual nights showing
microvariability, the spectral behavior varies from night to night
(see Fig. 13). A clear bluer-when-brighter chromatism is evident in
the long-term trend. The changes of color index with brightness
during the whole monitoring campaign are shown in Fig. 14. The solid
line is the unweighted least-squares fit to the data points, which
has a slope of 0.02 and a correlation coefficient of 0.753,
indicating a strong correlation between color and magnitude.
However, the color-magnitude correlation is complex in the three
observation periods.

Those nights showing strong correlation between color and magnitude
are shown in Fig. 15. In the first period of our observations (JD
2454765 - JD 2454770), the source displayed
 the strongest bluer-when-brighter (BWB) behavior. Clear microvariability was observed on all nights,
 with variability amplitude ranging from $\sim$ 0.04 - 0.2 (see Table 5). On the first night of observations of period 1, JD 2454765,
  the variability amplitude in the \emph{R} band is 0.183 mag and the correlation
  coefficient is 0.859. The mean \emph{R} magnitude was $\sim$ 13$^{m}$.12. On the next day, the variability amplitude was
  somewhat smaller, $\Delta$\emph{R} = 0.157 mag ($\overline{\emph{R}}$ $\sim$ 13$^{m}$.09) and the correlation
  coefficient r = 0.827.  On JD 2454767, the variability amplitude was
  the smallest among all days of observations in period 1, $\Delta$\emph{R}
    = 0.048 mag ($\overline{\emph{R}}$ $\sim$ 13$^{m}$.15) and r = 0.430. On JD 2454770, the amplitude
     is $\Delta$\emph{R} = 0.116 mag ($\overline{\emph{R}}$ $\sim$ 13$^{m}$.32) and r = 0.749. One can
    see a noticeable trend that the source showed a stronger BWB
    chromatism when the amplitude change became larger in period 1. However, this
    trend does not seem to be related to brightness. BWB chromatism
    can be found both when the source was at the brightest state ($\overline{\emph{R}}$ $\sim$ 13$^{m}$.09 on JD
    2454766) and the dimmest state ($\overline{\emph{R}}$ $\sim$ 13$^{m}$.32 on JD
    2454770) of period 1.

In the second period of our observations (JD 2454824 - JD 2454830),
BWB is still noticeable on the first two nights of observation. The
variability amplitude of these two nights were 0.127 and 0.137 mag
respectively (\emph{R} band), the corresponding mean \emph{R}
magnitudes were $\sim$ 13$^{m}$.36 and $\sim$ 13$^{m}$.21 and the
corresponding correlation coefficients were 0.563 and 0.618.
However, on subsequent nights with smaller variability amplitude, an
achromatic
     trend is noticed. On JD 2454826, JD 2454828 and JD 2454829, the
     amplitude changes in the \emph{R} band were 0.070, 0.043 and
     0.071 mag respectively, the corresponding correlation coefficients
     r were 0.150, -0.187 and 0.323, indicating no clear relationship
     between color and brightness. The source was at a rather dim
     state in these 3 days, with $\overline{\emph{R}}$ $\sim$ 13$^{m}$.34 on JD
    2454826, $\overline{\emph{R}}$ $\sim$ 13$^{m}$.50 on JD
    2454828 and $\overline{\emph{R}}$ $\sim$ 13$^{m}$.40 on JD
    2454829. The trend that the source displayed stronger BWB
    chromatism with increasing variability amplitude is still noticeable in period 2 but not as obvious as in period 1 and again
    the color change does not seem to be related to brightness.

    In the third period of observations (JD 2454865 - JD 2454873), each
night showed a
     variability
      magnitude of $>$ 0.1 mag. The source was observed for four nights in this period.
      It showed an achromatic trend on the first three nights of observations with $\Delta$\emph{R} = 0.112 ($\overline{\emph{R}}$ $\sim$ 13$^{m}$.35)
       on JD 2454865 - JD 2454866, $\Delta$\emph{R} = 0.152 ($\overline{\emph{R}}$ $\sim$ 12$^{m}$.99, the brightest state in the whole monitoring
       campaign) on JD 2454866 - JD 2454867 and $\Delta$\emph{R} = 0.245 ($\overline{\emph{R}}$ $\sim$ 13$^{m}$.34) on JD 2454871 - JD
       2454872. The corresponding correlation coefficients are $-$0.198, 0.333 and 0.115
       respectively, indicating no clear correlation between color and
       brightness. However, on JD 2454873, the source was at the
       dimmest state in the whole monitoring campaign with $\overline{\emph{R}}$ $\sim$
       13$^{m}$.64 and $\Delta$\emph{R} = 0.113, a BWB trend is noticeable with correlation coefficient r $=$
       0.516. Very obviously spectral change is neither related to brightness nor variability amplitude in period 3.

On the whole, we find no consistent relation between color vs
magnitude and also color vs variability amplitude on intranight
timescale. BWB chromatisim and achromatisim can be found when the
source is bright or faint, showing large variability amplitude or
small variability amplitude. This suggests that even on intranight
timescales, different mechanisms may be at work.

\section{Discussions}

The microvariability of S5 0716 + 714 has been observed by many
authors. Some ultra-rapid fluctuations on timescale $\leq$ 1.0 hour
were reported. Qian et al. (2002) recorded a brightness increase of
$\Delta$\emph{V} $\sim$ 0.78 mag in 9 mins in their 5.3-yr
monitoring programme. Xie et al. (2004) reported $\Delta$\emph{B}
$\sim$ 0.55 mag on a timescale of 36 mins. On longer timescales, Gu
et al. (2006) found the magnitude change of $\Delta$\emph{V} = 0.28
mag in $\sim$ 5 hours. In our monitoring programme, no ultra-rapid
fluctuations were detected. The shortest timescale was $\sim$ 2
hours, corresponding to $\Delta$R $\sim$ 0.046 mag on JD 2454826
while the longest timescale were $\sim$ 8 hours, corresponding to
$\Delta$R $\sim$ 0.245 mag, which happened on JD 2454871 - JD
2454872. In our whole monitoring campaign, the timescales are
generally a few hours with amplitude of $\sim$ 0.04 - 0.28 mag.

 Time lags between different optical bands have also been reported. Stalin et
al. (2006) found possible time lags of $\sim$ 6 and $\sim$ 13 mins
between the \emph{V} and \emph{R }bands in their 2 nights of
observations of the source. Qian et al. (2000) reported similar
results in which they found a time lag of $\sim$ 6 mins between the
\emph{V} and \emph{I} bands. From densely sampling data, Villata et
al. (2000) presented a time lag with a strict upper limit of $\sim$
10 mins between the \emph{B} and \emph{I} band. In this work, we
tried to find out the time lag between \emph{B} band and \emph{I}
band using the discrete cross-correlation function, DCF, suggested
by Edelson \& Krolik (1988) for unequally spaced data. It was
performed only on the day with high-quality data (error $\sim$ 0.003
and $\sim$ 0.005 for \emph{B} and \emph{I} band respectively) and
dense sampling rate (temporal resolution $\sim$ 5 minutes), which is
JD 2454826, the day with with a hint of periodicity in its light
curve. Fig. 16 presents the results of the calculations. The dashed
line indicates the centroid which was calculated using the method
proposed by Peterson (2001) and we found the barycenter using data
points located near the peak value, DCF$_{peak}$, specifically,
those greater than 0.8DCF$_{peak}$. The expression is:
\begin{equation}
DCF_{centroid} ={\frac{\sum \tau DCF(\tau)}{\sum DCF(\tau)}}
\end{equation}
We found a time lag of $\sim$ 11 minutes, with the \emph{B} band
leading \emph{I} band. However, given the sampling rate of $\sim$ 5
minutes and the negligible difference between the DCF values at -20
minutes and 0 minutes, this cannot be considered a convincing
measurement of a lag.  This result is consistent with Villata et al.
(2000), who presented a strict upper limit of $\sim$ 10 minutes to a
possible delay between \emph{B} - and \emph{I} - band variations
using high-quality, densely sampled data on a single night. This
result is expected in the shock-in-jet model but not in most
disc-based variability models (e.g. Wiita 2006).

In order to quantitatively analyze possible periods, we performed
structure function (SF) on JD 2454826, the only day with a
lightcurve that shows a hint of a period. SF, discussed fully by
Simonetti el al. (1985), is a common tool to search for
periodicities and timescales of variation . It identified a
timescale of 259 mins, 287 mins, 280 mins and 263 mins and a period
of 482 mins, 497 mins, 498 mins and 491 mins were found for
\emph{B}, \emph{V}, \emph{R} and \emph{I} band respectively. All
these results are consistent with visual inspection. The results are
shown in Fig. 17. Of course, since only one ``cycle'' is present
that night, the presence of a period is no more than speculative;
such a possible period does not appear to extend to the previous
night, for which there is also data.  Only observations made with
multiple ground based telescopes at different longitudes (or space
based telescopes) can convincingly find intranight periods that
substantially exceed $\sim$ 1 hr.

The spectral variability of blazars has been investigated by many
authors (e.g., Ghisellini et al. 1997; Fan \& Lin 1999; Romero et
al. 2000; Raiteri et al. 2003; Villata et al. 2000, 2004). Raiteri
et al. (2003) found different spectral behavior for the S5 0716 +
714 in short timescales. Sometimes the source was bluer when
brighter, sometimes the opposite and sometimes no spectral change
was seen. Stalin et al. (2006) found no clear evidence of color
variation with brightness in either their internight or intranight
monitoring of the source for a fortnight. Ghisellini et al. (1997)
and Wu et al. (2005) reported a BWB trend during fast flares but
this trend is insensitive in the long-term. However, Wu et al.
(2007) noticed that the source is bluer when brighter on both
intranight and internight timescales but this trend was not present
in the long-term data.

In our monitoring campaign, the source also showed different
spectral change behavior. On the long term, the source showed a
bluer-when-brighter behavior. Unlike previous studies which reported
consistent trends on spectral behavior on intranight timescales
(e.g. Ghisellini et al. 1997, Stalin et al. 2006, Wu et al. 2005),
no consistent spectral behavior is found among all those nights
displaying microvariability in our present study. The source
displayed BWB chromatism when it was either bright or dim, or when
showing large or small variability amplitudes. Achromatism was also
found when the source was in the same conditions, suggesting
different mechanisms for microvariability. The BWB behavior is
consistent with shock-in-jet model (Wagner et al. 1995). In this
model, shocks form at the base of the jet and propagate downstream,
accelerating electrons and compressing magnetic fields and resulting
in the observed variability. The model predicts a BWB phenomenon and
an irregular lightcurve, as observed in our cases showing this
phenomenon. In the case of JD 2454825 and JD 2454826, a symmetric
lightcurve was observed for the former case while a periodic
lightcurve for the latter. Such regular lightcurves are rarely
found. Wu et al. (2005) reported a single ``period'' of a sine-like
light curve on intranight timescales for two nights during their
observations of S5 0716+714. Using more sophisticated techniques
Gutpa et al. (2009) found multiple oscillations to be present during
5 nights out of the 20 highest quality light curves of a total
sample of 102 nights of data by Montagni et al. (2006).
 Unlike ours, their lightcurves are
sinelike with smooth turns while ours are sawtooth-like with sharp
turnoffs. The symmetric lightcurves on JD 2454825 showed a
correlation between color and magnitude, with the correlation
coefficient r = 0.616. So the variation may still be due to
intrinsic reasons. However, for the periodic lightcurve on JD
2454826, no strong correlation between color and magnitude is found,
with the correlation coefficient r = 0.150. Such achromatic
lightcurves may be produced by geometric effects like microlensing
or a lighthouse effect produced by different amounts of Doppler
boosting induced by a helical structure to the jet (e.g. Wagner \&
Witzel 1995). Microlensing predicts a strictly symmetric lightcurve,
which cannot explain the concave shape of the second halves of the
light curves in our case. Therefore a lighthouse effect is the most
probable mechanism for explaining periodic components in blazar
lightcurves as this type of variation is likely to be achromatic
(e.g., Camenzind \& Krockenberger 1992; Gopal-Krishna \& Wiita
1992).
\section{Summary}

We monitored S5 0716+714 during the period October 2008 to February
2009. The object remained active during the whole campaign, showing
microvariability on all nights of observations except for one. The
timescales range from $\sim$ 2 - 8 hours and variability amplitude
$\sim$ 0.04 - 0.28 mag. Four flares of amplitude of $\sim$ 0.4 mag,
0.5 mag, 0.3 mag and 0.75 mag were observed, each lasting for a few
days. The long-term spectral change is evidently
bluer-when-brighter. Among those nights displaying microvariability,
the source displayed different spectral behavior. Sometimes the
source seemed to become bluer with increasing variability amplitude
while sometimes not. In a bright state, it displayed a BWB
chromatism but this was also found when the source was in a dim
state. Achromatism was also observed when the source was both bright
and dim. The bluer-when-brighter behavior can be explained by
shock-in-jet model while the achromatic trend may be due to
geometric effects. A possible time lag of  $\sim$ 11 minutes between
the \emph{B} and \emph{I} bands was seen during one night.



\acknowledgments

This work is supported by the National Natural Science Foundation of
China (NSFC), through the Grants 10673001, 10778601, 10878007 and
10633010 the support from the program for New Century Excellent
Talents in University (NCET), and the Project sponsored by the
Scientific Research Foundation for the Returned Overseas Chinese
Scholars, State Education Ministry.

\clearpage



\begin{deluxetable}{rrr}
\tablecolumns{8} \tablewidth{0pc} \tablecaption{Observational data
for S5 0716+714 (\emph{B} filter)} \tablehead{ \colhead{Julian Date}
& \colhead{mag}   & \colhead{$\sigma$} } \startdata
2454824.05104 &  14.191 &  0.010 \\
2454824.05473 &  14.204 &  0.010 \\
2454824.05844 &  14.203 &  0.009 \\
2454824.06213 &  14.209 &  0.009 \\
2454824.06583 &  14.210 &  0.009 \\
2454824.06953 &  14.224 &  0.009 \\
2454824.07322 &  14.222 &  0.009 \\
2454824.07692 &  14.230 &  0.011 \\
2454824.08061 &  14.238 &  0.010 \\
2454824.08431 &  14.244 &  0.013 \\
\enddata

(This table is available in its entirety in a machine-readable form
in the online journal. A portion is shown here for guidance
regarding its form and content)
\end{deluxetable}

\begin{deluxetable}{rrr}
\tablecolumns{8} \tablewidth{0pc} \tablecaption{Observational data
for S5 0716+714 (\emph{V} filter)} \tablehead{ \colhead{Julian Date}
& \colhead{mag}   & \colhead{$\sigma$} }

\startdata
2454765.19653 &  13.600   &   0.015 \\
2454765.19877 &  13.605   &   0.015 \\
2454765.20096 &  13.604   &   0.015 \\
2454765.20315 &  13.615   &   0.015 \\
2454765.20534 &  13.613   &   0.015 \\
2454765.20751 &  13.611   &   0.015 \\
2454765.20970 &  13.614   &   0.015 \\
2454765.21189 &  13.617   &   0.015 \\
2454765.21407 &  13.615   &   0.015 \\
2454765.21625 &  13.617   &   0.015 \\
\enddata

(This table is available in its entirety in a machine-readable form
in the online journal. A portion is shown here for guidance
regarding its form and content)
\end{deluxetable}

\begin{deluxetable}{rrr}
\tablecolumns{8} \tablewidth{0pc} \tablecaption{Observational data
for S5 0716+714 (\emph{R} filter)} \tablehead{ \colhead{Julian Date}
& \colhead{mag}   & \colhead{$\sigma$} } \startdata
2454765.19751   &   13.190   &   0.013 \\
2454765.19987   &   13.188   &   0.011 \\
2454765.20206   &   13.190   &   0.011 \\
2454765.20424   &   13.194   &   0.011 \\
2454765.20642   &   13.186   &   0.011 \\
2454765.20861   &   13.194   &   0.011 \\
2454765.21080   &   13.191   &   0.011 \\
2454765.21297   &   13.194   &   0.011 \\
2454765.21516   &   13.191   &   0.011 \\
2454765.21735   &   13.199   &   0.011 \\

\enddata

(This table is available in its entirety in a machine-readable form
in the online journal. A portion is shown here for guidance
regarding its form and content)
\end{deluxetable}

\begin{deluxetable}{ccc}
\centering
 \tablecolumns{8} \tablewidth{0pc}
\tablecaption{Observational data for S5 0716+714 (\emph{I} filter)}
\tablehead{ \colhead{Julian Date} & \colhead{Differential mag(BL -
S$_{5}$)} & \colhead{$\sigma$} }
 \startdata
2454824.05376   &     -0.004   &    0.006 \\
2454824.05745   &      0.005   &    0.006 \\
2454824.06116   &      0.000   &    0.006 \\
2454824.06485   &     -0.009   &    0.006 \\
2454824.06855   &     -0.021   &    0.006 \\
2454824.07225   &     -0.020   &    0.006 \\
2454824.07594   &     -0.022   &    0.006 \\
2454824.07964   &     -0.008   &    0.006 \\
2454824.08333   &     -0.001   &    0.035 \\
2454824.08703   &      0.012   &    0.009 \\

\enddata

(This table is available in its entirety in a machine-readable form
in the online journal. A portion is shown here for guidance
regarding its form and content)
\end{deluxetable}
\begin{deluxetable}{ccccccc}
\centering
 \tablecolumns{7} \tablewidth{0pc}
\tablecaption{Results of microvariability of the BL Lacertae S5
0716+714} \tablehead{ \colhead{Date}(dd.mm.yyyy) & \colhead{Band} &
\colhead{Amplitude}(mag) & \colhead{N} &
 \colhead{C(V/N)} & \colhead{$\sigma$} & \colhead{r} }
\startdata
 25.10.2008   &      V   &    0.200   &  94   &      9.03(V)   &    0.008  &  0.859  \\
 25.10.2008   &      R   &    0.183   &  92   &      13.90(V)  &    0.005  &  ...  \\
 26.10.2008   &      V   &    0.163   &  82   &      17.09(V)  &    0.003  &  0.827  \\
 26.10.2008   &      R   &    0.157   &  84   &      16.30(V)  &    0.003  &  ...  \\
 27.10.2008   &      V   &    0.048   &  63   &      4.08(V)   &    0.003  &  0.430  \\
 27.10.2008   &      R   &    0.042   &  62   &      4.00(V)   &    0.003  &  ...  \\
 30.10.2008   &      V   &    0.133   &  80   &      14.24(V)  &    0.003  &  0.749  \\
 30.10.2008   &      R   &    0.116   &  79   &      14.55(V)  &    0.003  &  ...  \\
 23.12.2008   &      B   &    0.150   & 107    &      6.83(V)   &   0.005  &  0.563  \\
 23.12.2008   &      V   &    0.138   & 107    &      7.80(V)   &   0.004  &  ...  \\
 23.12.2008   &      R   &    0.127   & 107    &      7.02(V)   &   0.004  &  ...  \\
 23.12.2008   &      I   &    ...     & 107    &      1.66(N)   &   0.015  &  ...  \\
 24.12.2008   &      V   &    0.132   & 110    &      15.47(V)  &   0.003  &  0.618  \\
 24.12.2008   &      R   &    0.137   & 110    &      14.09(V)  &   0.003  &  ...  \\
 24.12.2008   &      I   &    0.139   & 109    &      3.88(V)   &   0.010  &  ...  \\
 25.12.2008   &      B   &    0.089   &  127   &      6.47(V)   &  0.003   &   0.150 \\
 25.12.2008   &      V   &    0.080   &  128   &      5.51(V)   &  0.004   &   ... \\
 25.12.2008   &      R   &    0.070   &  126   &      4.45(V)   &  0.004   &   ... \\
 25.12.2008   &      I   &    0.074   &  128   &      3.51(V)   &  0.005   &   ... \\
 27.12.2008   &      B   &    0.058   &  36   &      3.42(V)   &  0.005    &  -0.187   \\
 27.12.2008   &      V   &    0.045   &  51   &      2.67(V)   &  0.006    &   ...  \\
 27.12.2008   &      R   &    0.043   &  50   &      2.68(V)   &  0.006    &   ...  \\
 27.12.2008   &      I   &    ...     &  51   &      2.03(N)   &  0.008    &   ...  \\
 28.12.2008   &      V   &    0.069   &  148   &      5.14(V)   &  0.003    &  0.323  \\
 28.12.2008   &      R   &    0.071   &  148   &      4.36(V)   &  0.004    &  ...  \\
 $^\ast$29.12.2008   &      V   &    ...     &  84   &      1.98(N)   &  0.003   &  ...   \\
 $^\ast$29.12.2008   &      R   &    ...     &  84   &      2.17(N)   &  0.003   &  ...  \\
 $^\ast$29.12.2008   &      I   &    ...     &  84   &      1.61(N)   &  0.005   &  ...  \\
 03.02.2009   &      V   &    0.108   &  178   &      4.86(V)   &   0.005   & -0.198  \\
 03.02.2009   &      R   &    0.112   &  177   &      5.10(V)   &   0.005   &  ... \\
 04.02.2009   &      B   &    0.181   & 126    &      5.42(V)   &  0.006    &  0.333 \\
 04.02.2009   &      V   &    0.157   & 127    &      5.42(V)   &  0.006    &  ... \\
 04.02.2009   &      R   &    0.152   & 127    &      4.67(V)   &  0.006    &  ... \\
 04.02.2009   &      I   &    ...     & 125    &      1.97(N)   &  0.006    &  ... \\
 09.02.2009   &      B   &    0.276   & 102    &      7.60(V)   &  0.011    &  0.115 \\
 09.02.2009   &      V   &    0.274   & 100    &      8.93(V)   &  0.009    &  ... \\
 09.02.2009   &      R   &    0.245   & 102    &      6.64(V)   &  0.011    &  ... \\
 09.02.2009   &      I   &    0.225   & 102    &      6.46(V)   &  0.011    &  ... \\
 10.02.2009   &      B   &    0.121   & 86    &      7.87(V)   &  0.004     & 0.516  \\
 10.02.2009   &      V   &    0.116   & 86    &      8.26(V)   &  0.004     &  ... \\
 10.02.2009   &      R   &    0.113   & 85    &      8.10(V)   &  0.004     &  ... \\

\enddata
\end{deluxetable}

\begin{figure}
\includegraphics[angle=90,width=1.1\hsize,height=1.0\hsize]{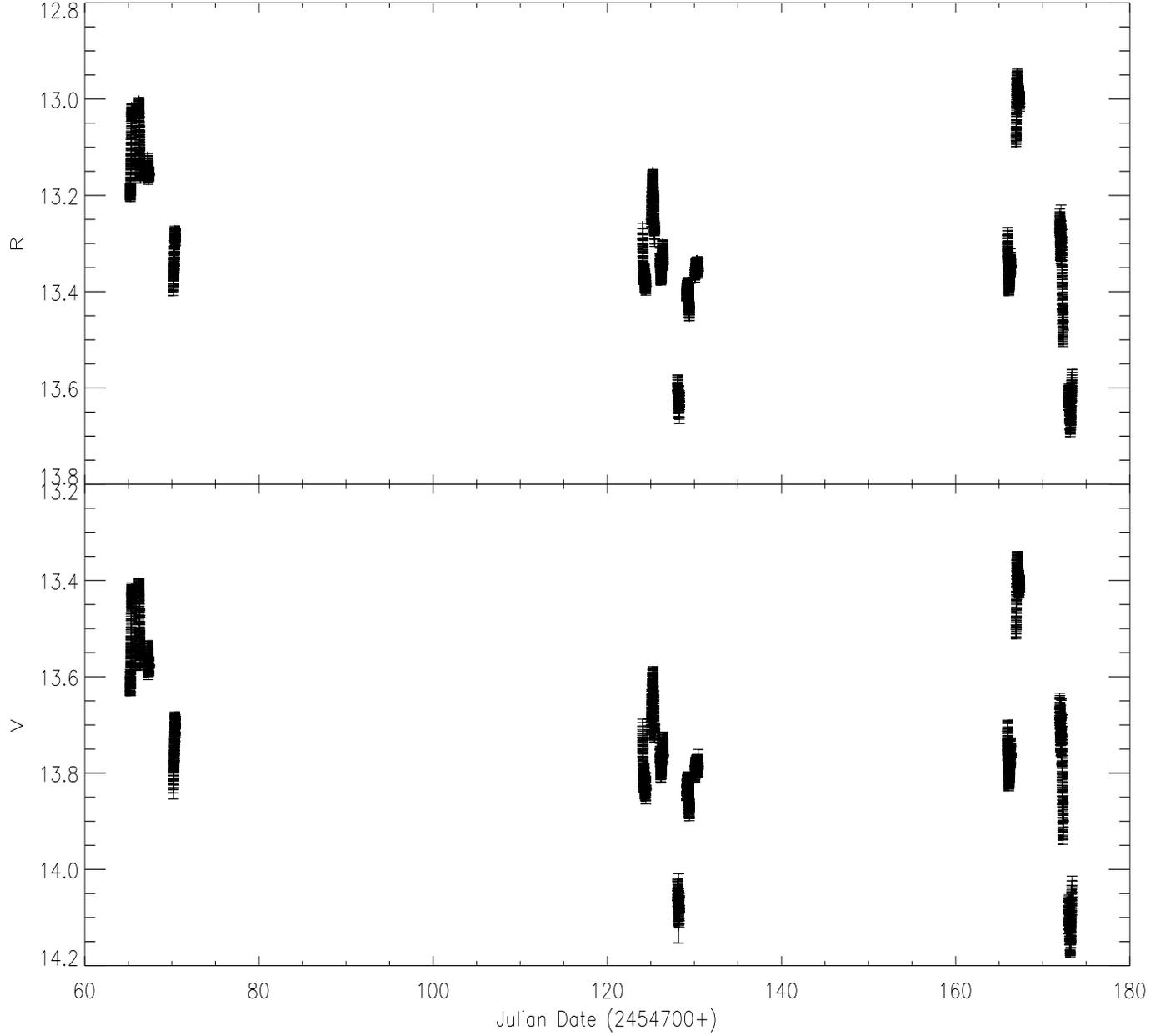}
\caption{Light curves of S5 0716+714 in the \emph{VR} bands from
October 25, 2008 to February 10, 2009. \label{fig1}}
\end{figure}

\clearpage

\begin{figure}
\includegraphics[angle=90,width=0.8\hsize,height=0.7\hsize]{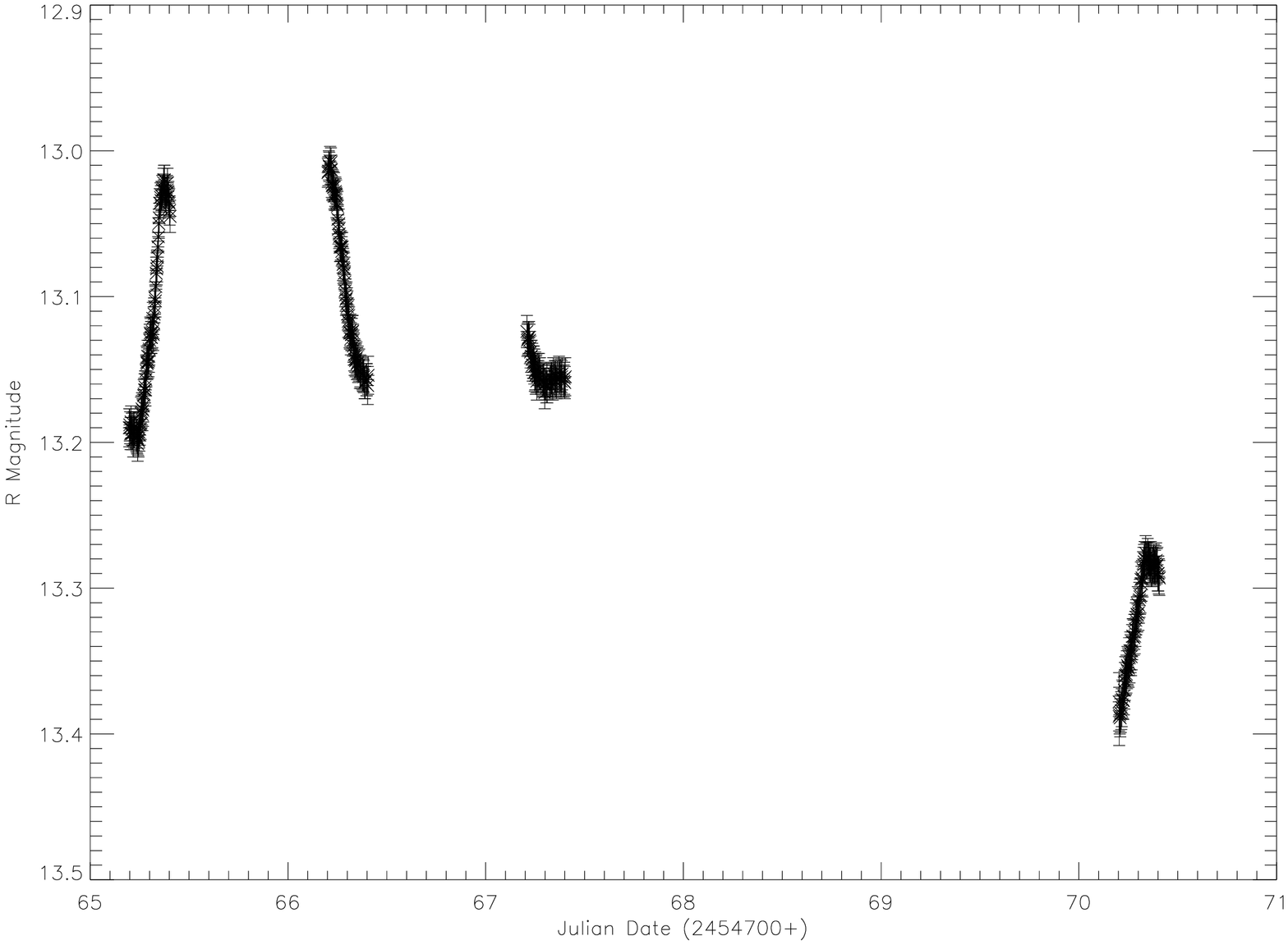}
\includegraphics[angle=90,width=0.8\hsize,height=0.7\hsize]{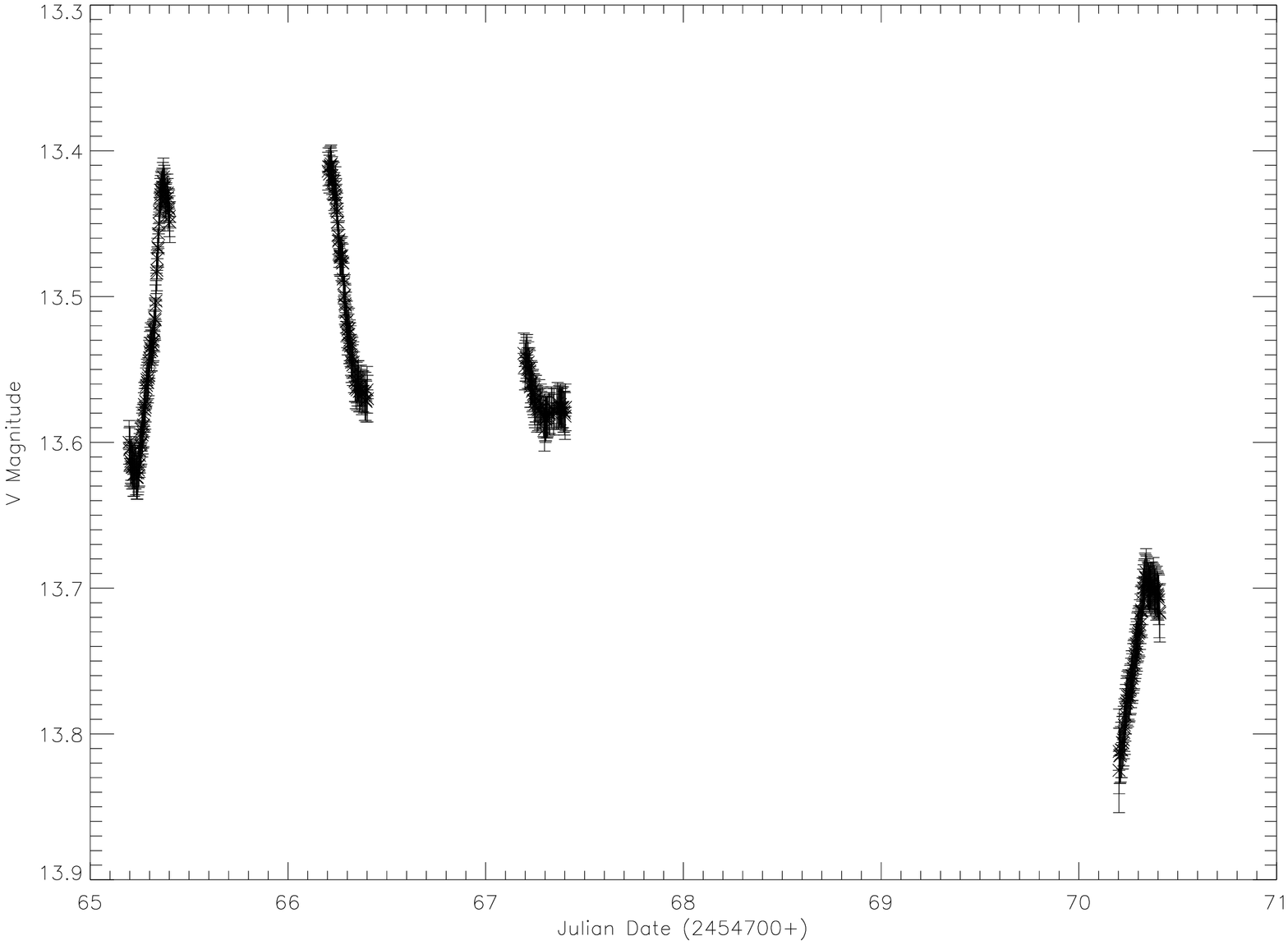}
\caption{Light curves of S5 0716+714 in the \emph{V}(bottom) and
\emph{R}(top) bands in period 1. \label{fig1}}
\end{figure}
\clearpage

\begin{figure}
\includegraphics[angle=90,width=0.8\hsize,height=0.7\hsize]{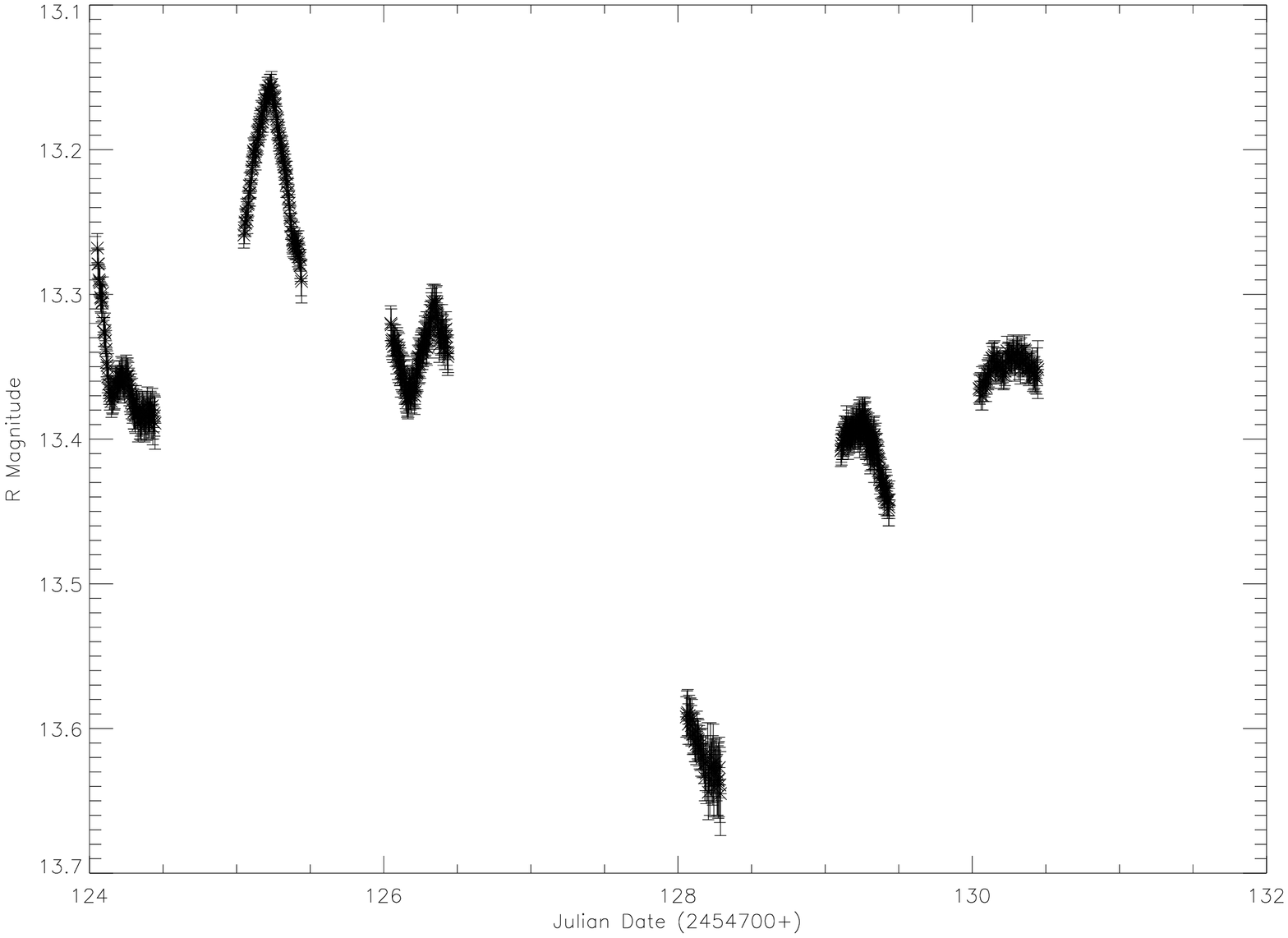}
\includegraphics[angle=90,width=0.8\hsize,height=0.7\hsize]{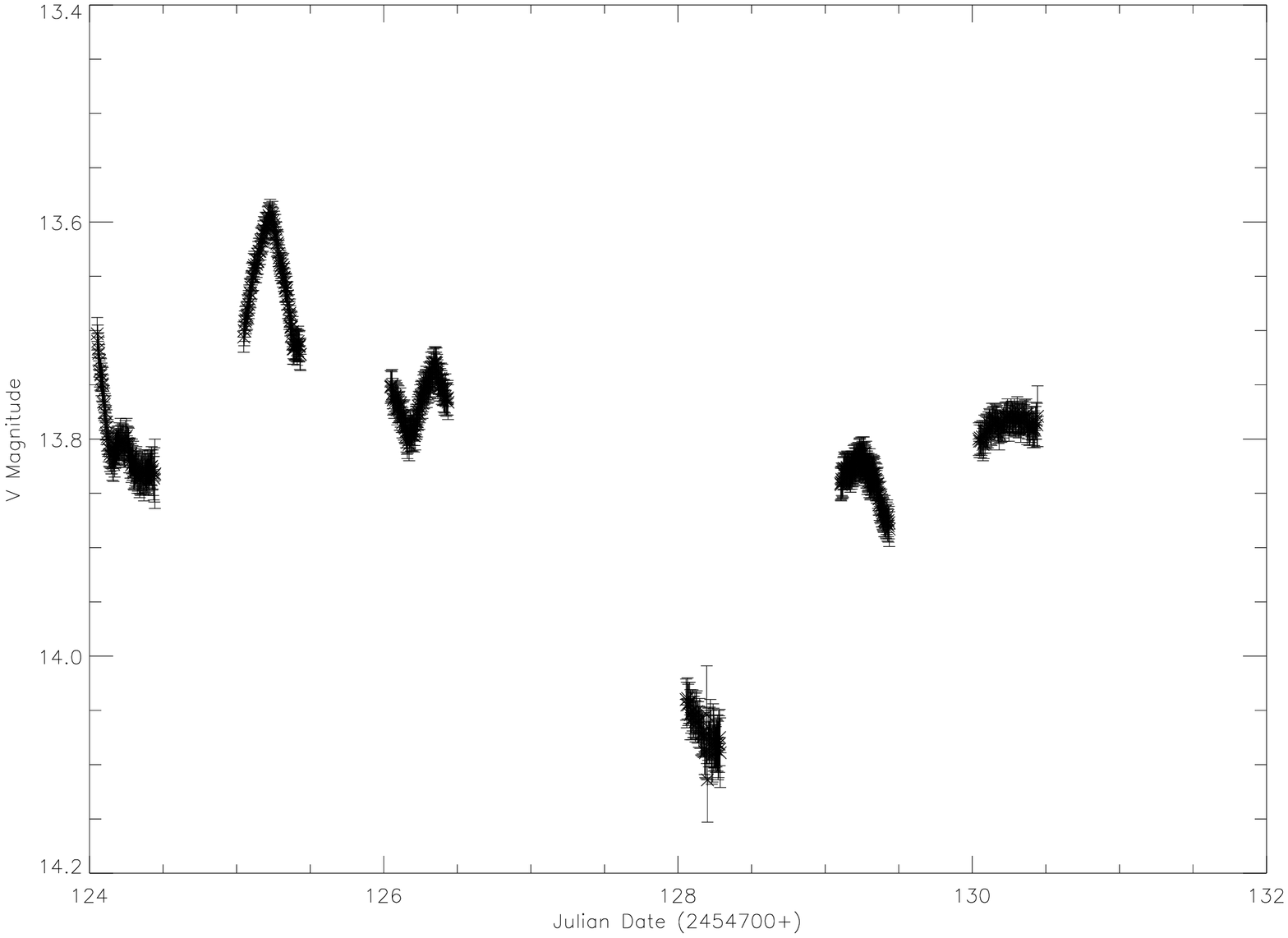}
\caption{Light curves of S5 0716+714 in the \emph{V}(bottom) and
\emph{R}(top) bands in period 2. \label{fig1}}
\end{figure}
\clearpage

\begin{figure}
\includegraphics[angle=90,width=0.8\hsize,height=0.7\hsize]{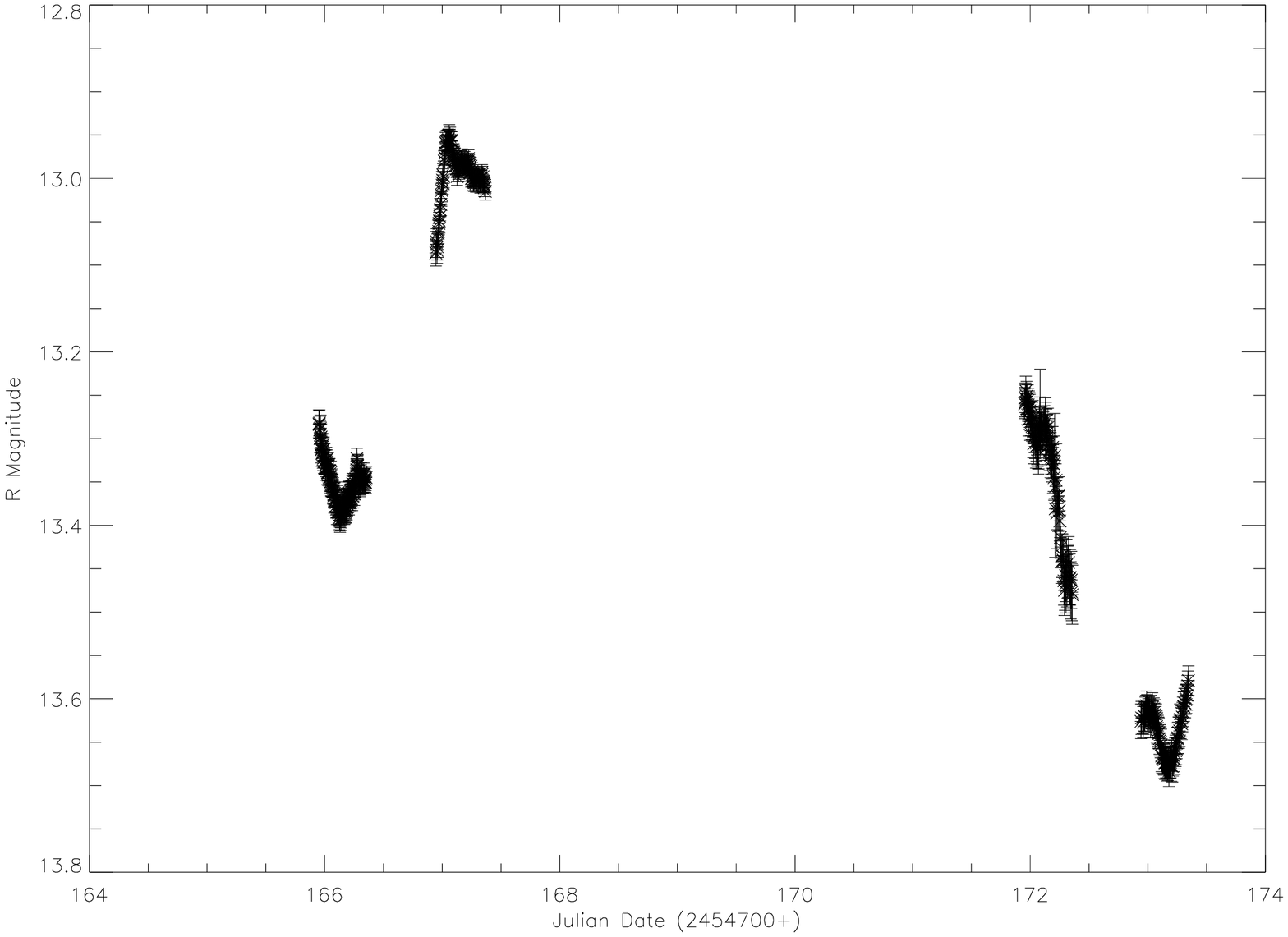}
\includegraphics[angle=90,width=0.8\hsize,height=0.7\hsize]{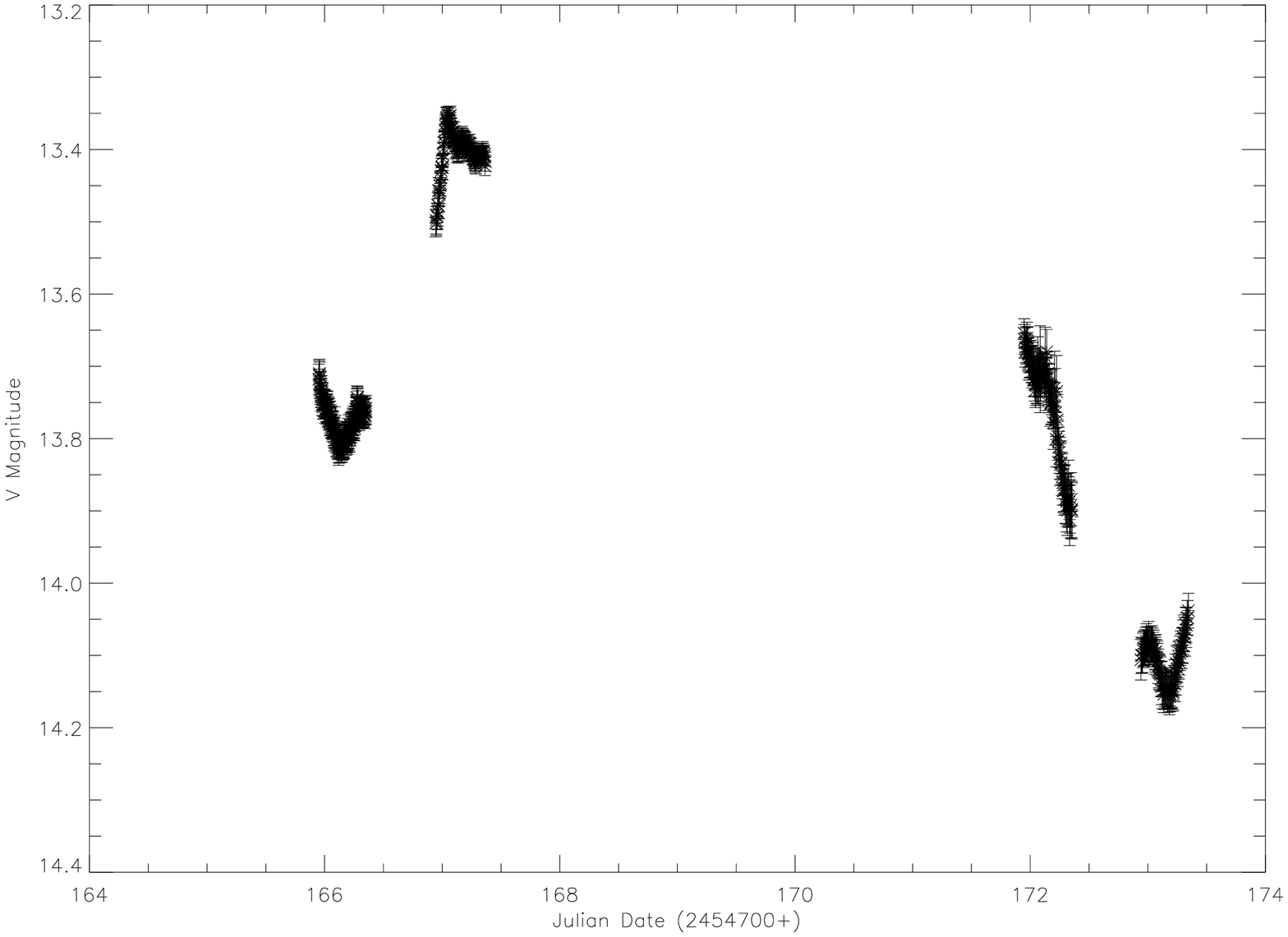}
\caption{Light curves of S5 0716+714 in the \emph{V}(bottom) and
\emph{R}(top) bands in period 3. \label{fig1}}
\end{figure}



\begin{figure}
\includegraphics[angle=90,width=0.5\hsize,height=0.4\hsize]{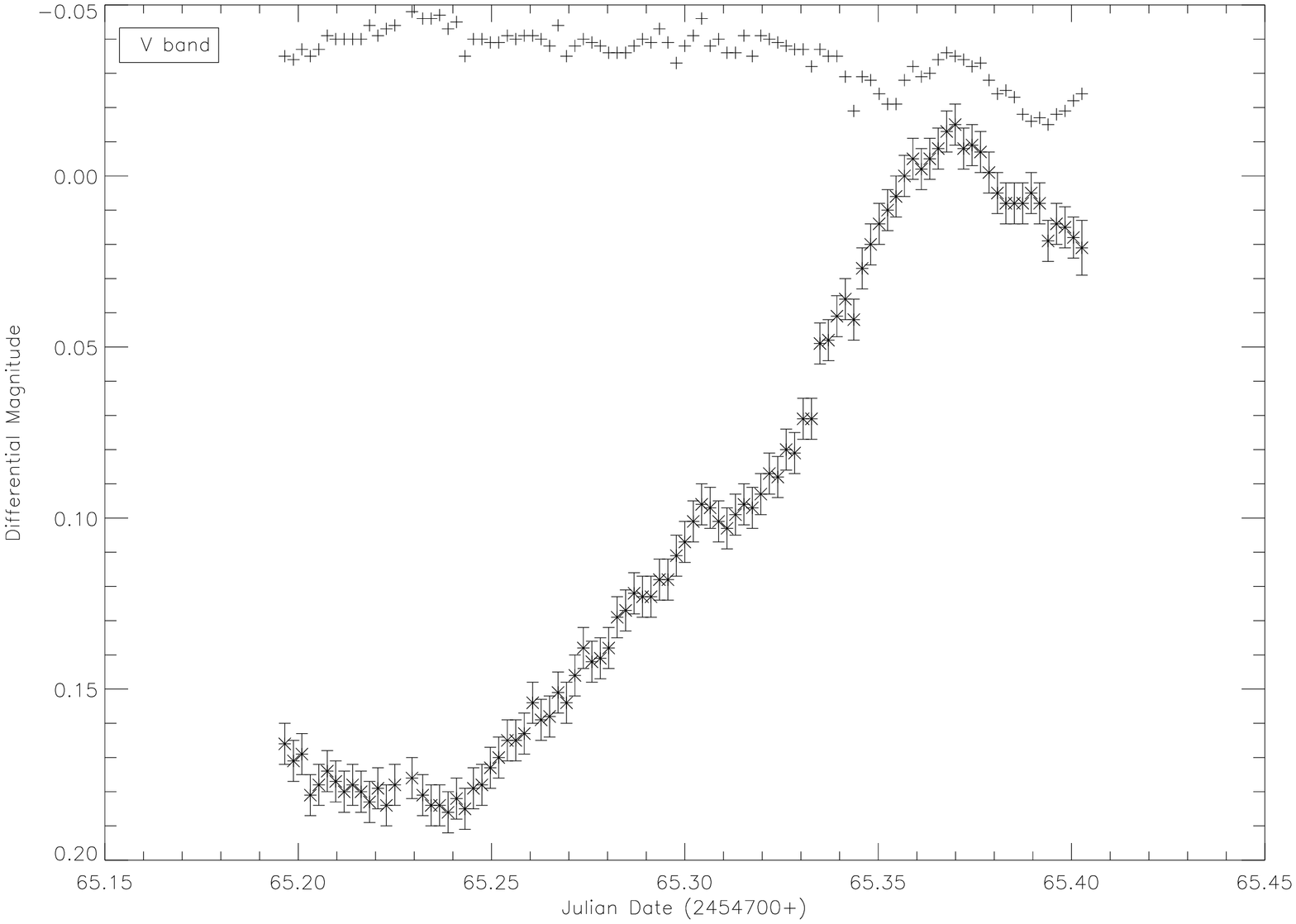}
\includegraphics[angle=90,width=0.5\hsize,height=0.4\hsize]{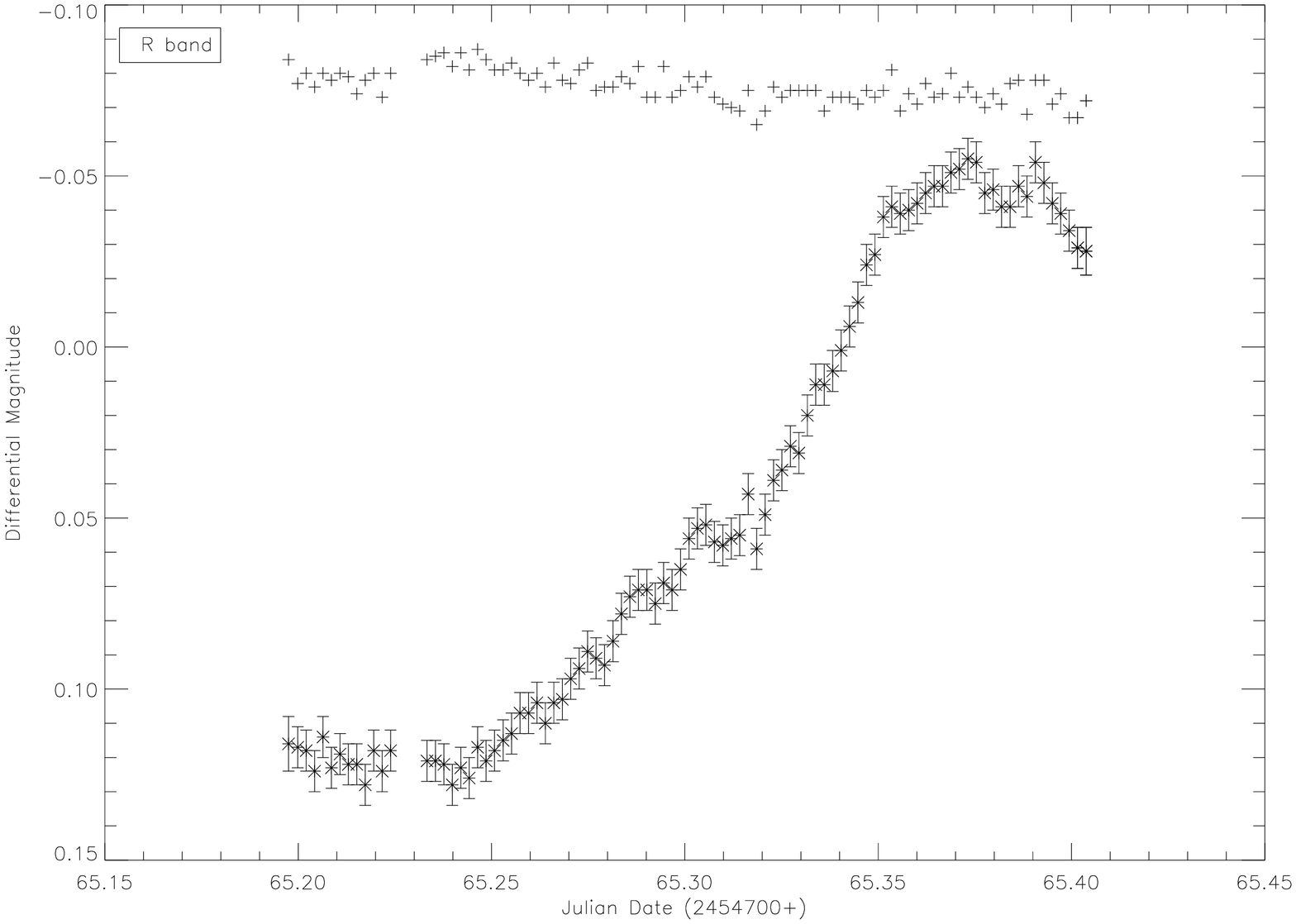}
\includegraphics[angle=90,width=0.5\hsize,height=0.4\hsize]{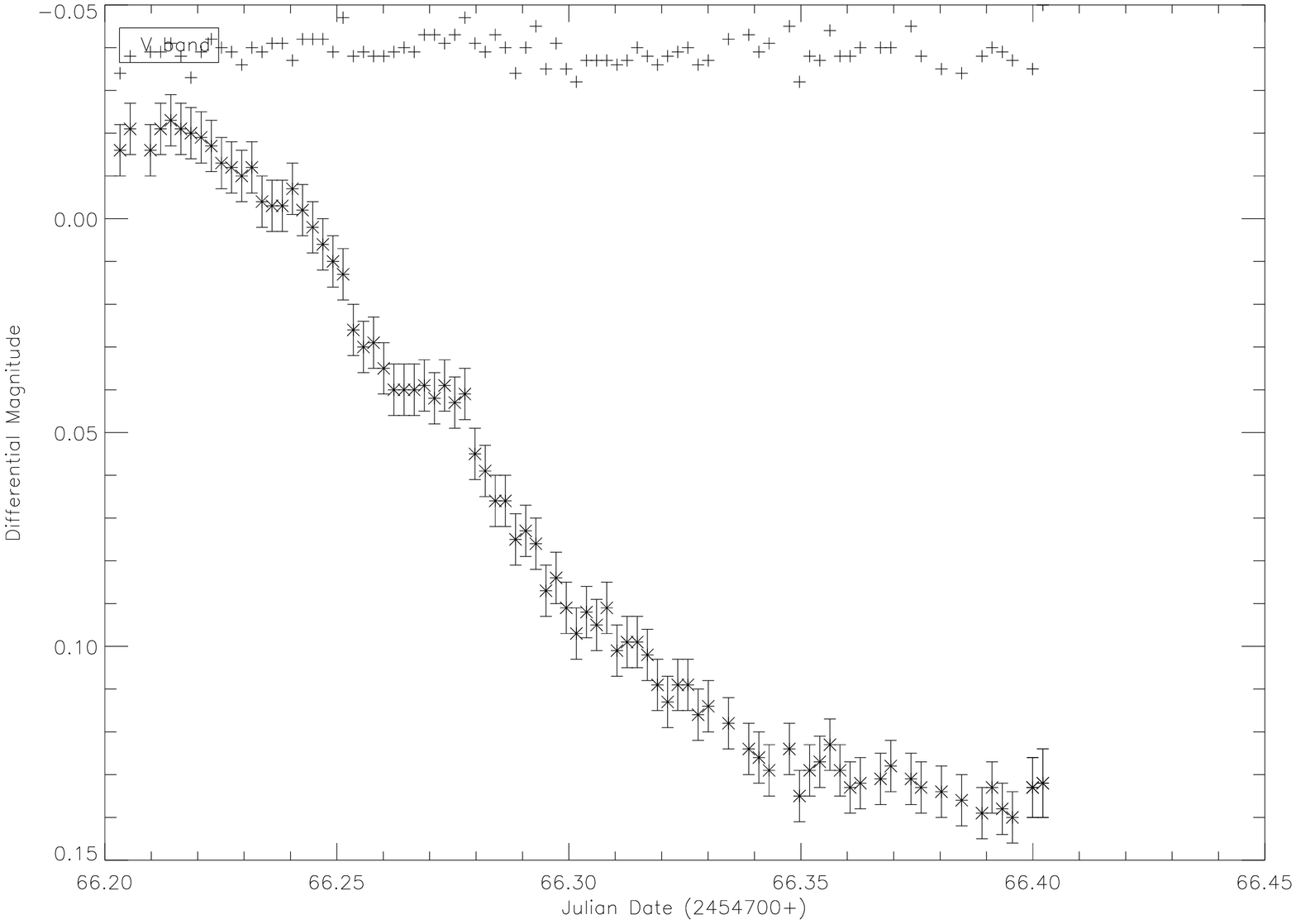}
\includegraphics[angle=90,width=0.5\hsize,height=0.4\hsize]{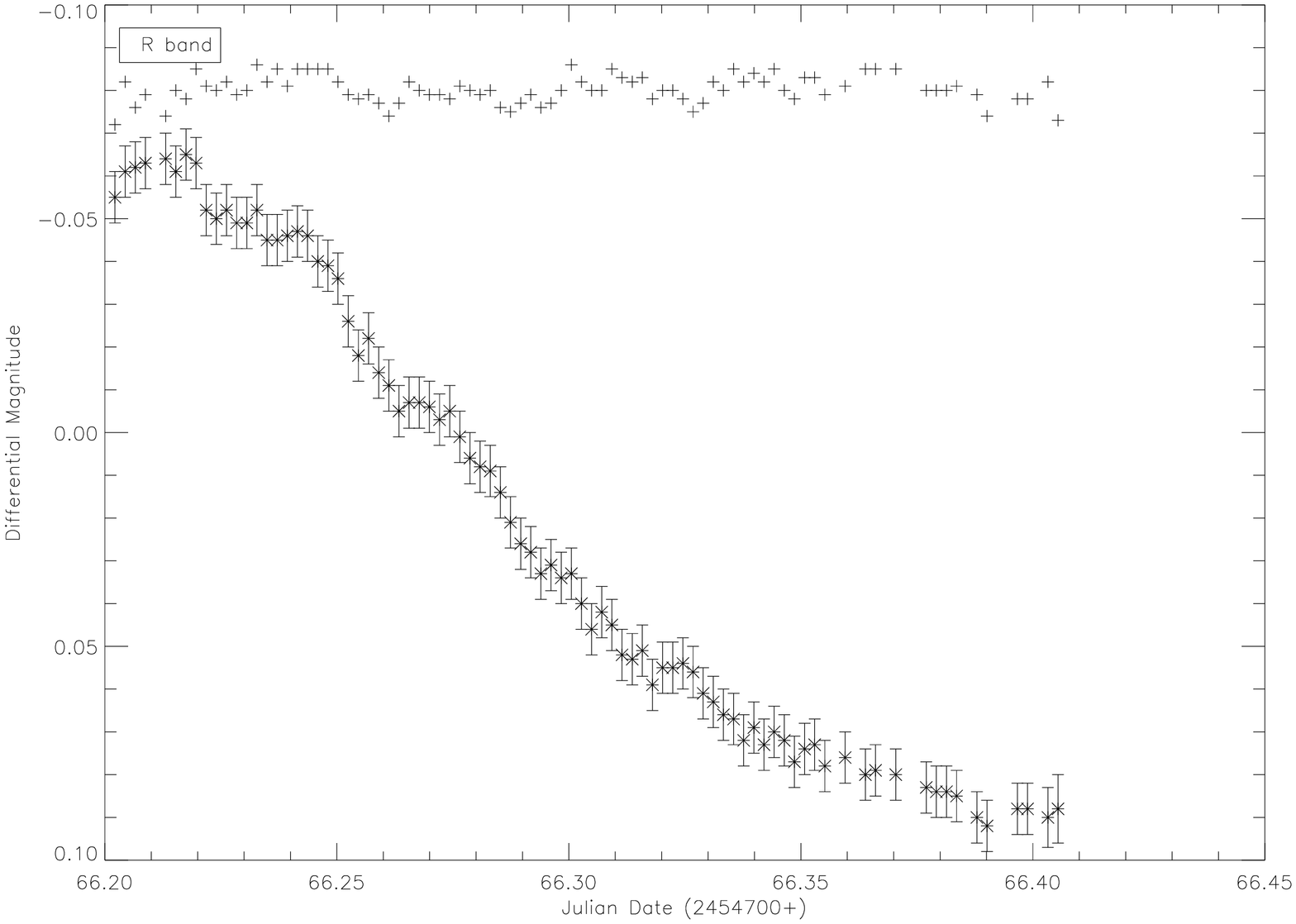}
\includegraphics[angle=90,width=0.5\hsize,height=0.4\hsize]{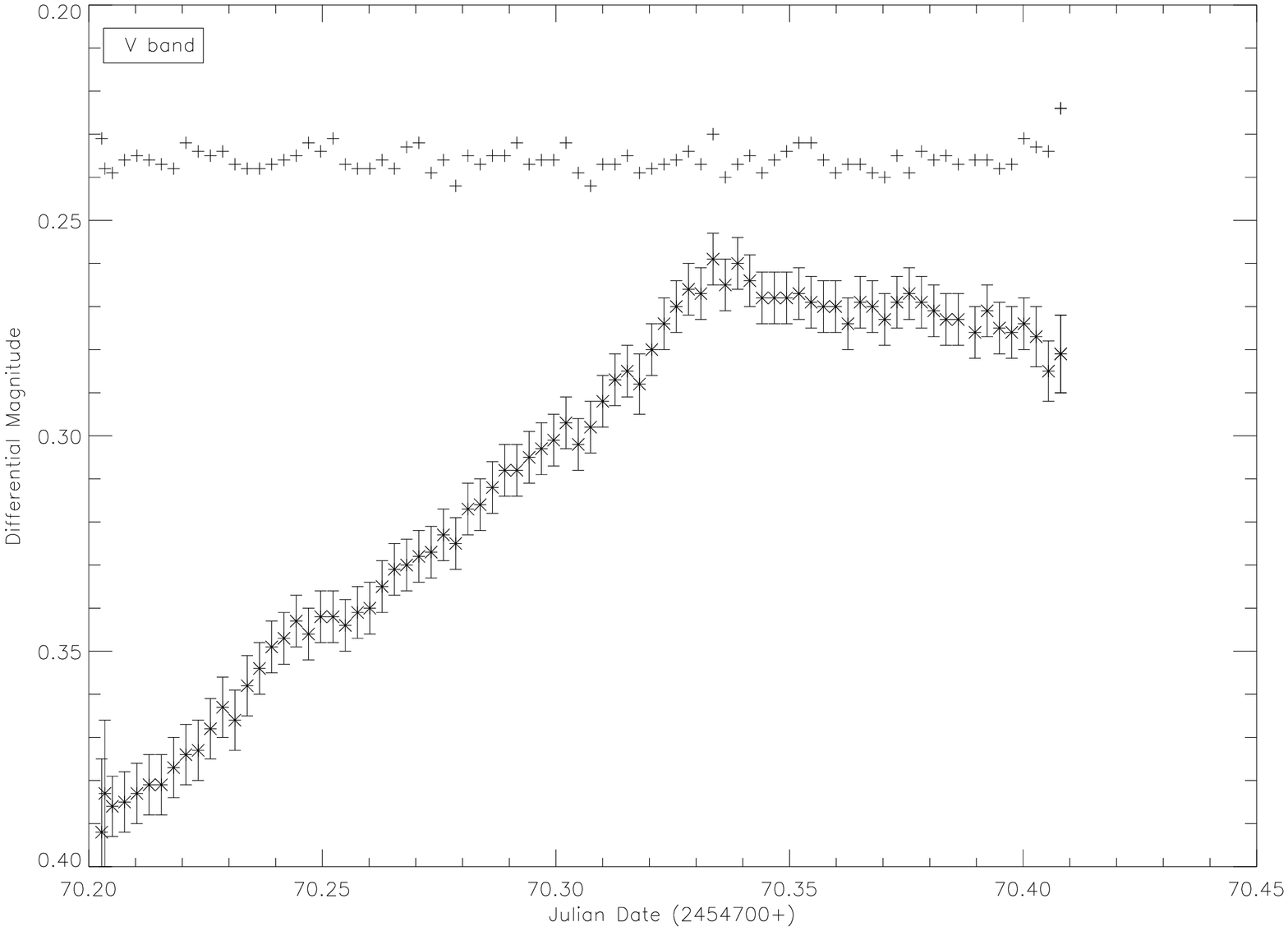}
\includegraphics[angle=90,width=0.5\hsize,height=0.4\hsize]{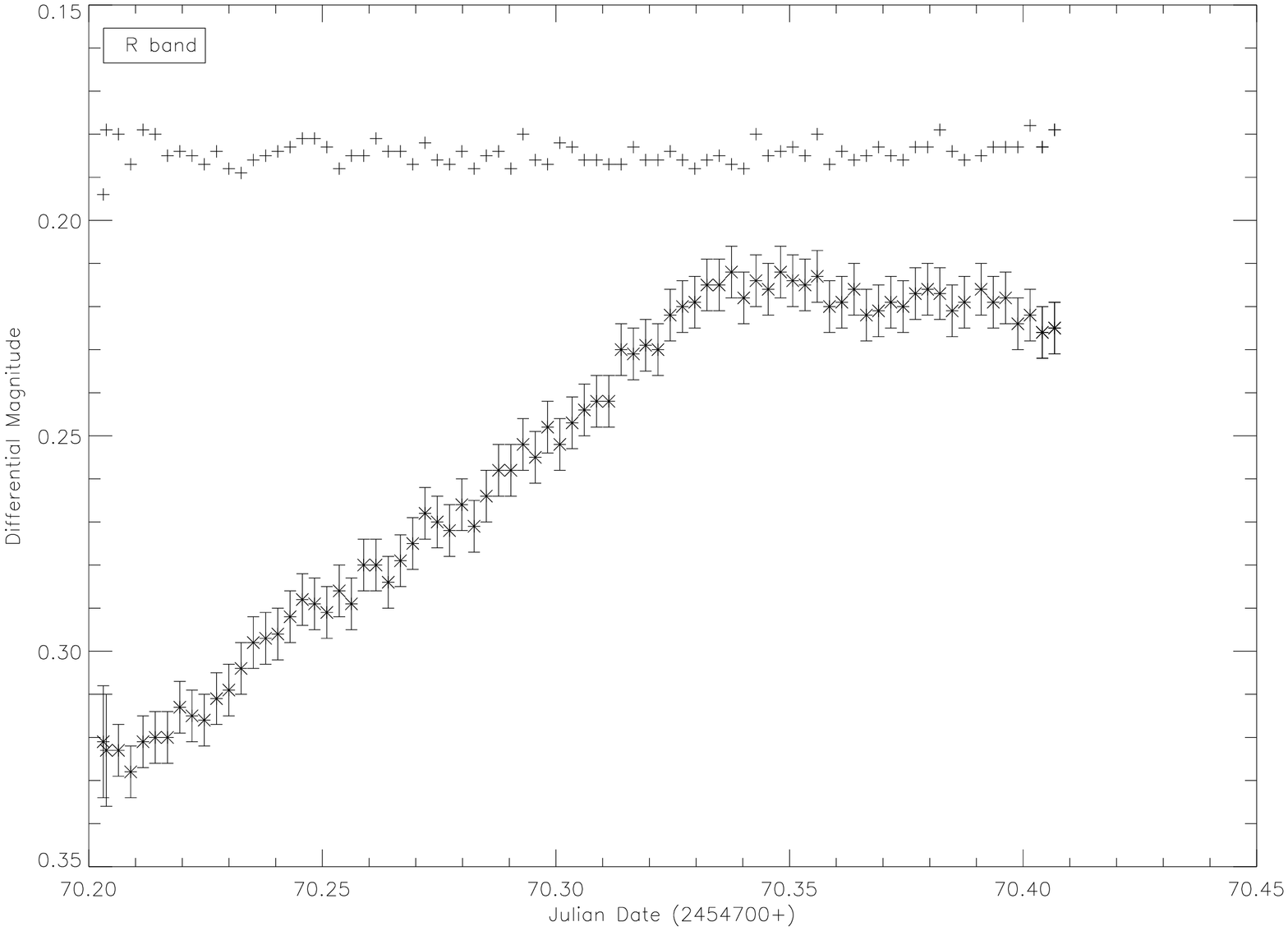}
\caption{Intranight differential light curves of period 1 in
\emph{V} and \emph{R} bands. The observation date are JD 2454765
(top left for \emph{V} band and top right for \emph{R} band), JD
2454766 (middle left for \emph{V} band and middle right for \emph{R}
band) and JD 2454770 (bottom left for \emph{V} band and bottom right
for \emph{R} band). Crosses represent the differential magnitude
between the source and star 5 while plus signs represent the
differential magnitude between star 5 and star 6 shifted by an
arbitrary offset.}
\end{figure}

\begin{figure}
\includegraphics[angle=90,width=0.5\hsize,height=0.4\hsize]{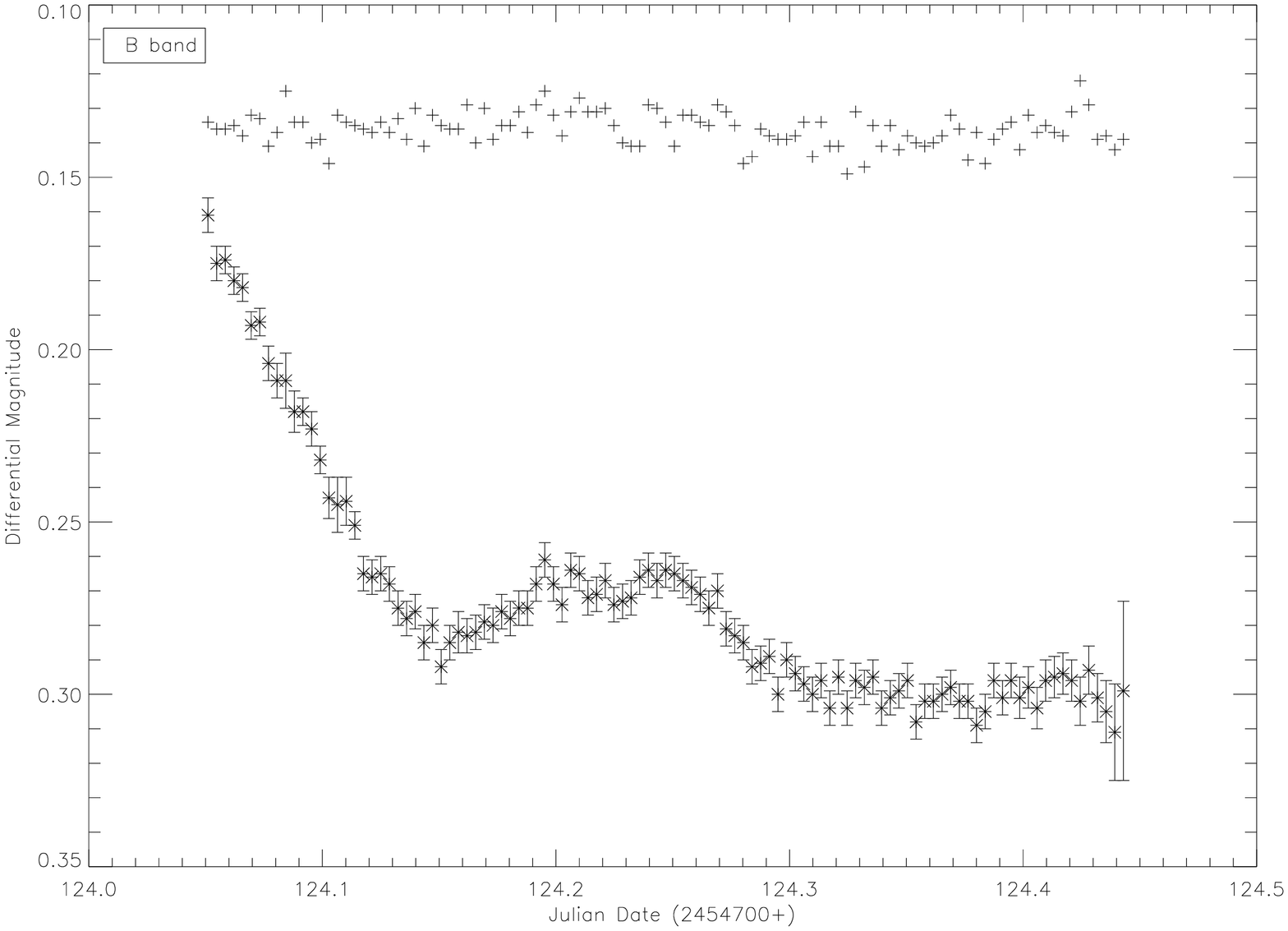}
\includegraphics[angle=90,width=0.5\hsize,height=0.4\hsize]{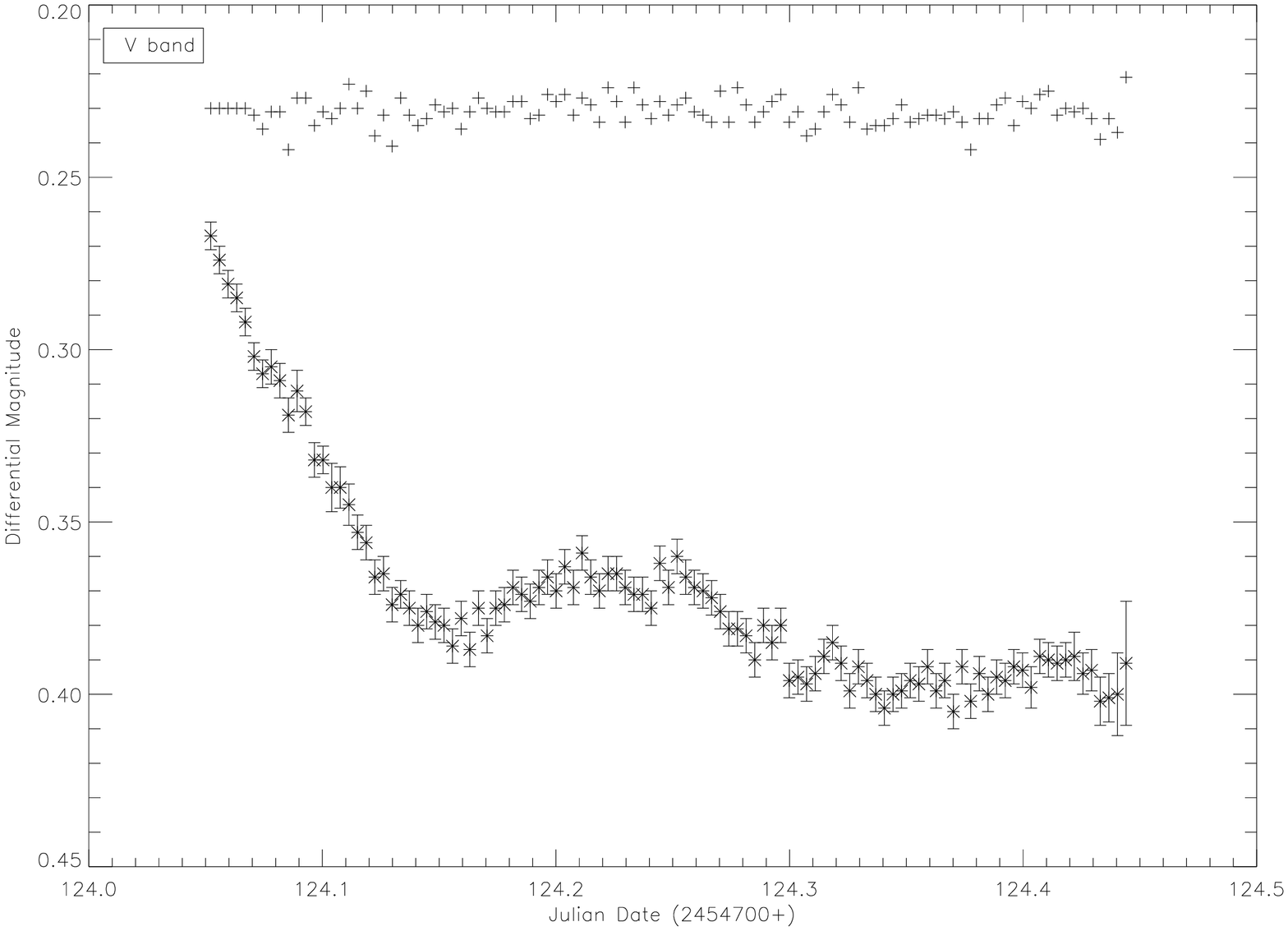}
\includegraphics[angle=90,width=0.5\hsize,height=0.4\hsize]{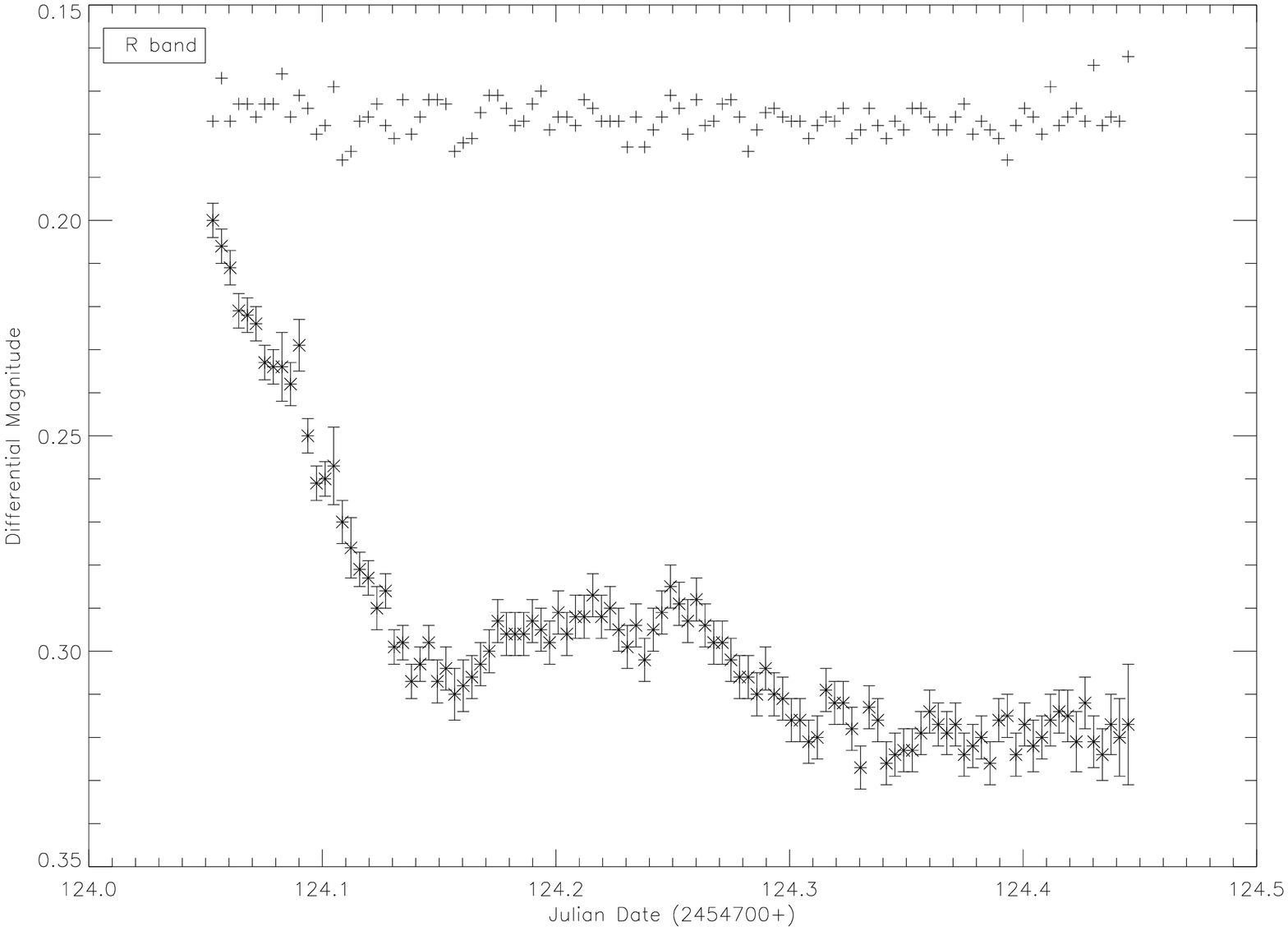}
\includegraphics[angle=90,width=0.5\hsize,height=0.4\hsize]{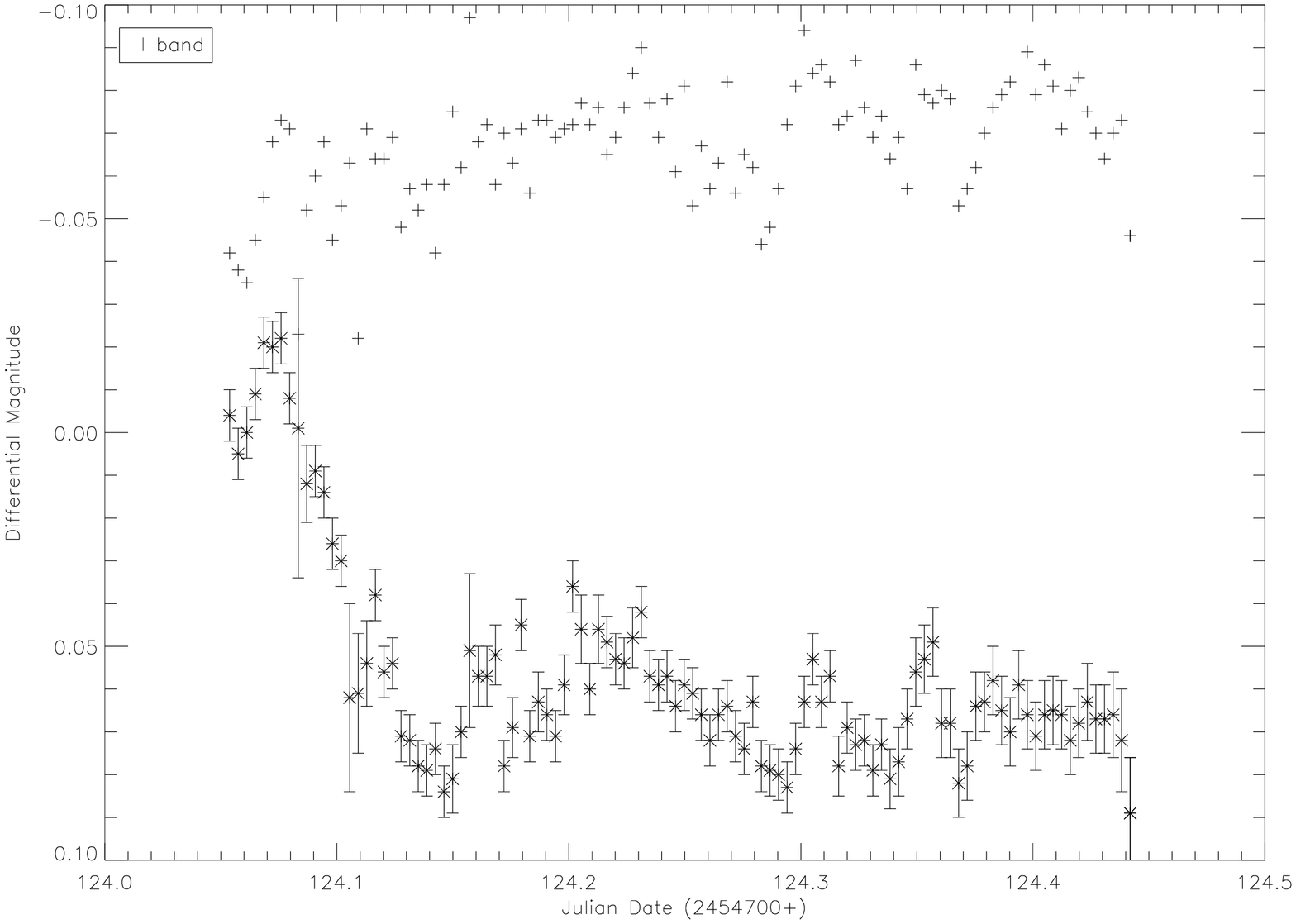}
\caption{Intranight differential light curves on JD 2454824 in
\emph{B}(top left), \emph{V}(top right), \emph{R}(bottom left) and
\emph{R}(bottom right) bands. Crosses represent the differential
magnitude between the source and star 5 while plus signs represent
the differential magnitude between star 5 and star 6 shifted by an
arbitrary offset.}
\end{figure}

\begin{figure}
\includegraphics[angle=90,width=0.5\hsize,height=0.4\hsize]{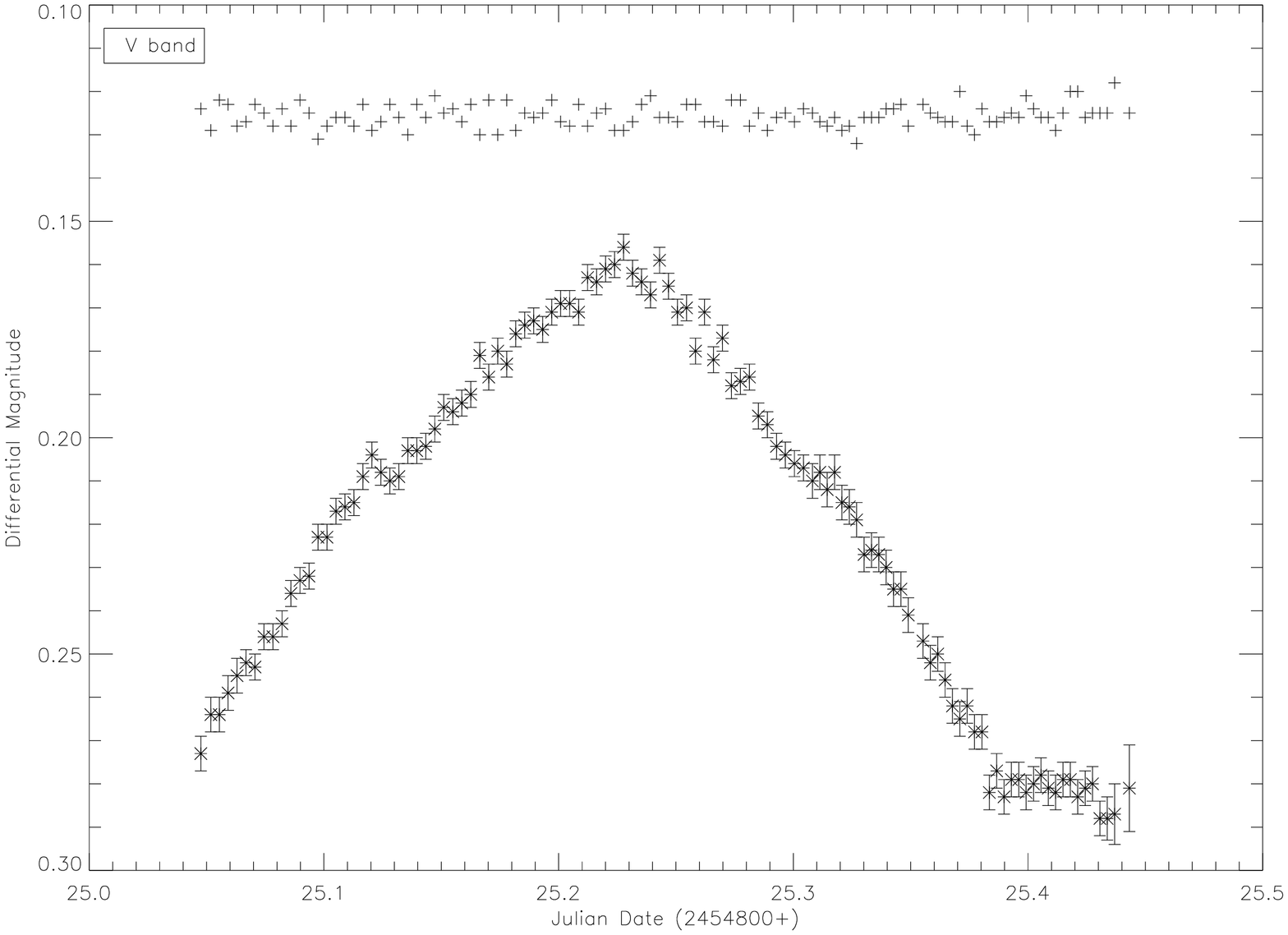}
\includegraphics[angle=90,width=0.5\hsize,height=0.4\hsize]{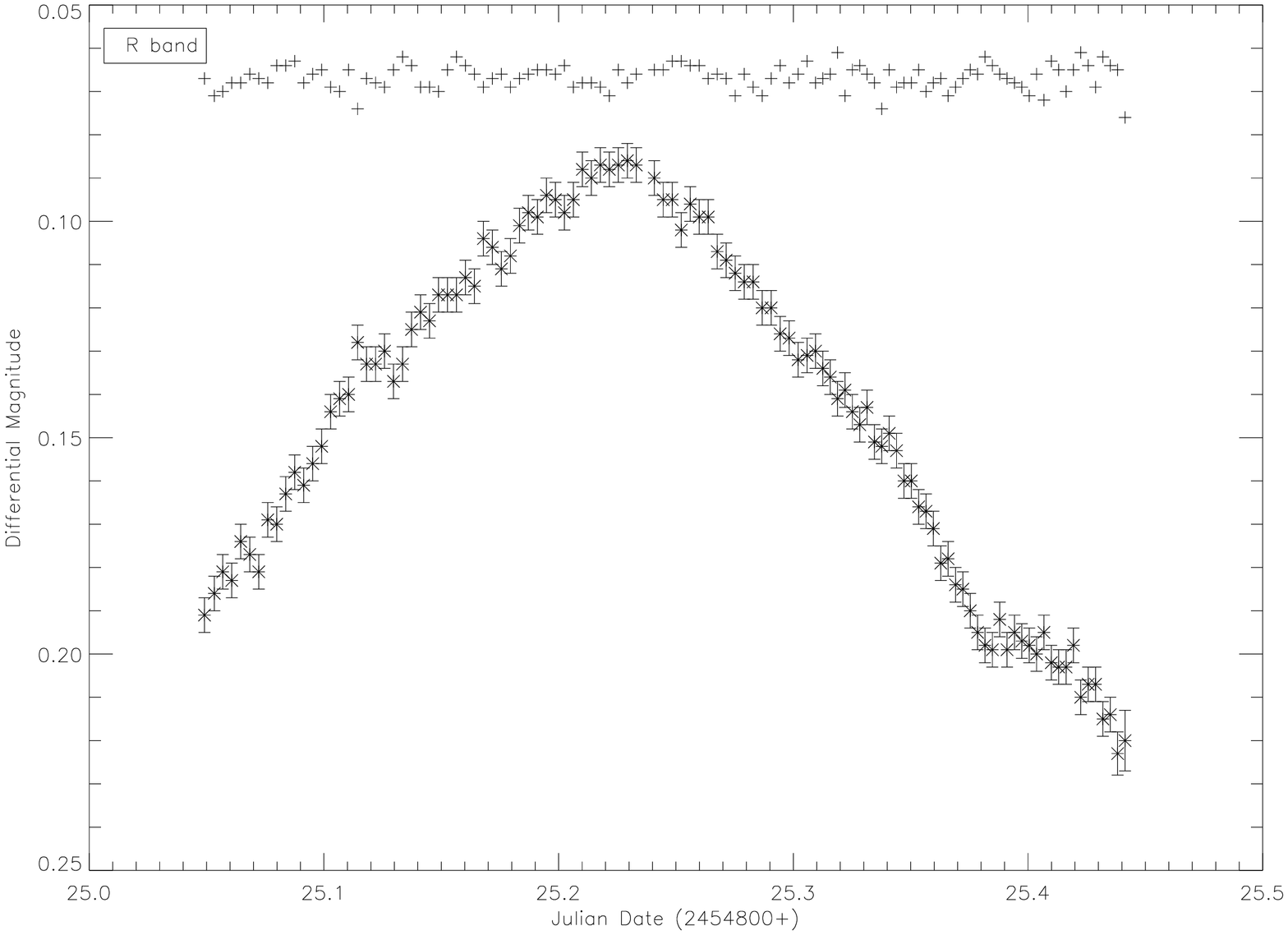}
\includegraphics[angle=90,width=0.5\hsize,height=0.4\hsize]{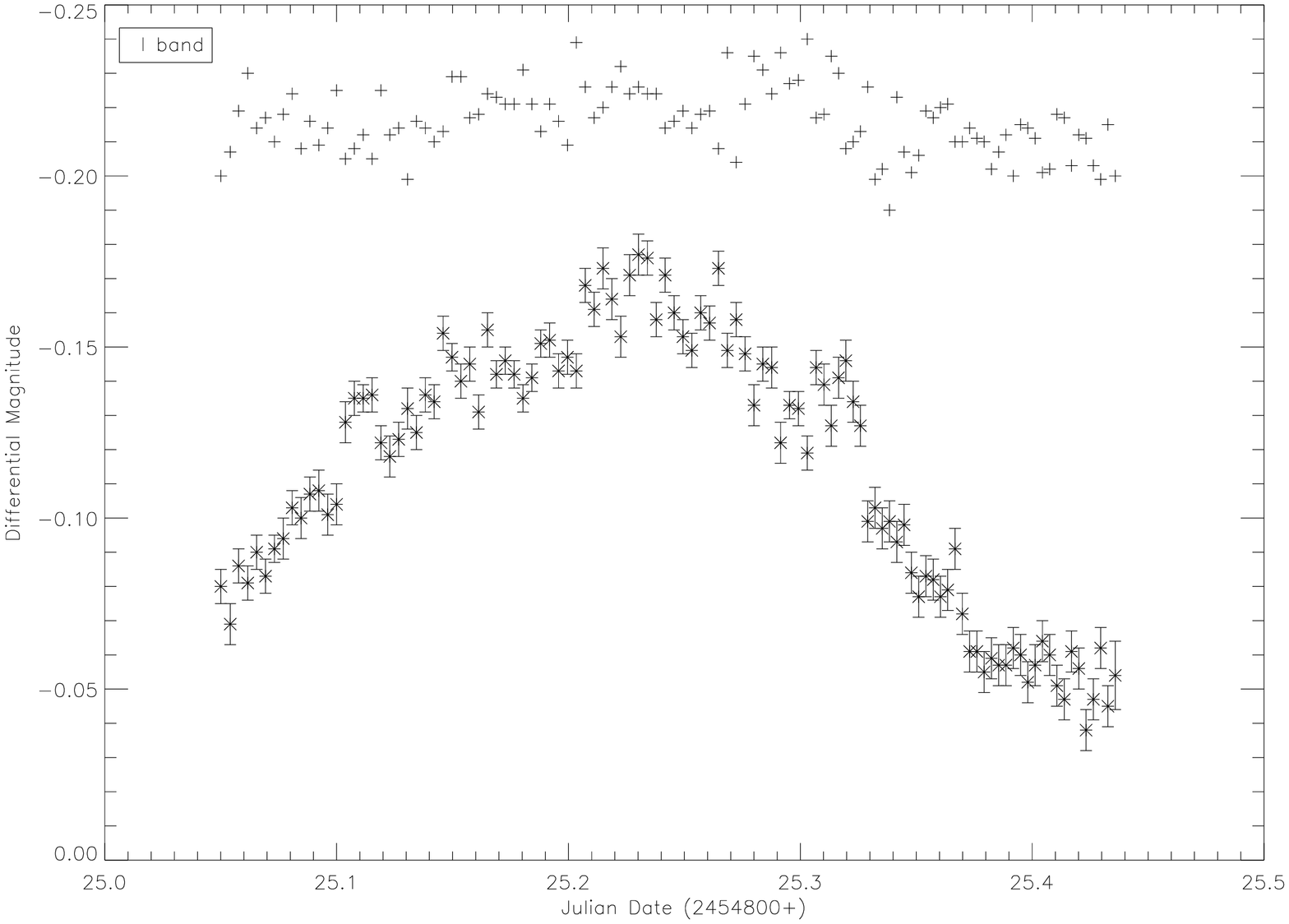}
\caption{Intranight differential light curves on JD 2454825 in
\emph{V}(top left), \emph{R}(top right) and \emph{I}(bottom left)
bands. Crosses represent the differential magnitude between the
source and star 5 while plus signs represent the differential
magnitude between star 5 and star 6 shifted by an arbitrary offset.}
\end{figure}

\begin{figure}
\includegraphics[angle=90,width=0.5\hsize,height=0.4\hsize]{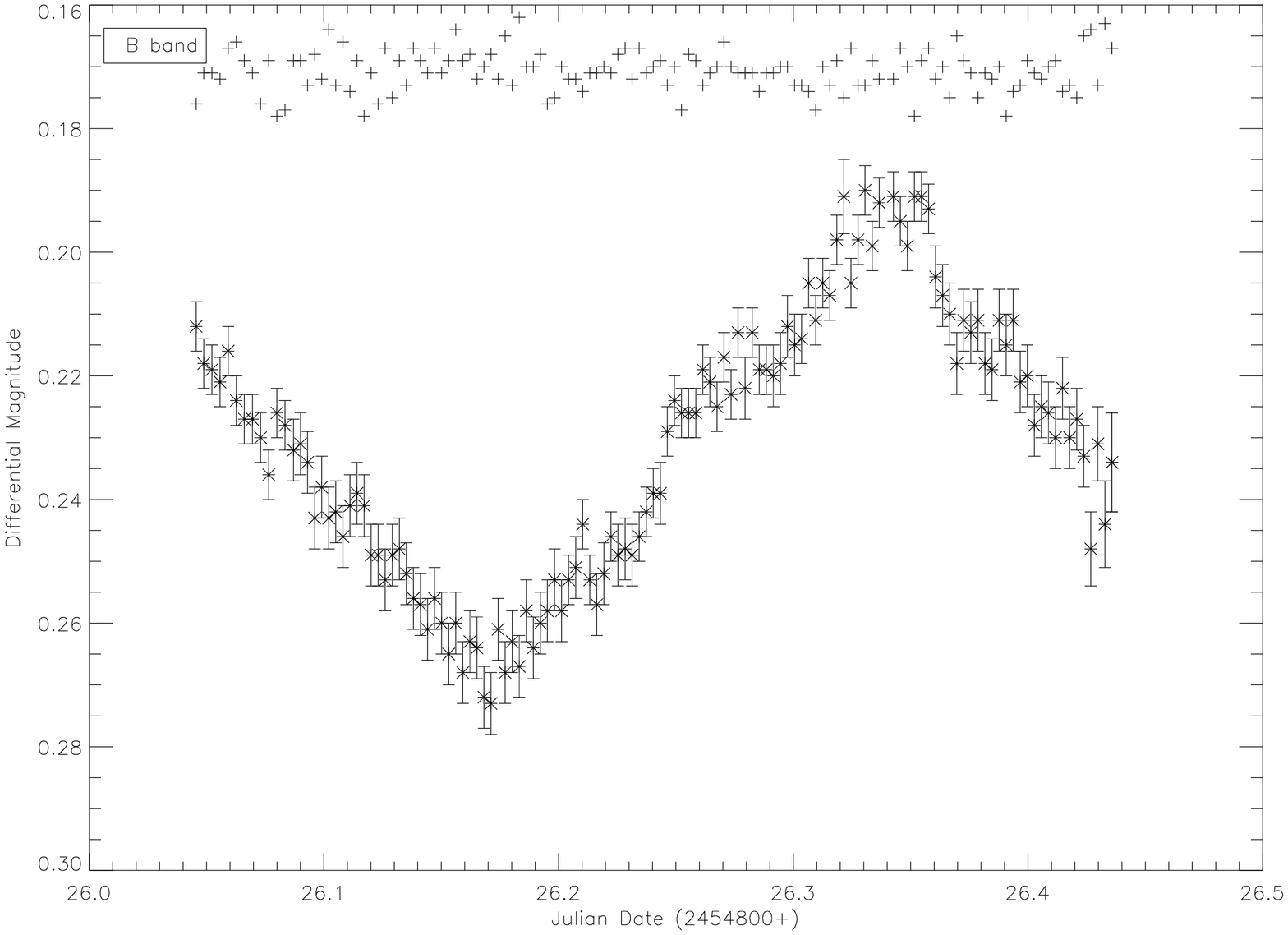}
\includegraphics[angle=90,width=0.5\hsize,height=0.4\hsize]{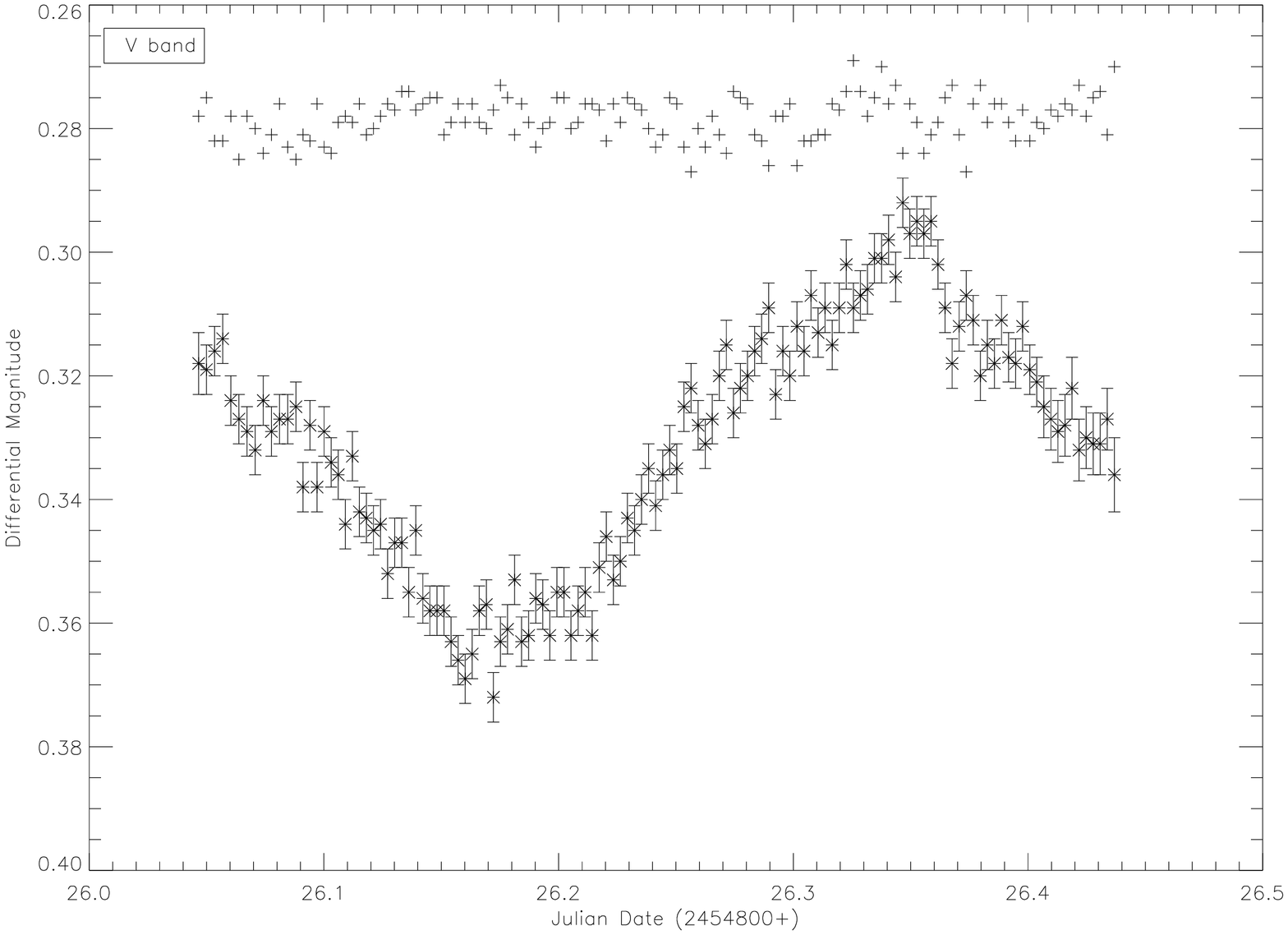}
\includegraphics[angle=90,width=0.5\hsize,height=0.4\hsize]{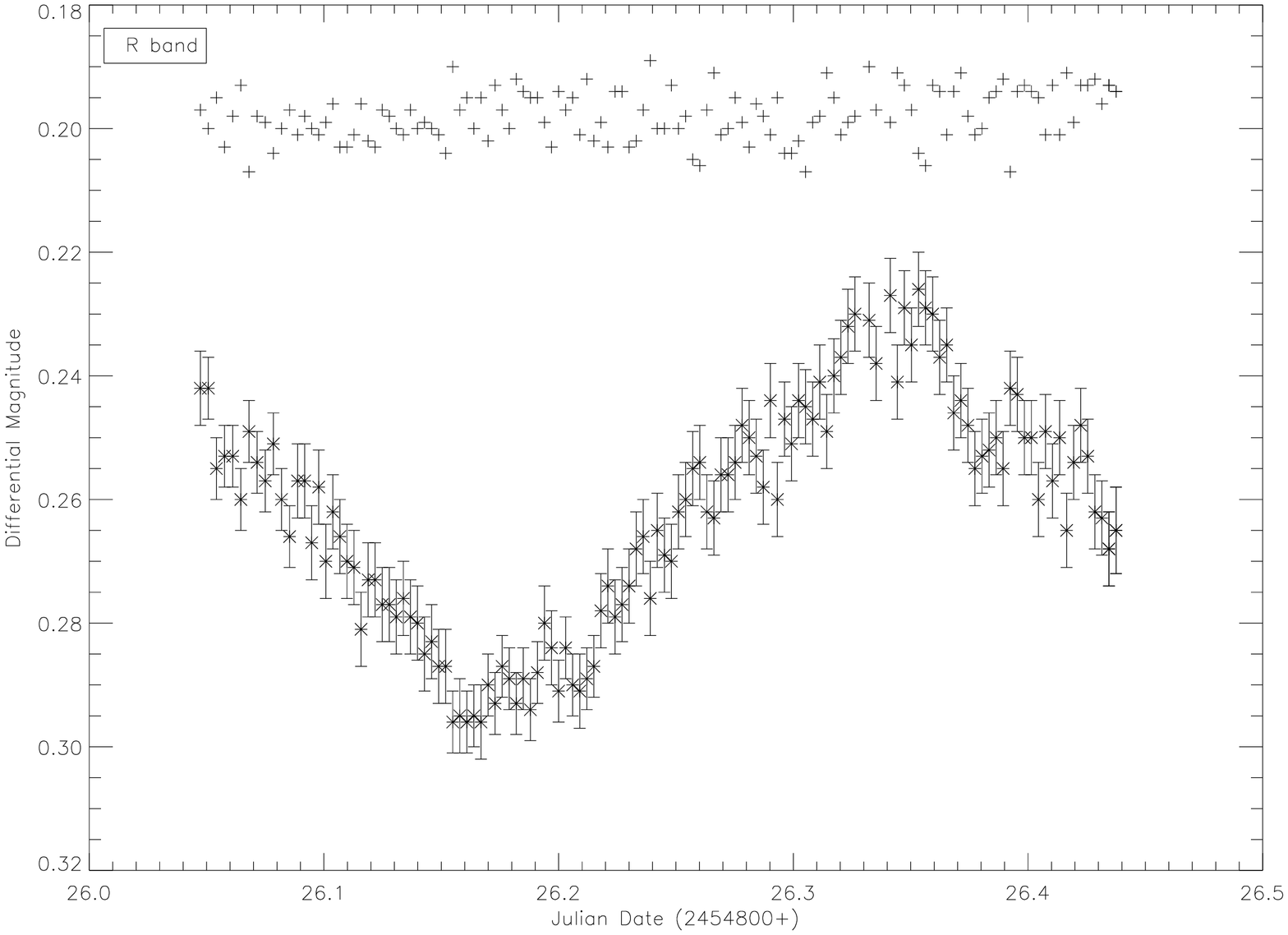}
\includegraphics[angle=90,width=0.5\hsize,height=0.4\hsize]{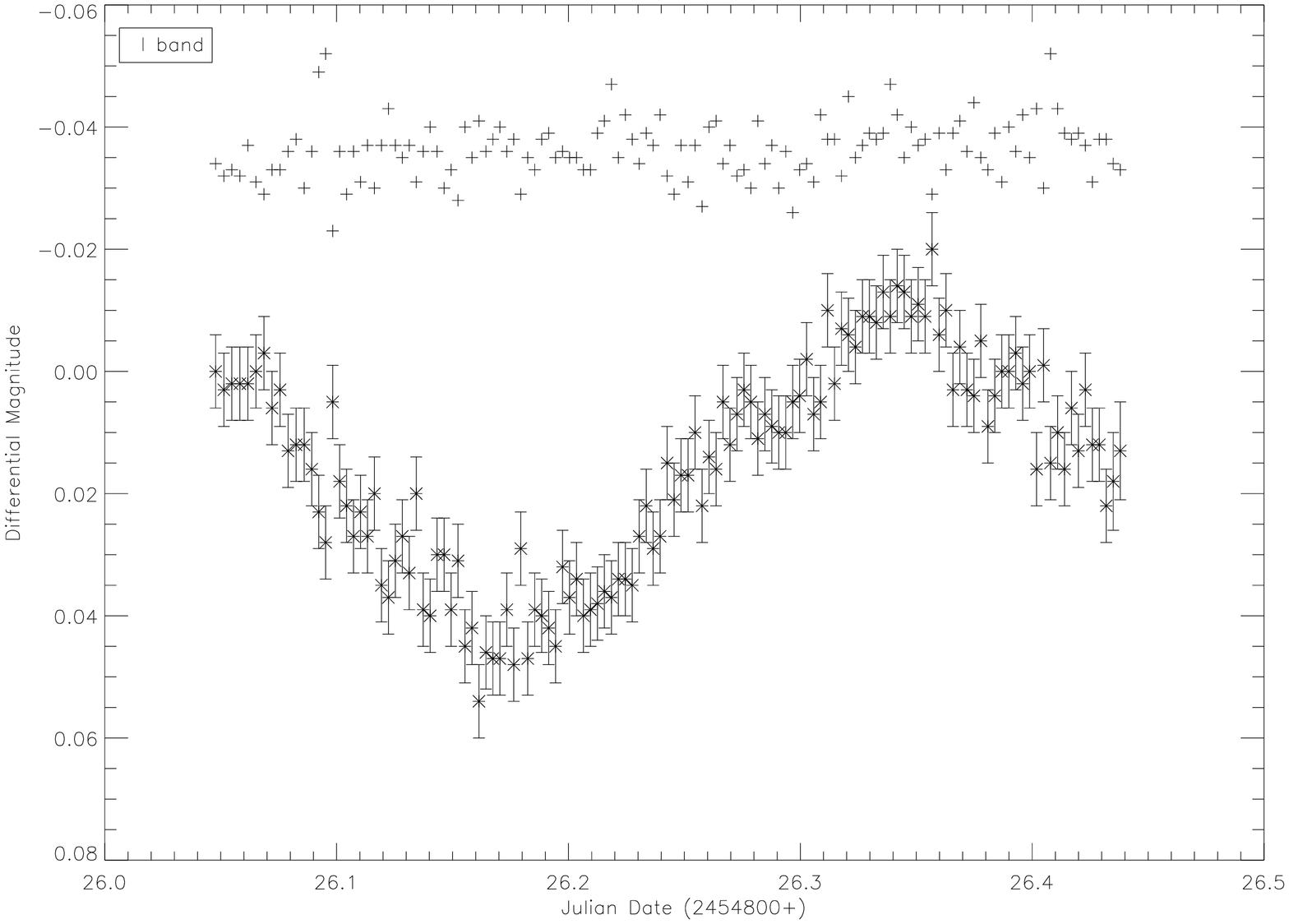}
\caption{Intranight differential light curves on JD 2454826 in
\emph{B}(top left), \emph{V}(top right), \emph{R}(bottom left) and
\emph{I}(bottom right) bands. Crosses represent the differential
magnitude between the source and star 5 while plus signs represent
the differential magnitude between star 5 and star 6.}
\end{figure}

\begin{figure}
\includegraphics[angle=90,width=0.5\hsize,height=0.4\hsize]{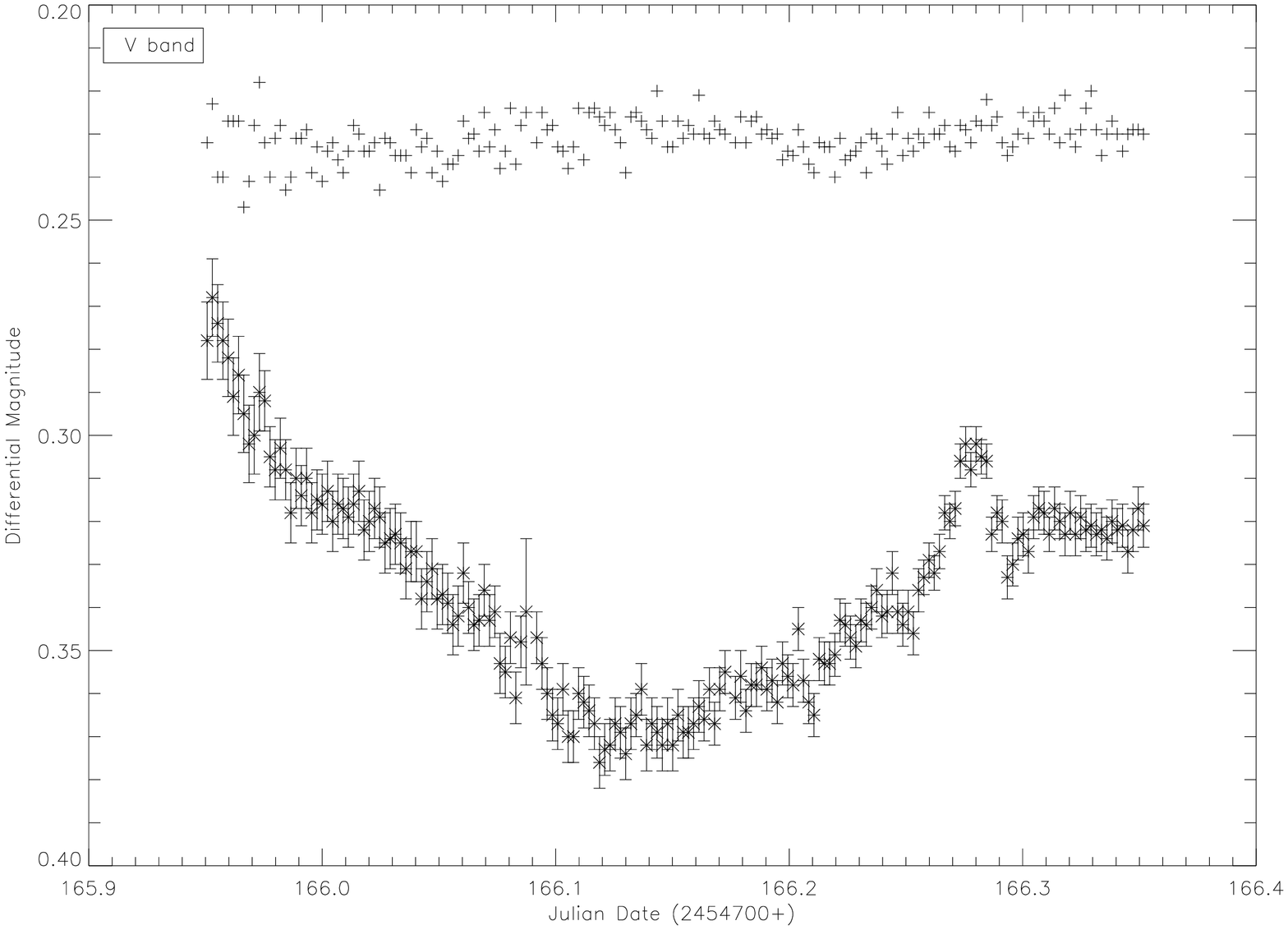}
\includegraphics[angle=90,width=0.5\hsize,height=0.4\hsize]{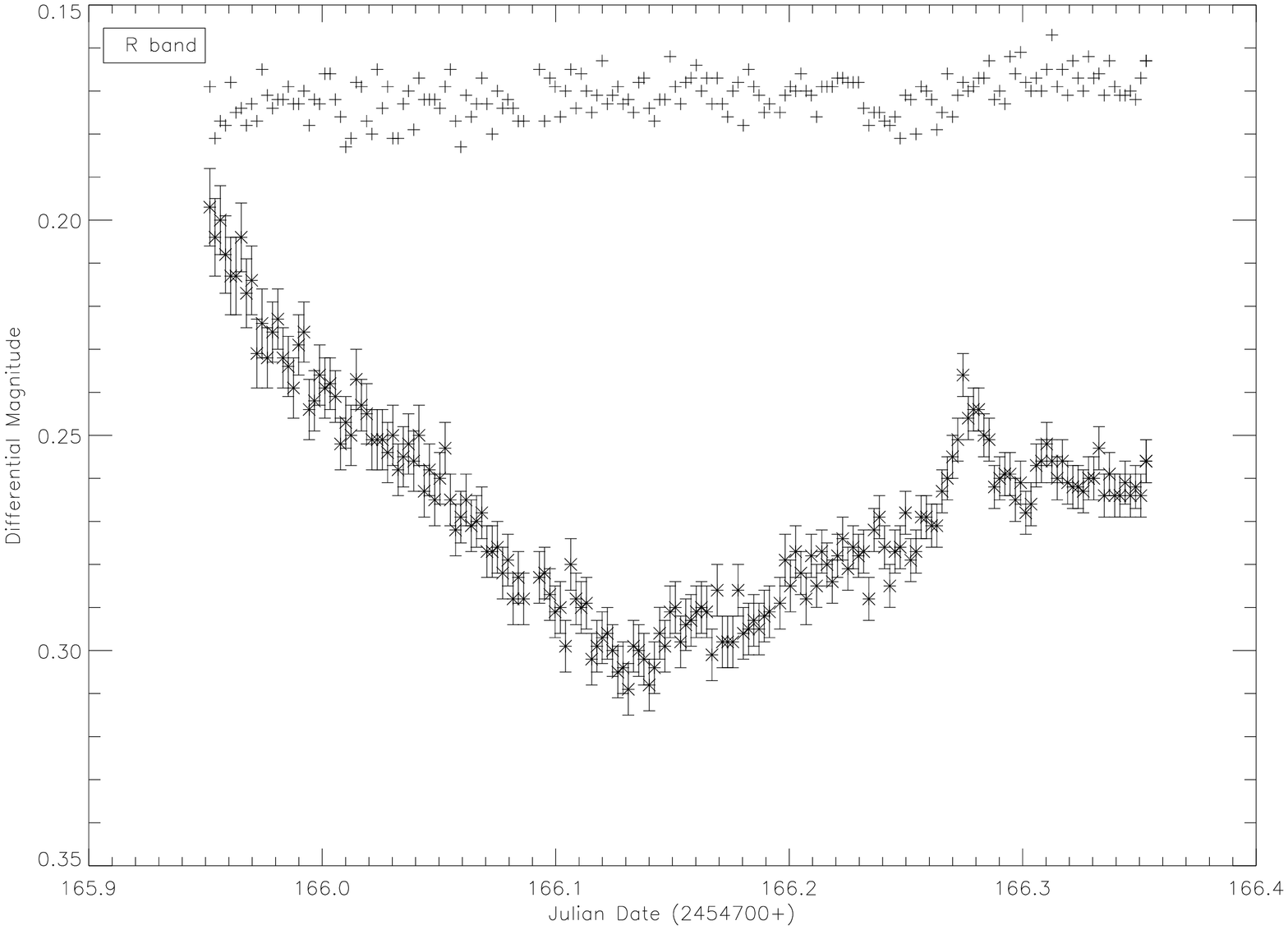}
\caption{Intranight differential light curves on JD 2454865 -
2454866 in \emph{V}(left) and \emph{R}(right) bands. Crosses
represent the differential magnitude between the source and star 5
while plus signs represent the differential magnitude between star 5
and star 6.}
\end{figure}

\begin{figure}
\includegraphics[angle=90,width=0.5\hsize,height=0.4\hsize]{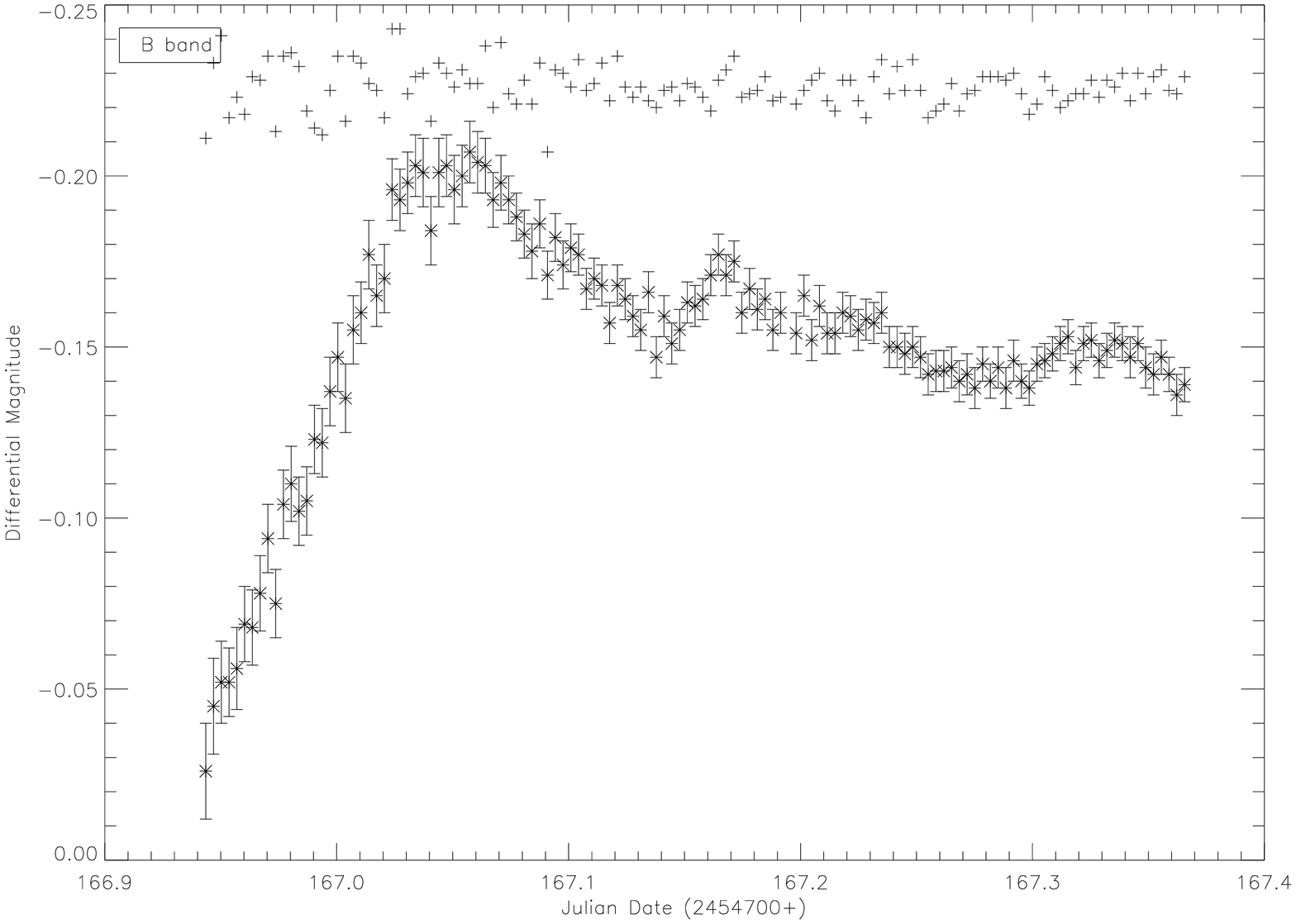}
\includegraphics[angle=90,width=0.5\hsize,height=0.4\hsize]{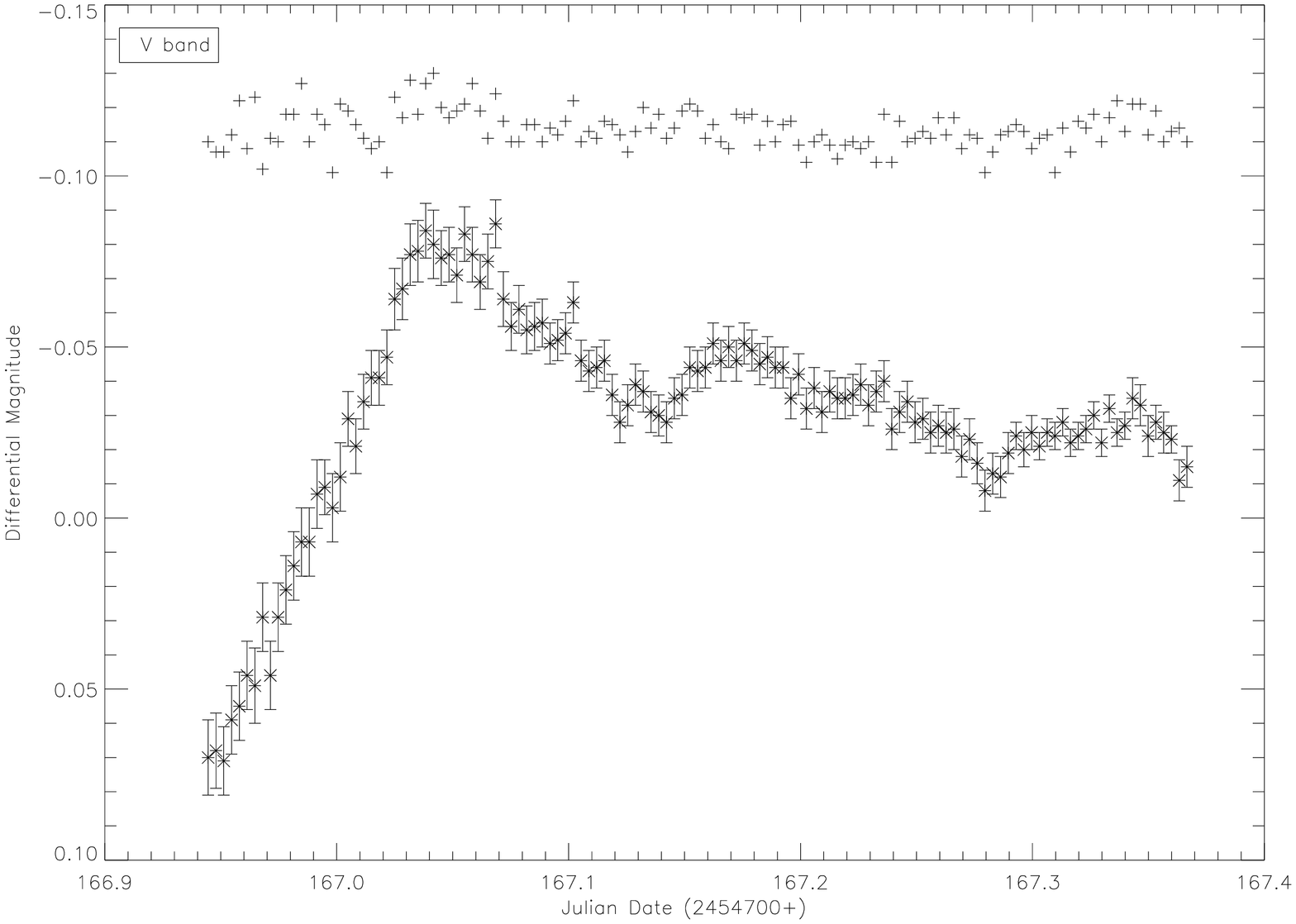}
\includegraphics[angle=90,width=0.5\hsize,height=0.4\hsize]{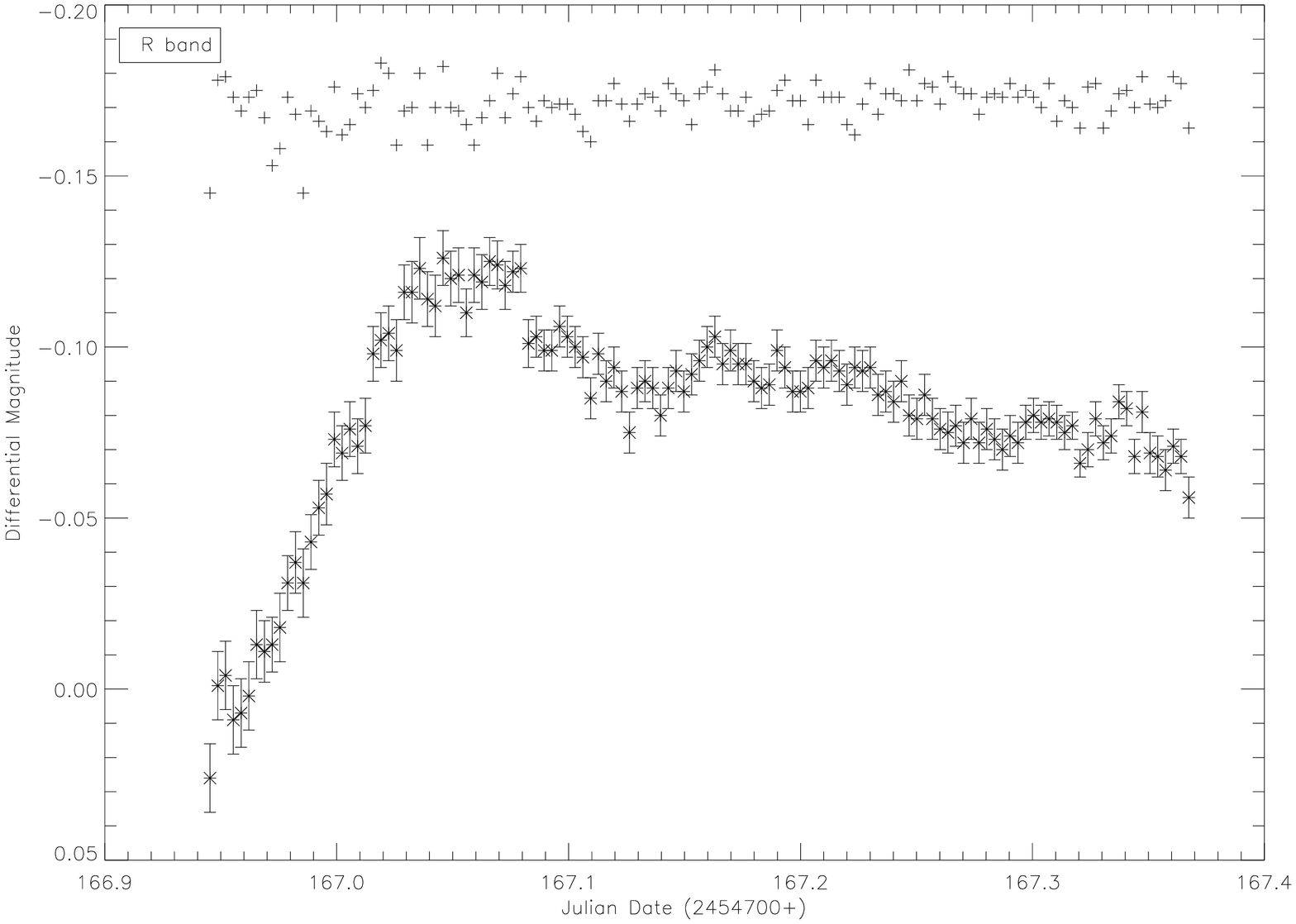}
\caption{Intranight differential light curves on JD 2454866 - JD
2454867 in \emph{B}(top left), \emph{V}(top right) and
\emph{R}(bottom left) bands. Crosses represent the differential
magnitude between the source and star 5 while plus signs represent
the differential magnitude between star 5 and star 6.}
\end{figure}

\begin{figure}
\includegraphics[angle=90,width=0.5\hsize,height=0.4\hsize]{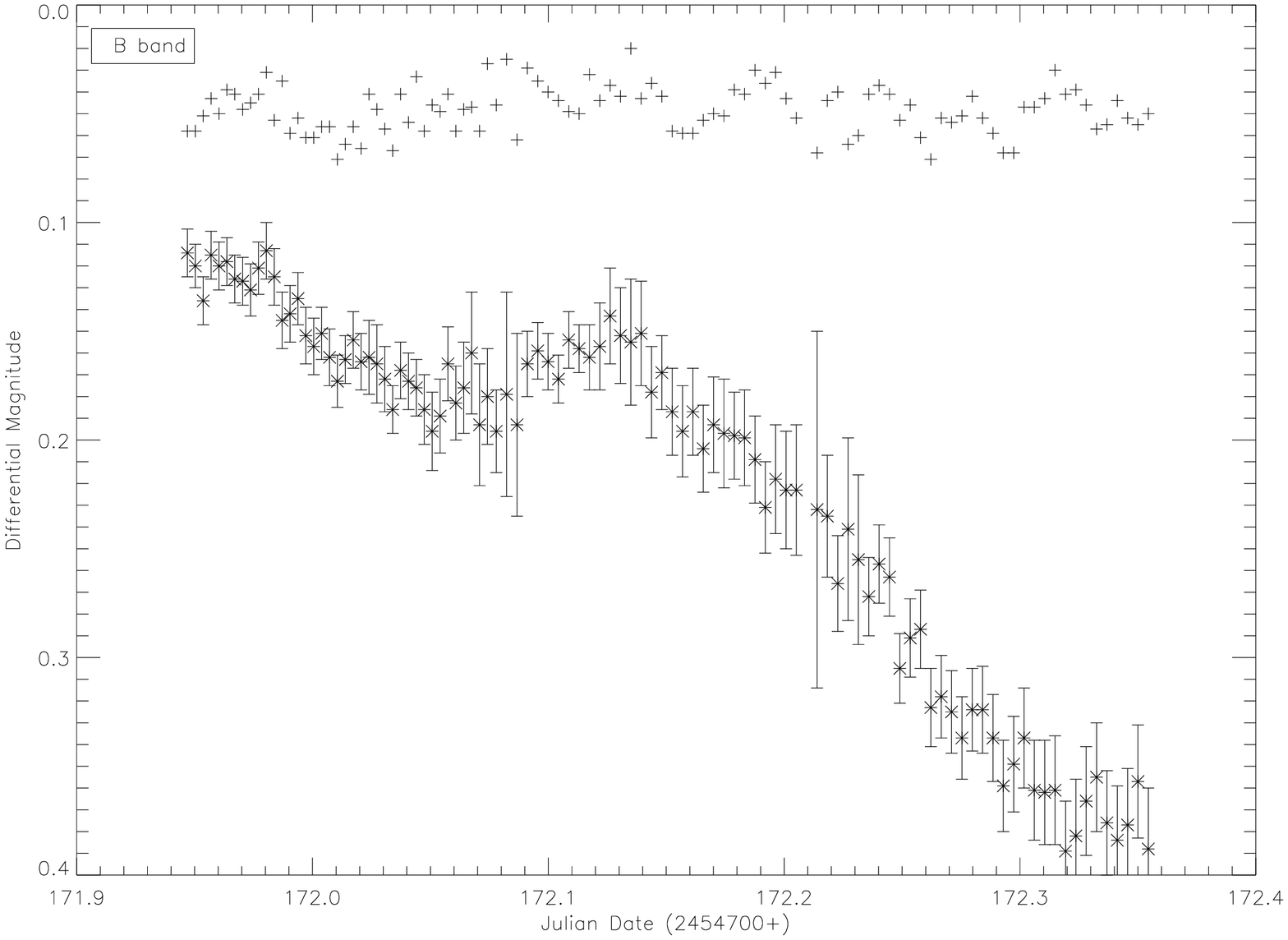}
\includegraphics[angle=90,width=0.5\hsize,height=0.4\hsize]{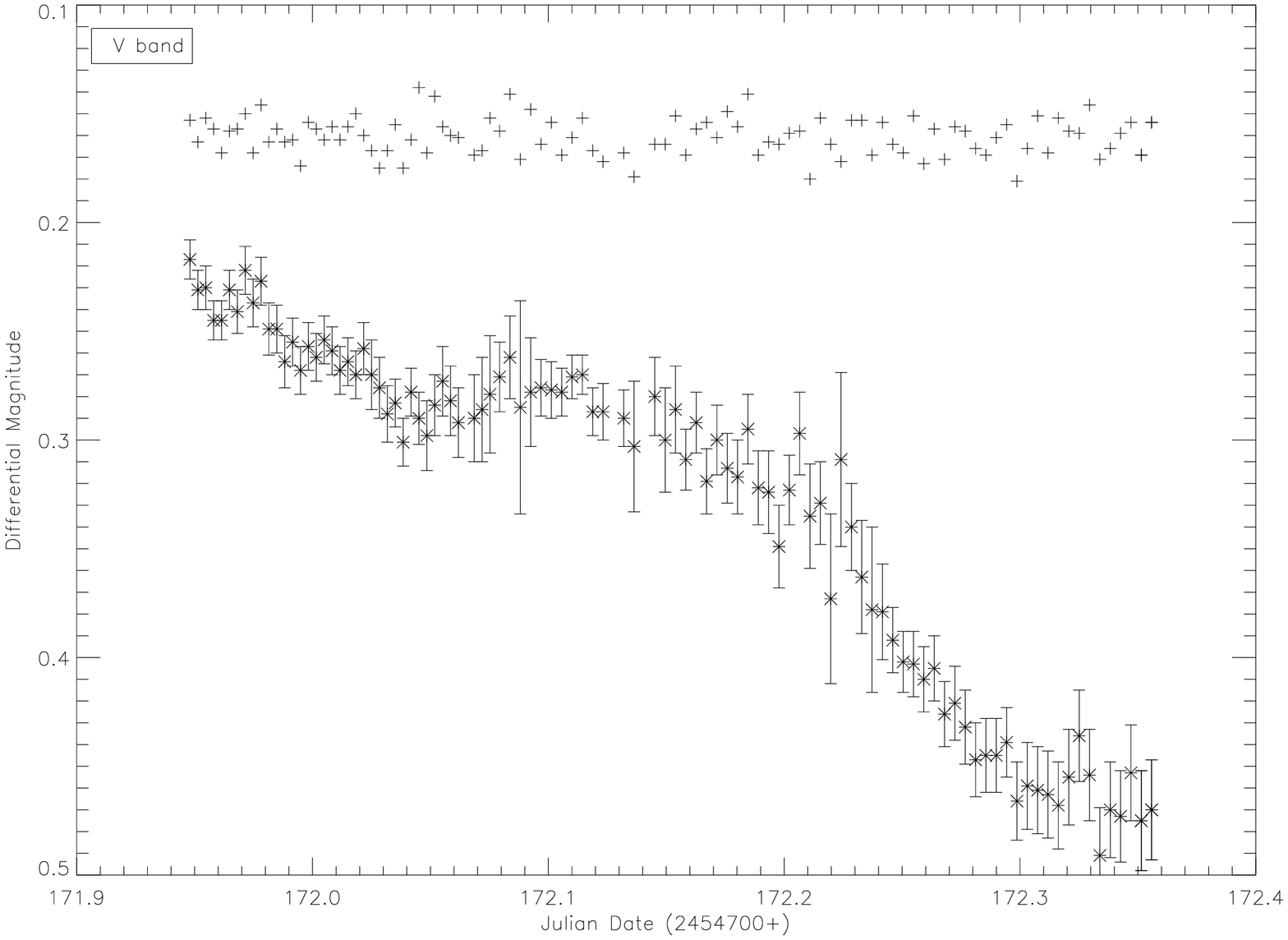}
\includegraphics[angle=90,width=0.5\hsize,height=0.4\hsize]{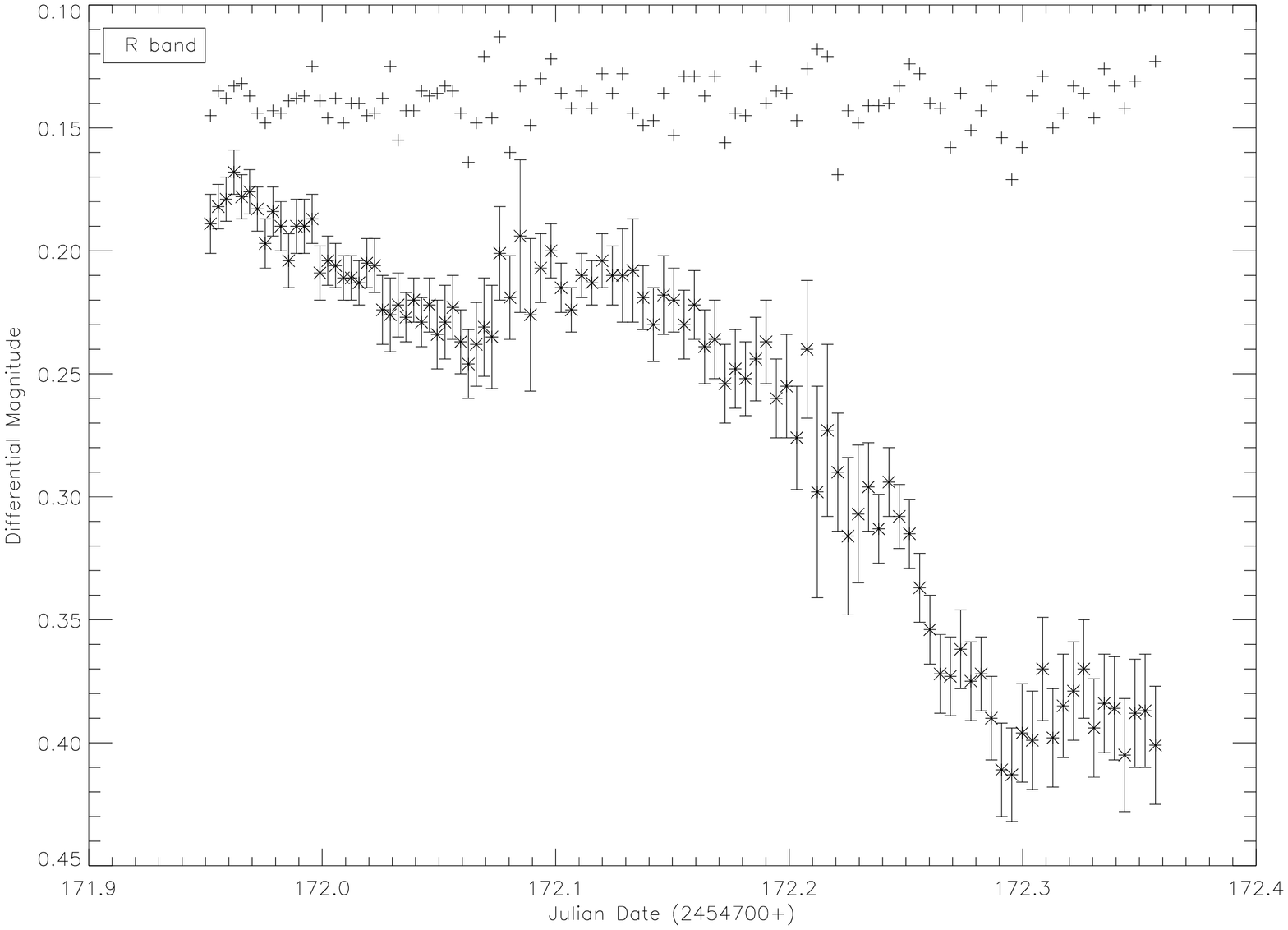}
\includegraphics[angle=90,width=0.5\hsize,height=0.4\hsize]{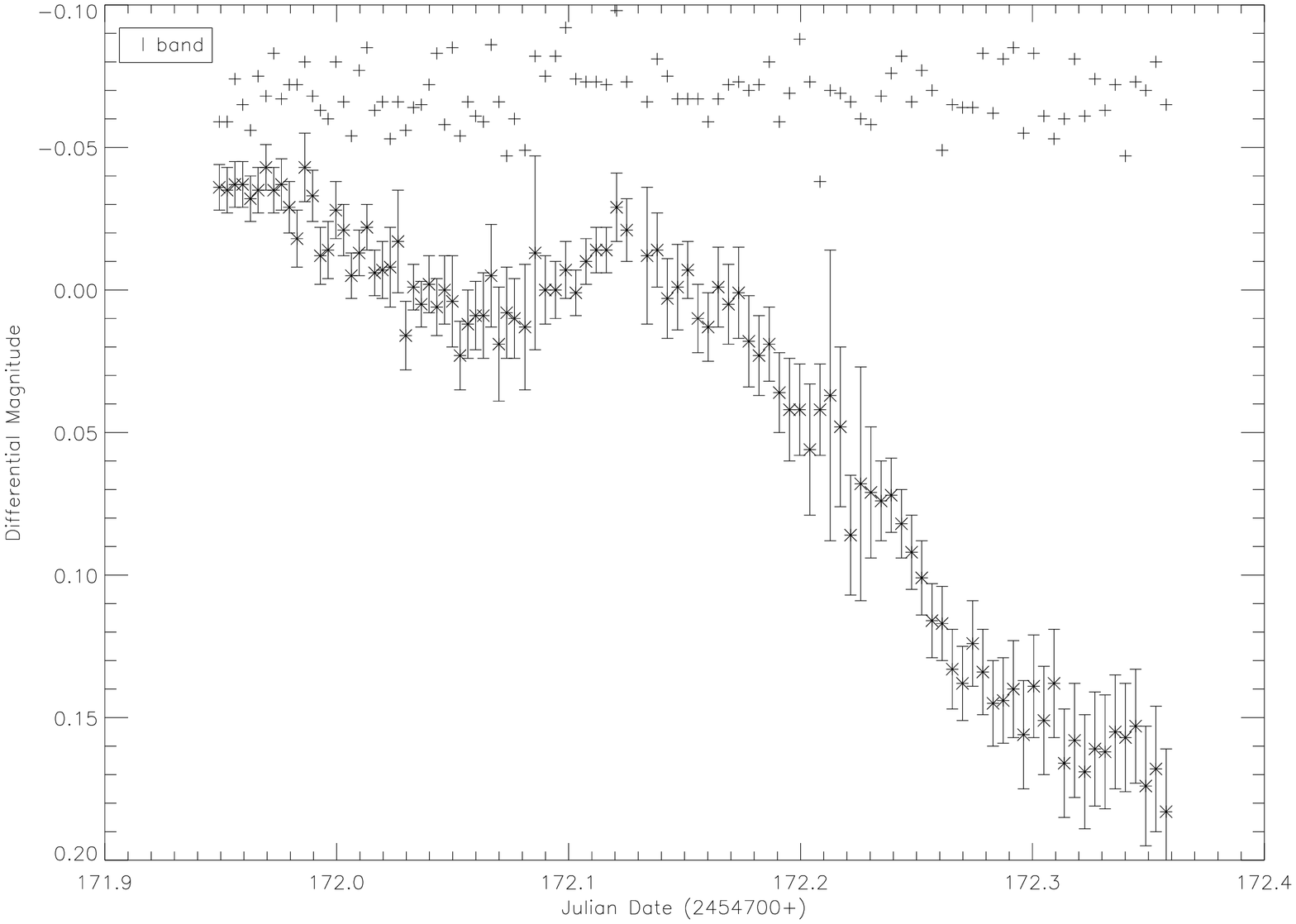}
\caption{Intranight differential light curves on JD 2454871 - JD
2454872 in \emph{B}(top left), \emph{V}(top right), \emph{R}(bottom
left) and \emph{I}(bottom right) bands. Crosses represent the
differential magnitude between the source and star 5 while plus
signs represent the differential magnitude between star 5 and star
6.}
\end{figure}

\begin{figure}
\includegraphics[angle=90,width=0.5\hsize,height=0.4\hsize]{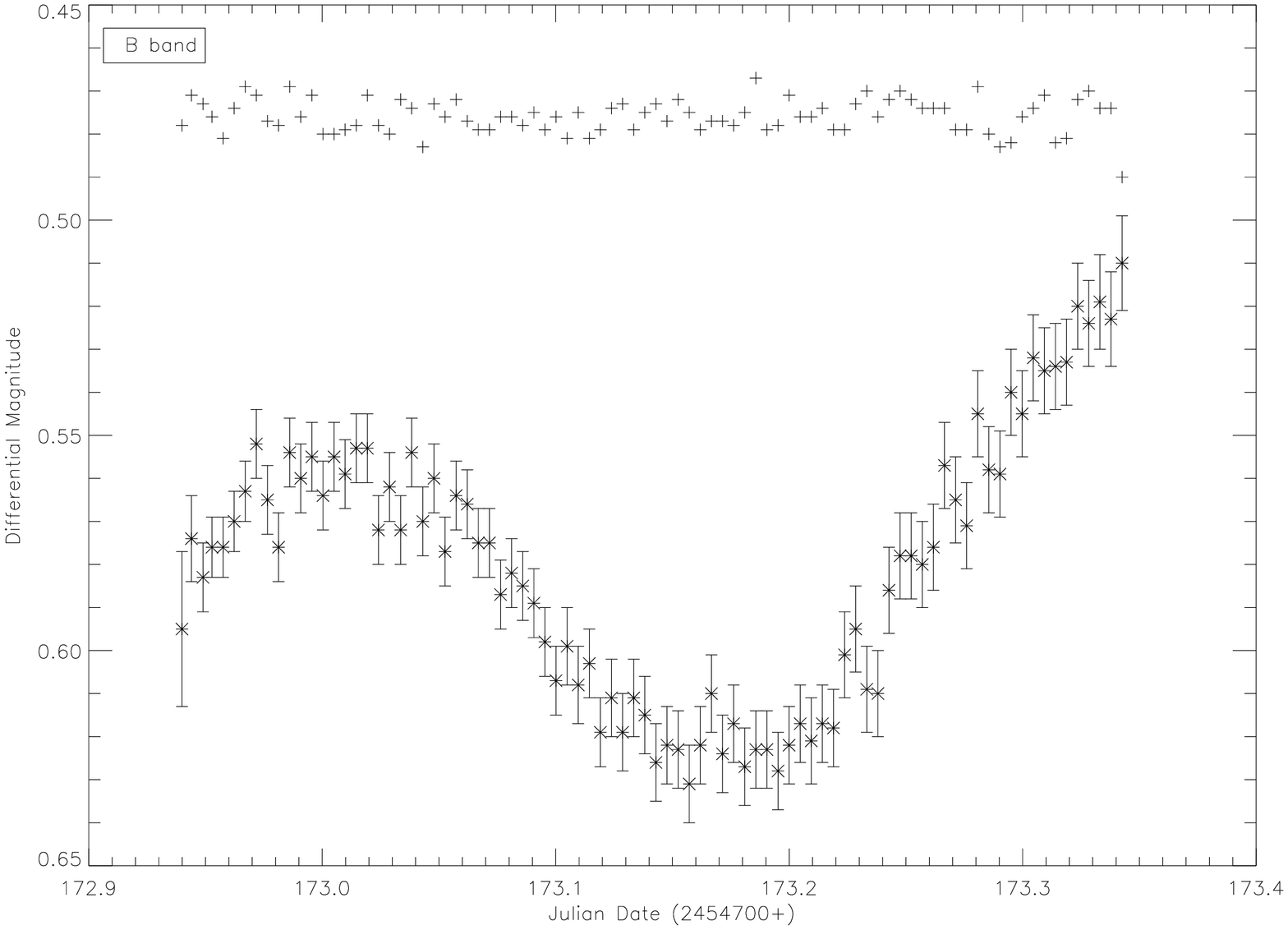}
\includegraphics[angle=90,width=0.5\hsize,height=0.4\hsize]{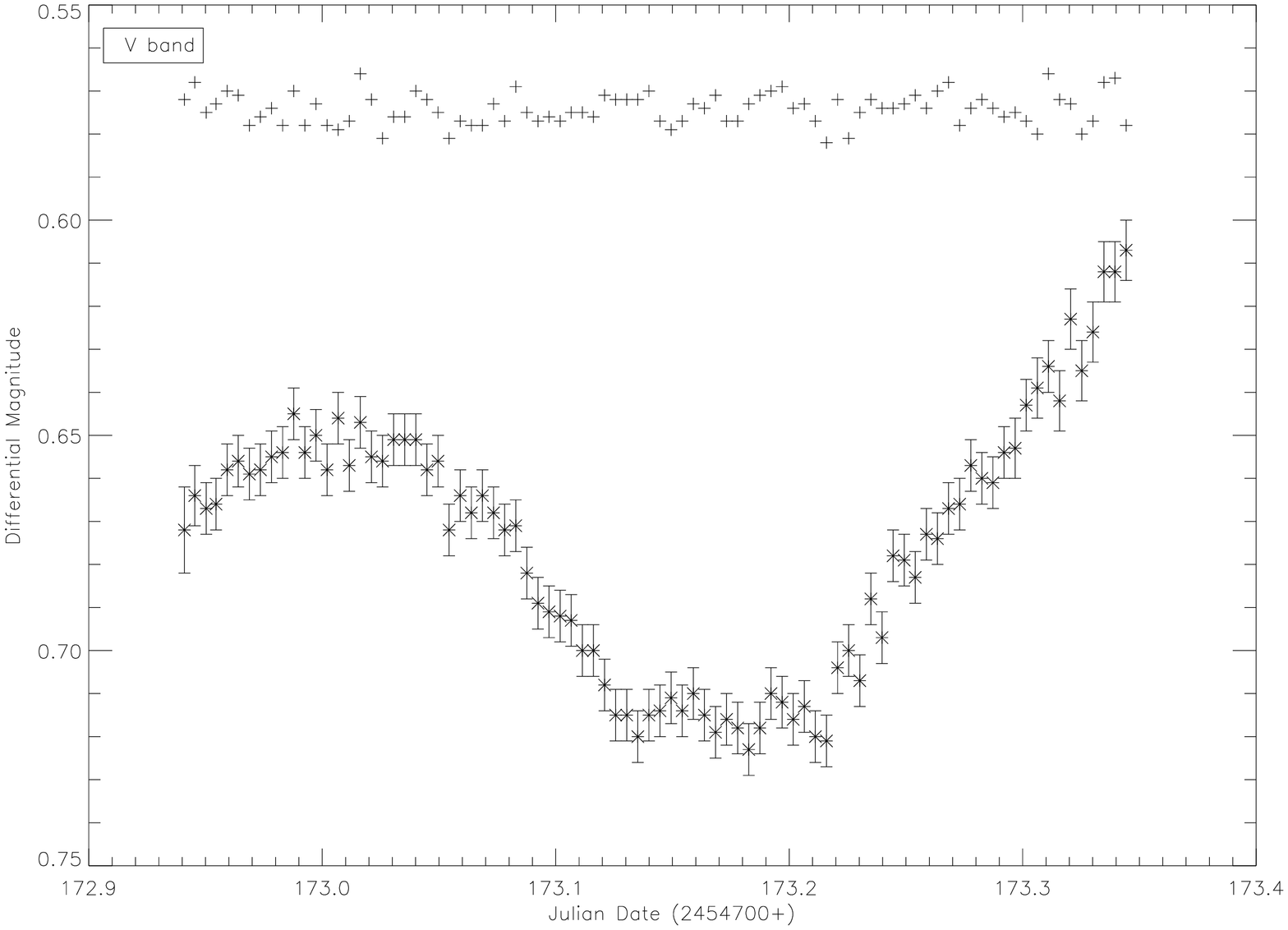}
\includegraphics[angle=90,width=0.5\hsize,height=0.4\hsize]{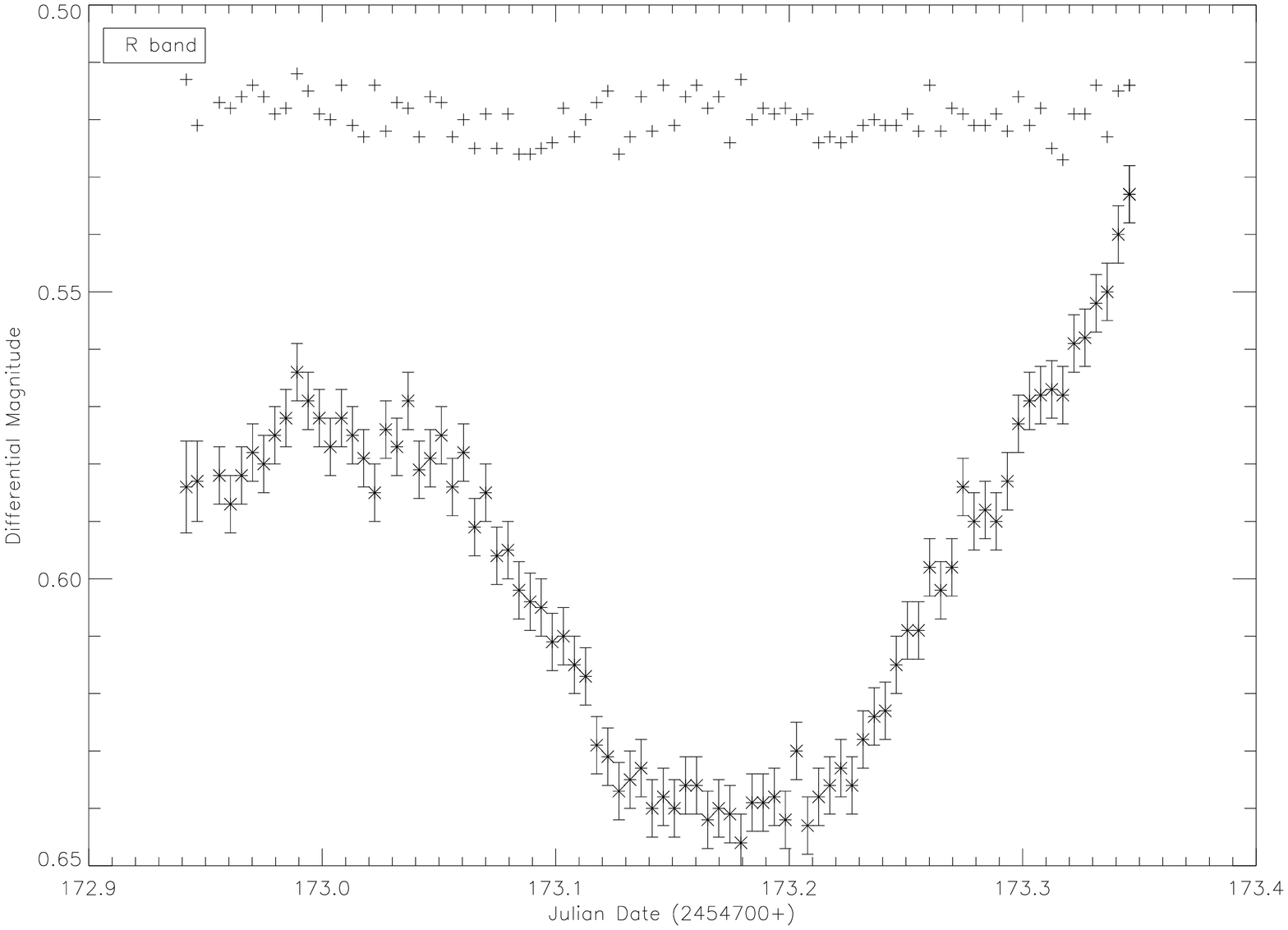}
\caption{Intranight differential light curves on JD 2454872 - JD
2454873 in \emph{B}(top left), \emph{V}(top right) and
\emph{R}(bottom left) bands. Crosses represent the differential
magnitude between the source and star 5 while plus signs represent
the differential magnitude between star 5 and star 6.}
\end{figure}

\begin{figure}[c]
\begin{center}
\includegraphics[angle=90,width=1.1\hsize,height=0.9\hsize]{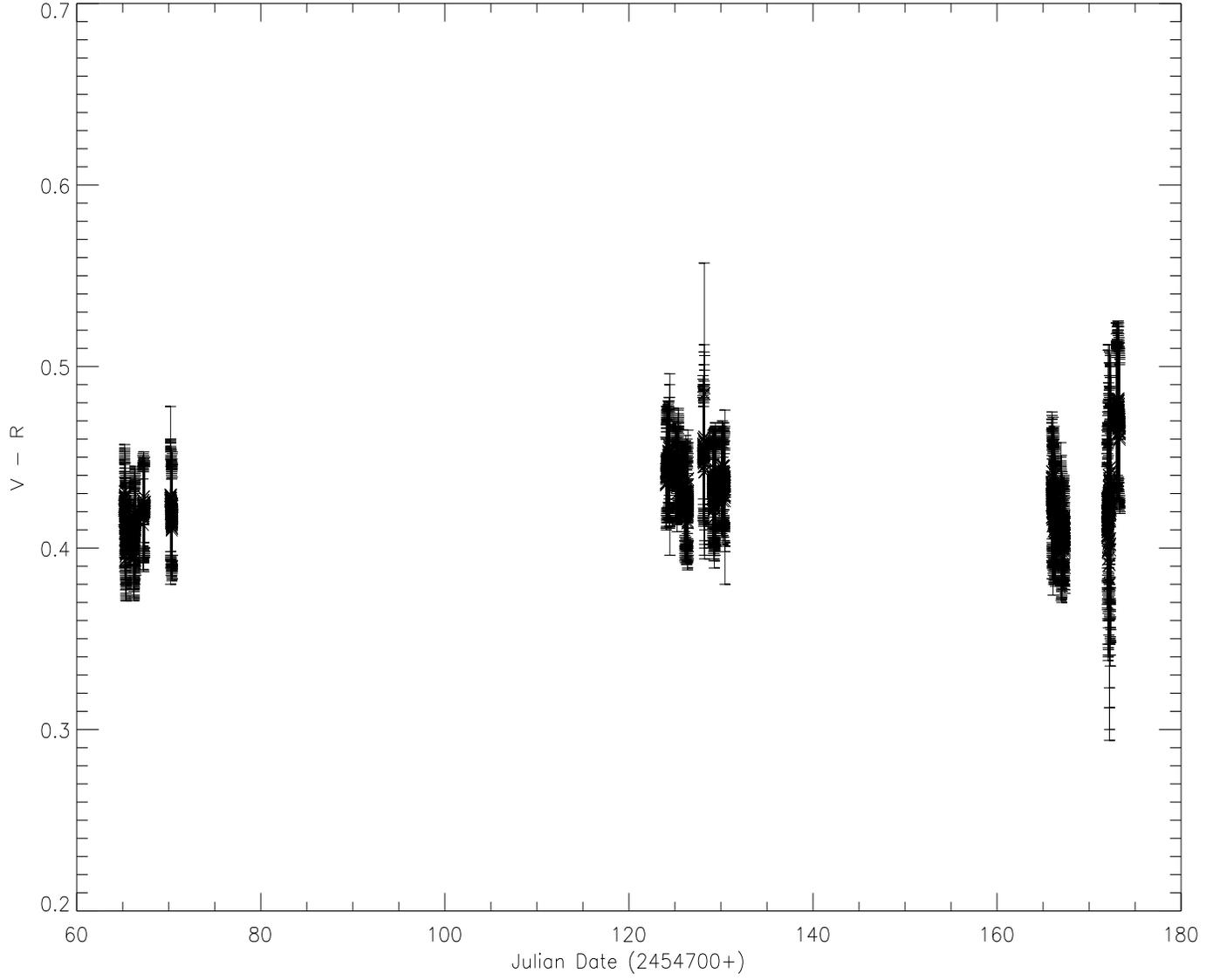}
\caption{ Color vs. time in the whole monitoring campaign. }
\end{center}
\end{figure}

\begin{figure}[c]
\begin{center}
\includegraphics[angle=90,width=1.1\hsize,height=0.9\hsize]{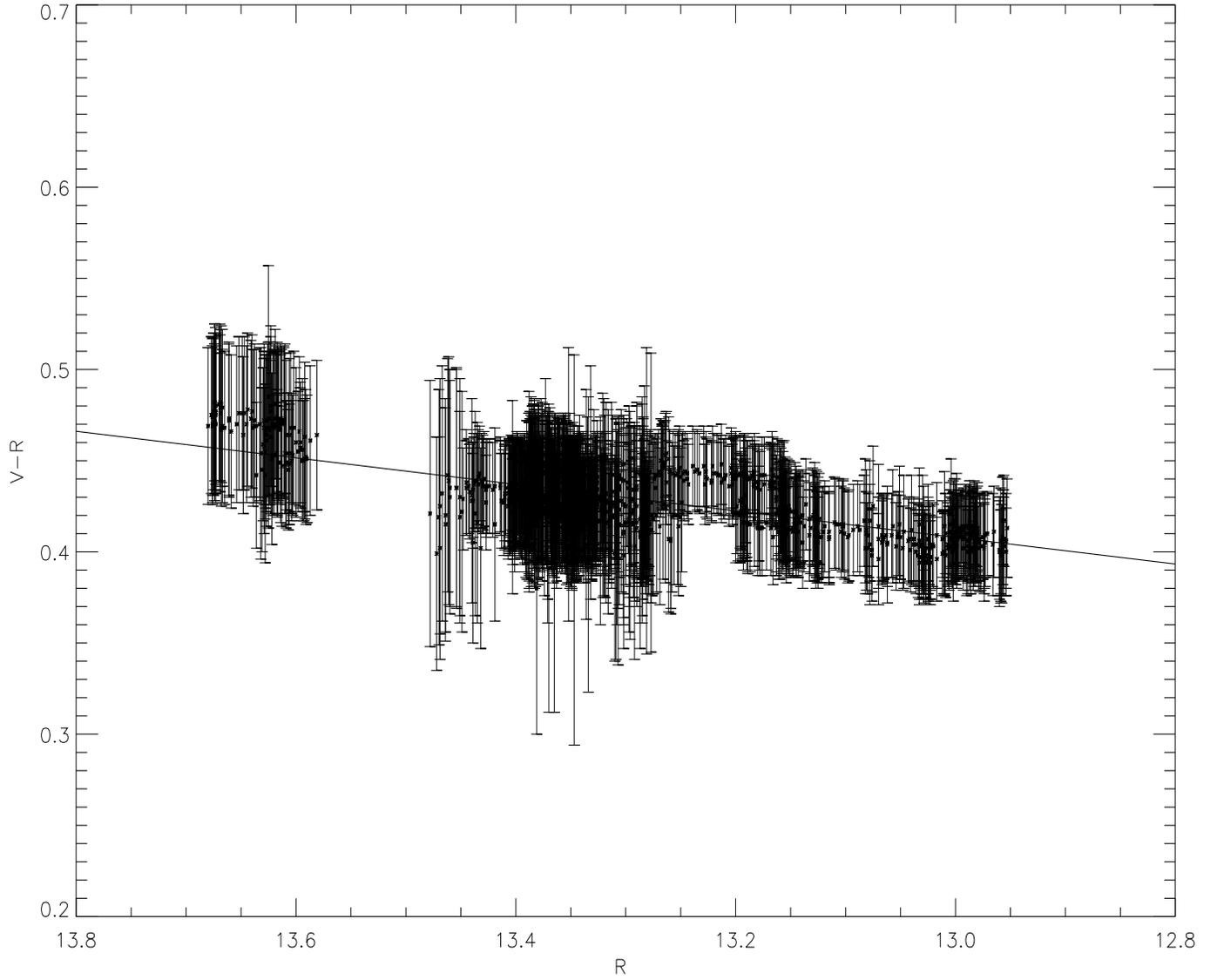}
\caption{Color vs. magnitude in the whole monitoring campaign. The
solid line is the linear fit to the points. The Pearson correlation
coefficient is 0.753, indicating strong correlations between color
and magnitude. }
\end{center}
\end{figure}

\begin{figure}

\includegraphics[angle=90,width=0.5\hsize,height=0.6\hsize]{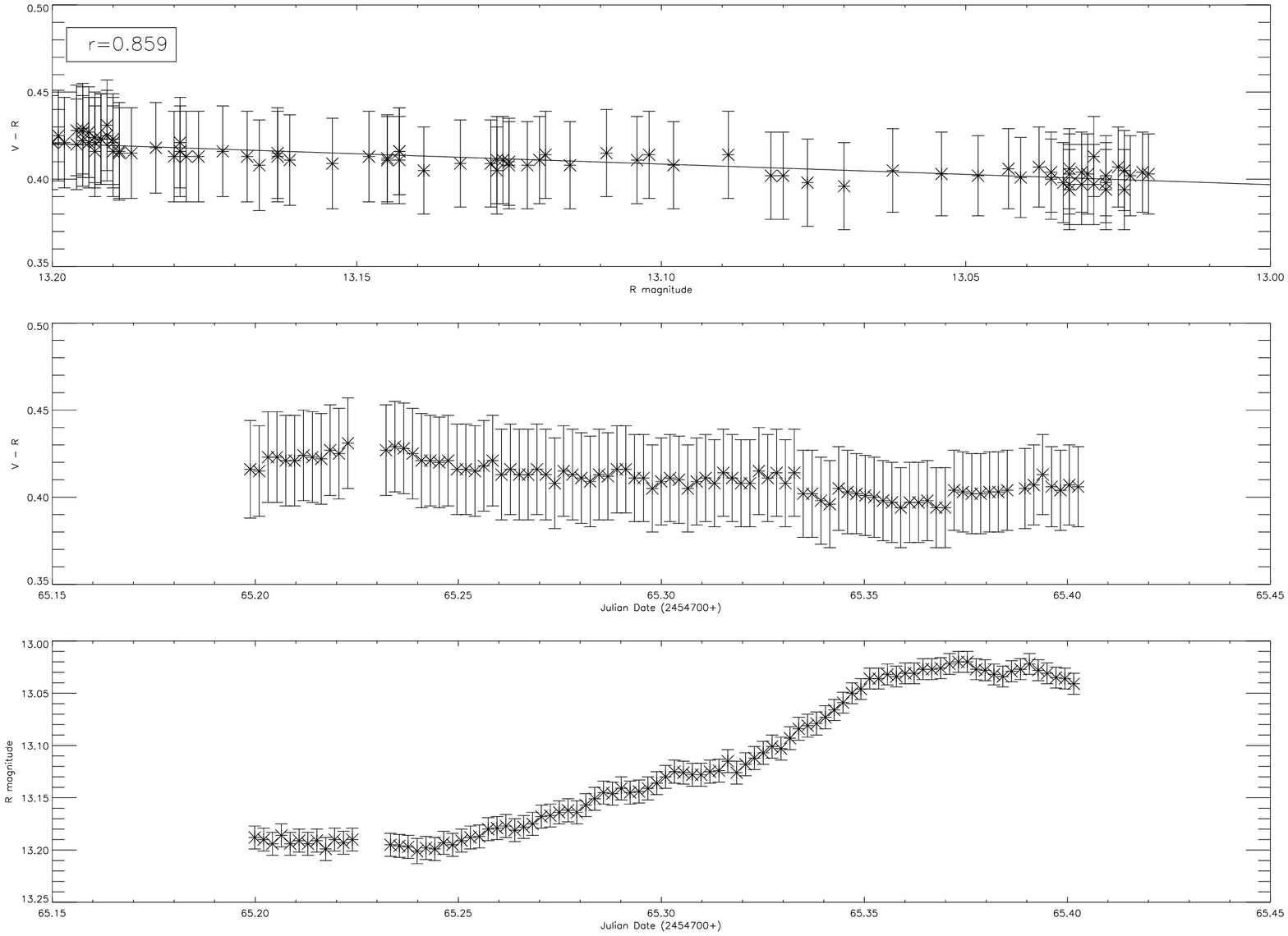}
\includegraphics[angle=90,width=0.5\hsize,height=0.6\hsize]{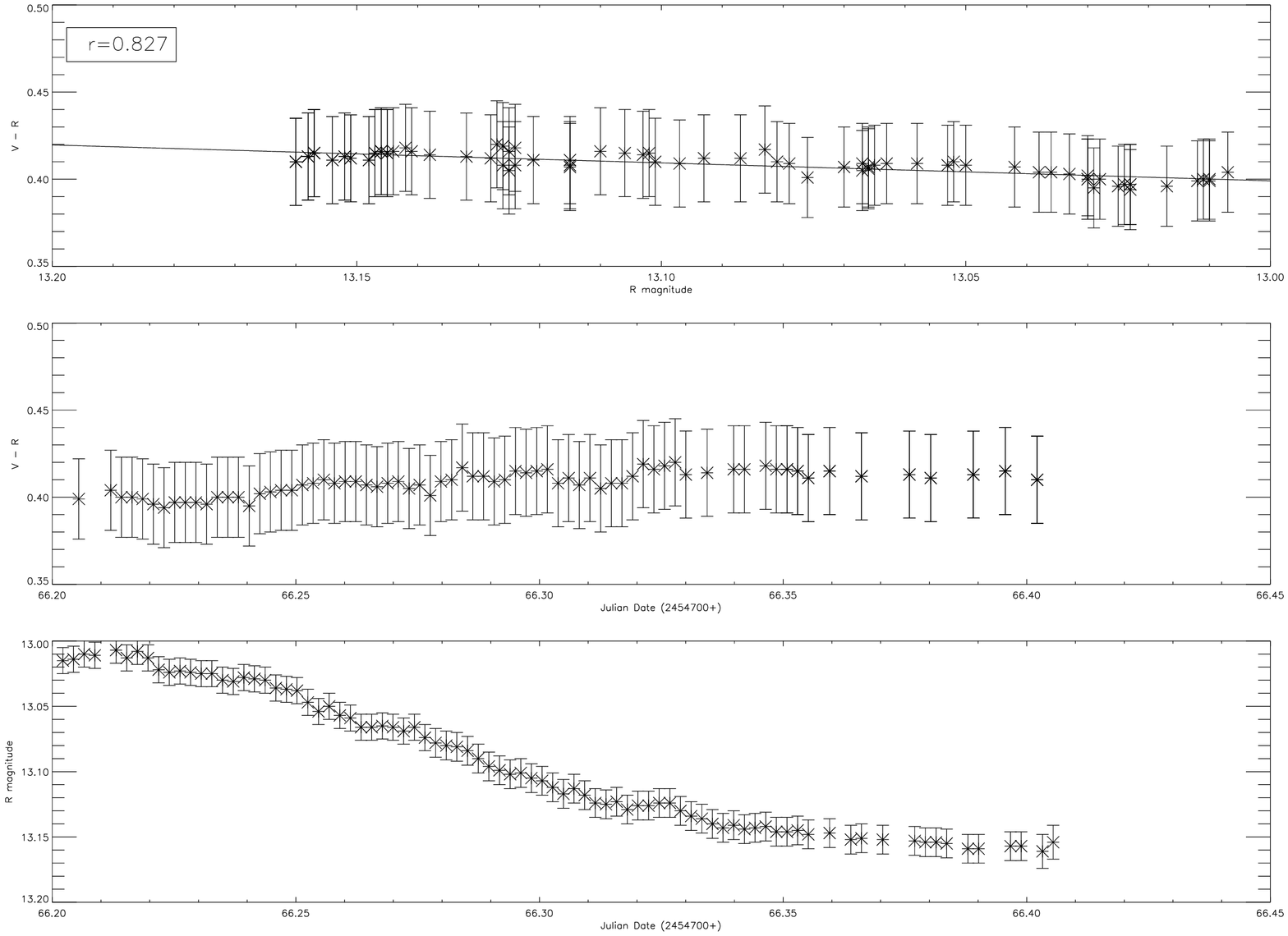}
\includegraphics[angle=90,width=0.5\hsize,height=0.6\hsize]{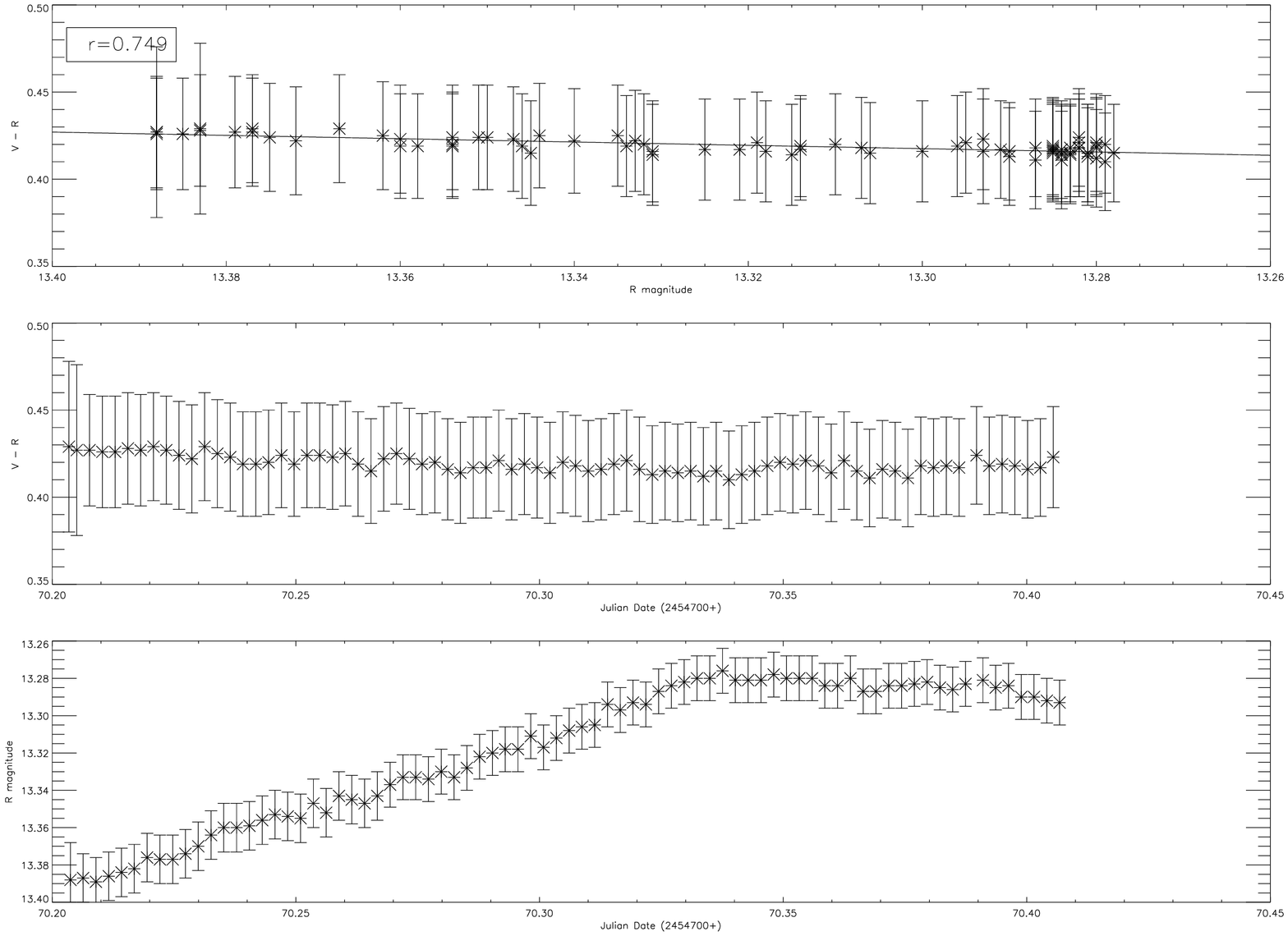}
\includegraphics[angle=90,width=0.5\hsize,height=0.6\hsize]{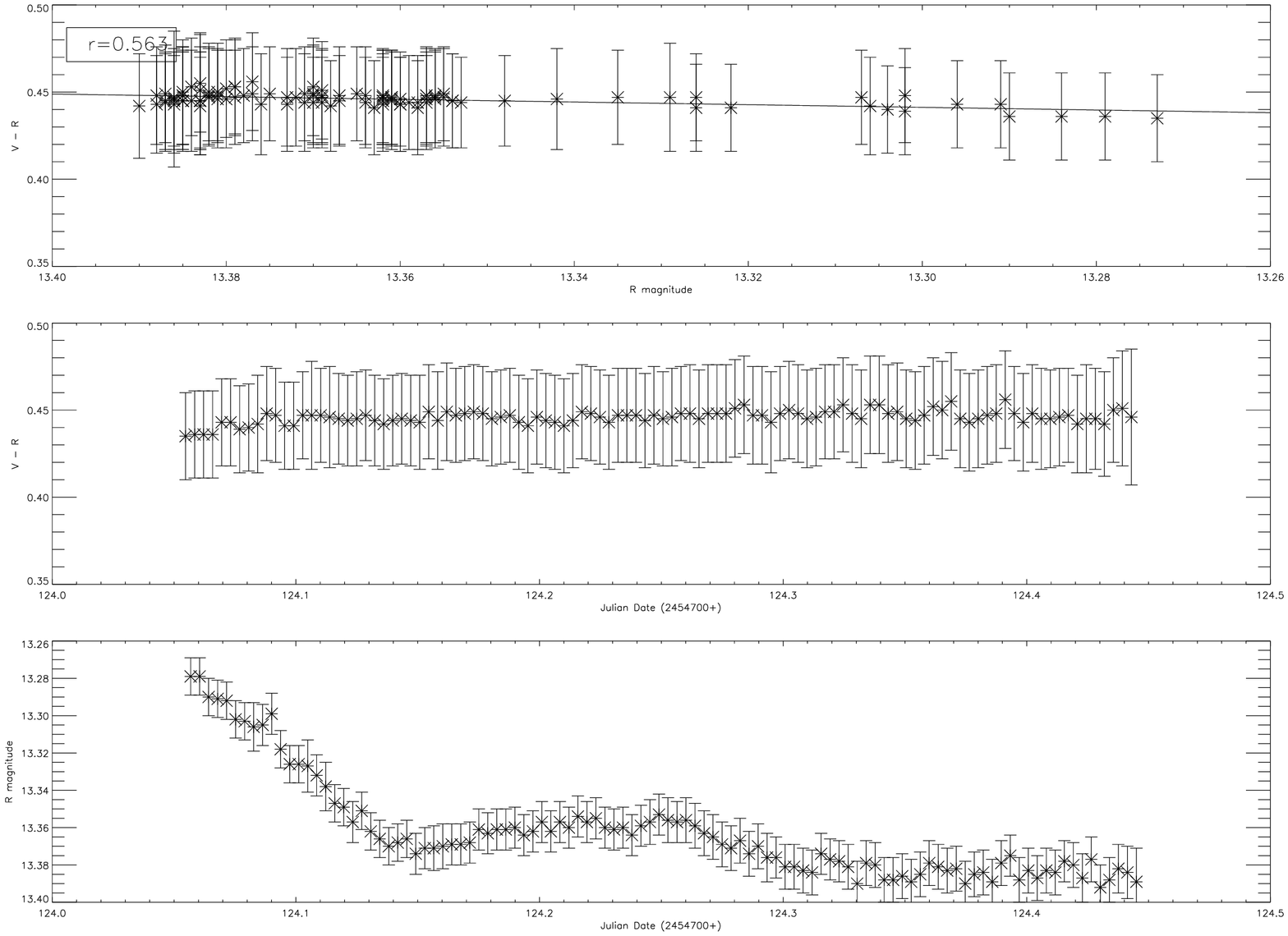}

\caption{Some examples of S5 0716 + 714 showing the relationship
between the \emph{V} - \emph{R} color index vs source brightness
(top panel), color index vs time (middle panel) and the
corresponding brightness change with time (bottom panel). Solid
lines are the best fit to the data points. r indicates the linear
Pearson correlation coefficient of the best fit. Date of
observations are 25-10-2008 (top left), 26-10-2008 (top right),
30-10-2008(bottom left), 23-12-2008 (bottom right).}

\end{figure}
\begin{figure}
\includegraphics[angle=90,width=1.0\hsize,height=0.7\hsize]{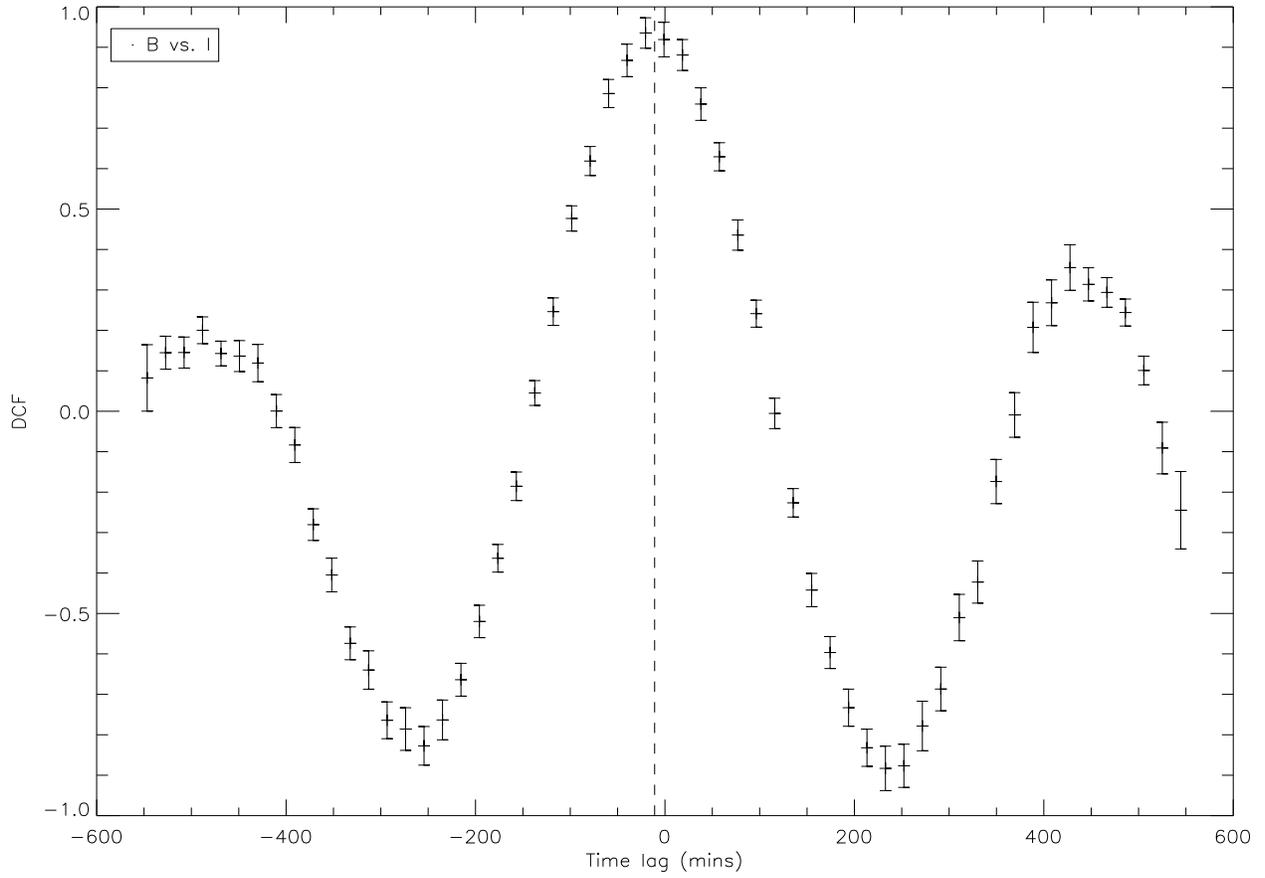}

\caption{Discrete correlation functions between \emph{B} and
\emph{I} bands. The dashed line indicates the centroid.}
\end{figure}

\begin{figure}
\includegraphics[angle=90,width=0.5\hsize,height=0.4\hsize]{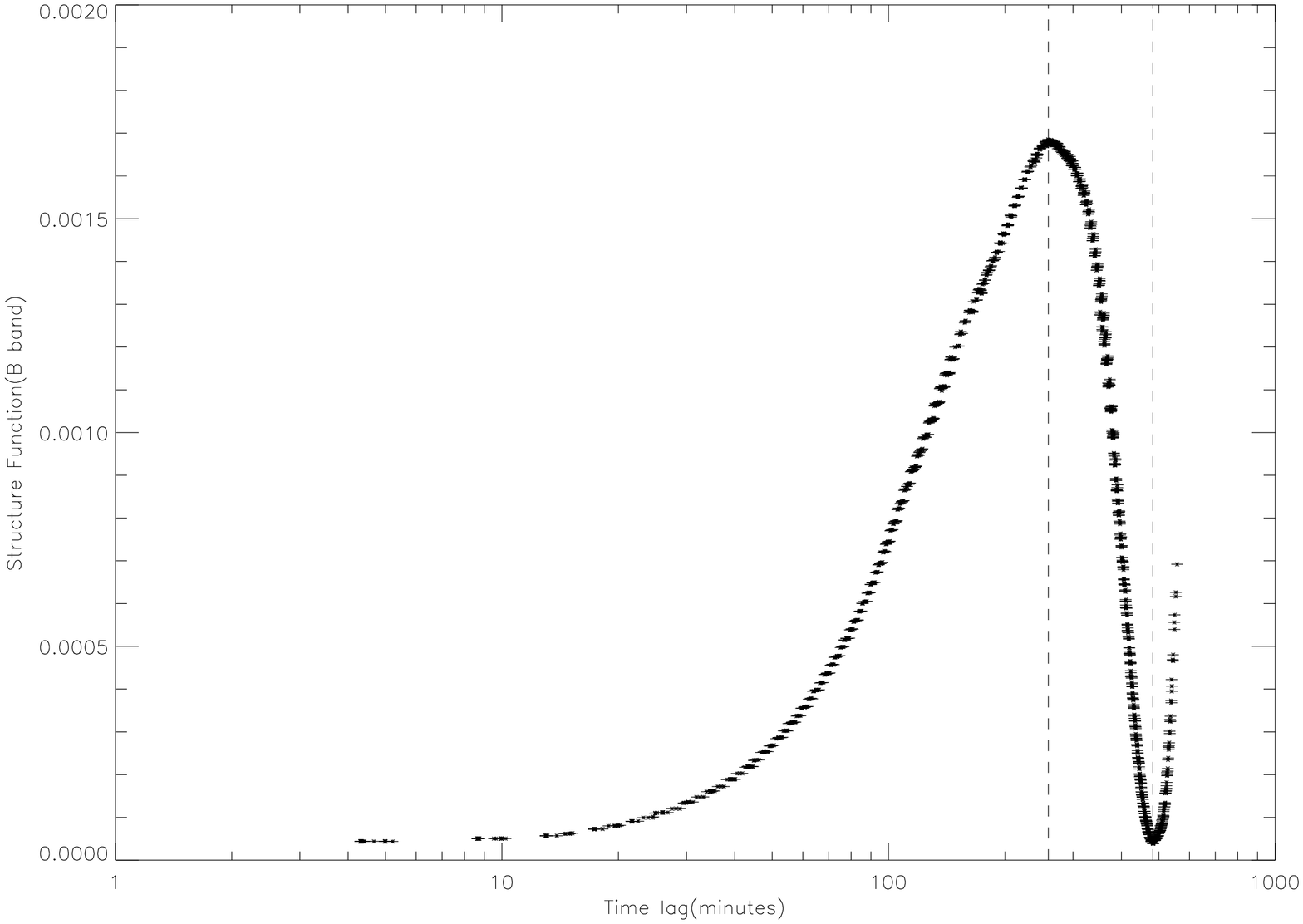}
\includegraphics[angle=90,width=0.5\hsize,height=0.4\hsize]{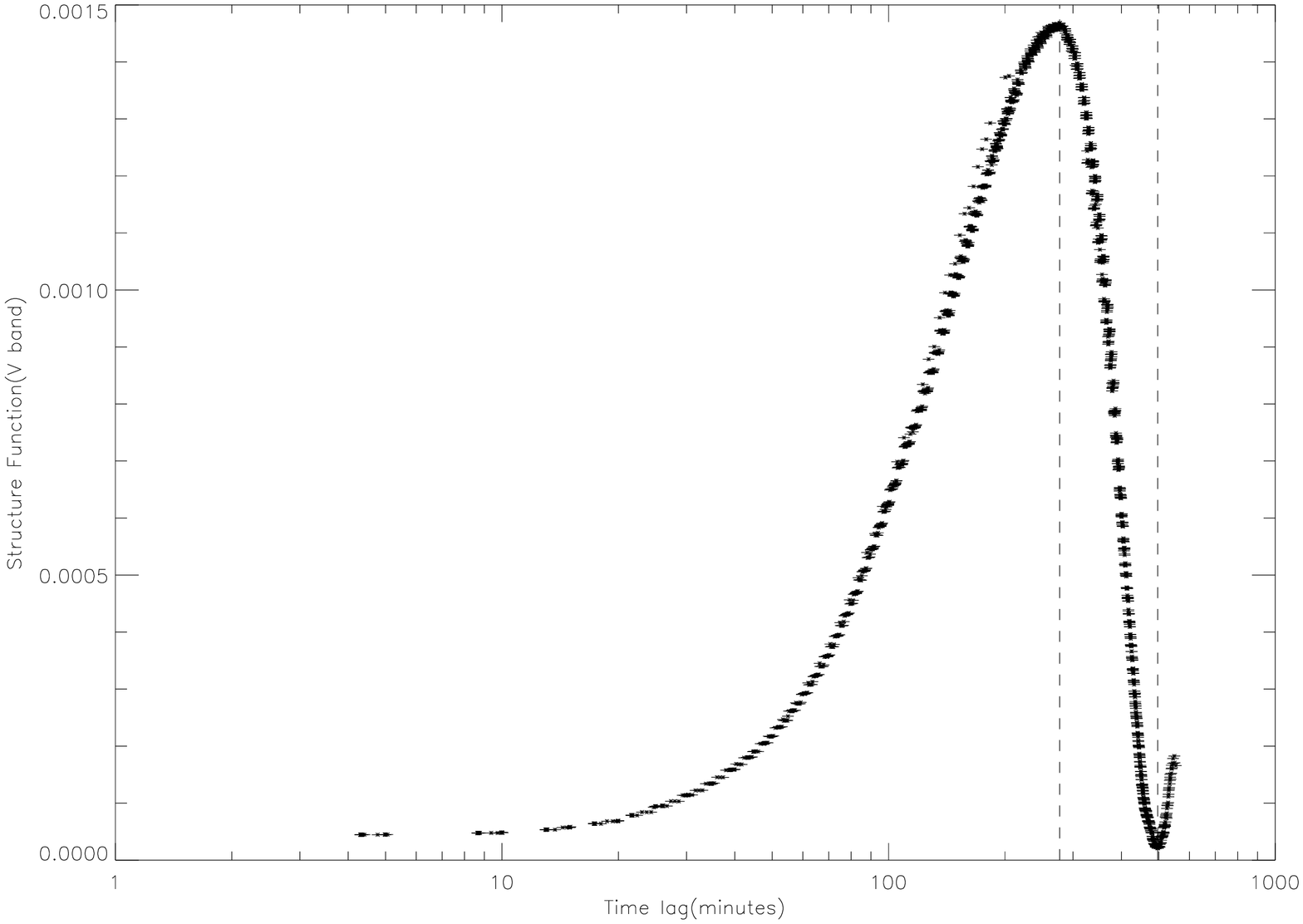}
\includegraphics[angle=90,width=0.5\hsize,height=0.4\hsize]{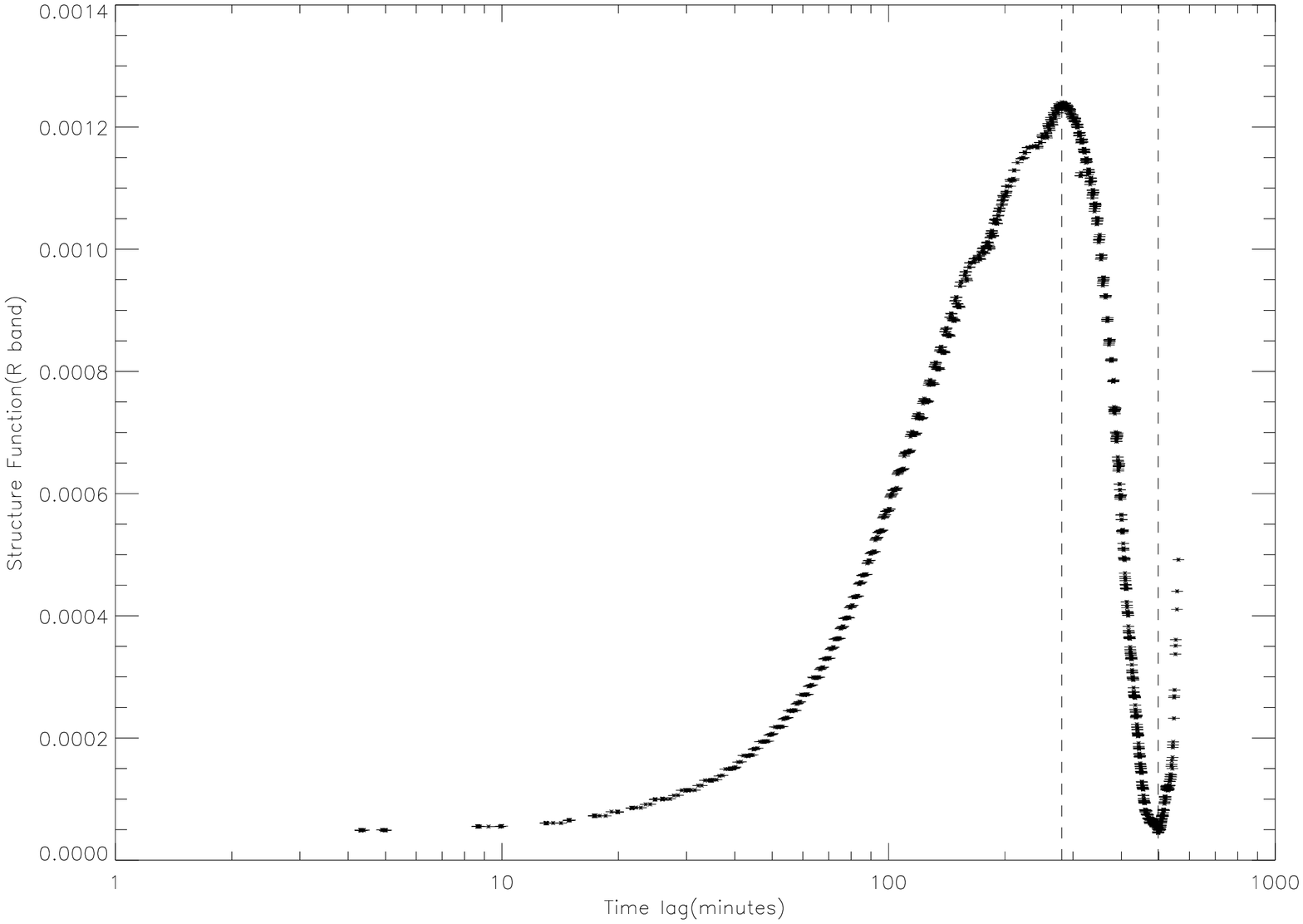}
\includegraphics[angle=90,width=0.5\hsize,height=0.4\hsize]{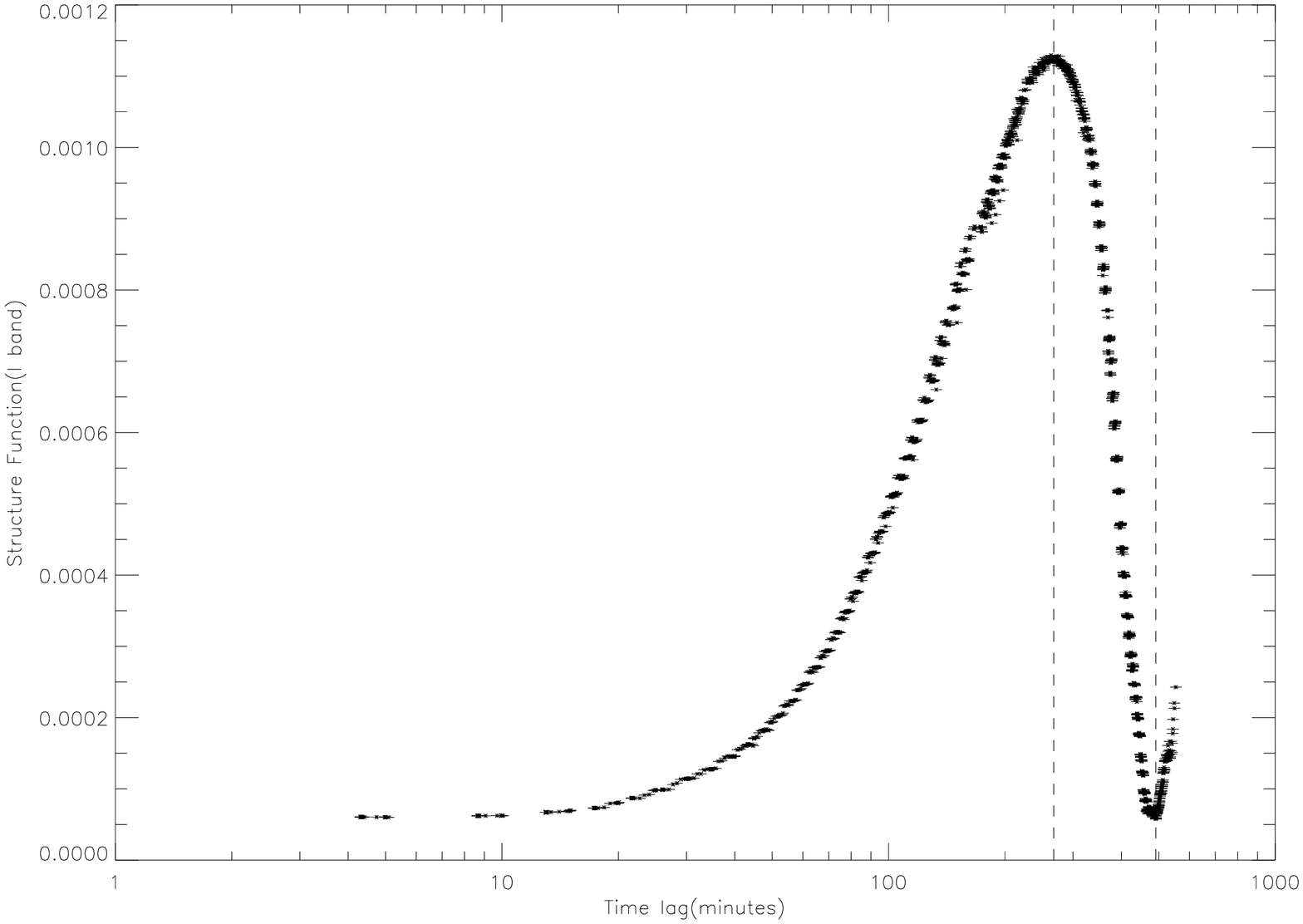}
\caption{Structure function of S5 0716+714 on JD 2454826 in
\emph{B}(top left), \emph{V}(top right), \emph{R}(bottom left) and
\emph{I}(bottom right) bands. The dashed lines indicate timescales
at the maxima and periods at the minima of the structure function.}
\end{figure}

\clearpage




\end{document}